\documentclass[pre,showkeys,11pt]{revtex4-1}
\usepackage{lineno}
\usepackage{amssymb}
\usepackage{amsmath}
\usepackage{mathrsfs}
\usepackage{array}
\usepackage{subfigure}
\usepackage{graphicx}
\usepackage{multirow}
\usepackage{placeins}

\def\etaD{\kappa_{\text{D}}}
\def\betaD{\bar\kappa_{\text{D}}}

\def\Rvoi{R_{\text{SC}}}
\def\Rmoy{R_{\text{ag}}}
\def\Rquad{R_{\text{quad}}}

\def\GS{\mathcal{S}}
\def\beij{\langle \eta_{ij} \rangle}
\def\avCCN{\langle CCN \rangle}

\def\Lmax{L_{\text{max}}}
\def\bLmax{\bar L_{\text{max}}}

\begin{document}

\title{Structure of sticky-hard-spheres random-aggregates: the viewpoint of contact coordination and tetrahedra}
\date{}
\author{Bl\'etry, M.}
\email[Corresponding author ]{bletry@icmpe.cnrs.fr}

\author{Russier, V.}
\author{Barb\'e, E.}
\affiliation{ICMPE-CNRS, 2-8 rue Henri Dunant, 94320 Thiais, France}
\author{Bl\'etry, J}
\affiliation{Honorary Professor, Bahia Blanca University, Argentina}

\begin{abstract}
We study more than $10^4$ random aggregates of $10^6$ monodisperse sticky hard spheres each, generated by various static algorithms. Their packing fraction varies from 0.370 up to 0.593. 
These aggregates are shown to be based on two types of disordered structures: random regular polytetrahedra and random aggregates, the former giving rise to $\delta$ peaks on pair distribution functions. 
Distortion of structural (Delaunay) tetrahedra is studied by two parameters, which show some similarities and some differences in terms of overall tendencies. Isotropy of aggregates is characterized by the nematic order parameter.
The overall structure is then studied by distinguishing spheres in function of their contact coordination number (CCN). 
Distributions of average CCN around spheres of a given CCN value show trends that depend on packing fraction and building algorithms. 
The radial dependency of the average CCN turns out to be dependent upon the CCN of the central sphere and shows discontinuities that resemble those of the pair distribution function.
Moreover, it is shown that structural details appear when the CCN is used as pseudo chemical parameter, such as various angular distribution of bond angles, partial pair distribution functions, Ashcroft-Langreth and Bhatia-Thornton partial structure factors. These allow distinguishing aggregates with the same values of packing fraction or average tetrahedral distortion or even similar global pair distribution function, indicative of the great interest of paying attention to contact coordination numbers to study more precisely the structure of random aggregates.
\end{abstract}

\keywords{Sticky hard spheres ; random packings ; contact coordination number ; Delaunay tesselation ; tetrahedra distorsion ; distribution function of isocoordinated spheres ; Ashcroft-Langreth partial structure factor ; Bhatia-Thornton partial structure factor }
\maketitle

\section{Introduction}
Random aggregates of monodisperse spheres are of great interest to simulate various physical systems, such as amorphous solids, liquids, powders, etc. In particular, a fine characterization of their local or longer range structural properties is of interest to classify the various types of random aggregates of spheres that can exist and study their properties. 

Many different approaches allow building random aggregates of spheres. Broadly, they may be divided into two categories. The first one consists of algorithms for which all spheres are introduced at once and then the system relaxes towards some more or less disordered structural state (eg molecular dynamics \cite{AW59}, Lubachevsky-Stillinger algorithm \cite{LS90}, Jodrey-Tory algorithm \cite{JT85} or Monte-Carlo relaxation of chains of hard-sphere \cite{KL08}). This family of algorithms produces aggregates whose properties mimic what was found in many disordered systems, notably the dependency of the contact coordination number with packing fraction and they are also able to produce aggregates with packing fraction equal or even superior to the random close packing (RCP) value ($\gamma_{RCP} \approx 0.64$) or equal to or lower than the random loose packing value (RLP, $\gamma_{RLP} \approx 0.555$ see \cite{OL90}). However, $\gamma_{RLP}$ presents a strong dependency on interparticles interactions \cite{DYZY06} which suggests that no geometrical property must be associated to the RLP transition, whereas the RCP state might have structural or geometrical constraints that set an upper bound to $\gamma_{RCP}$, such as the proportion of spheres involved in quasi perfect tetrahedra, as observed by Anikeenko and Medvedev \cite{AM07}.

The second large category of algorithms introduces spheres in the aggregate one by one and sets their position definitely at once (see, e.g., \cite{JPM92}). Aggregates built by such methods have two important differences with those of the first category: first, their average contact coordination number remains close to 6, second, to the best of our knowledge, it is impossible to produce aggregates with packing fraction higher than 0.6. Hence, the latter family of algorithms gives access to random aggregates with somewhat different structural properties from the former. Moreover, using sticky hard spheres, contact neighbours are rigorously defined by Dirac $\delta$ functions and this allows studying various properties related to the coordination number of spheres in a simple and non ambiguous manner.

In \cite{BB15}, a first study of several hundreds sequentially built aggregates was conducted. It was found that for the lowest packing fractions, $\delta$ peaks appear on the pair distribution function, which corresponds to the formation of a disordered polytetrahedral structure in the random packing aggregate. 

The present study extends the results obtained in \cite{BB15} by looking at structural tetrahedra, isotropy, contact coordination numbers, partial distribution functions as well as partial structure factors, corresponding to pairs of spheres with equal or different contact coordination numbers. It also introduces a new family of random aggregates formed by sphere added in regular building tetrahedra, which appears essential to understand the composite nature of other sequentially built random aggregates.

\section{Studied aggregates}
More than 10000 aggregates, each one containing $10^6$ spheres with radius $r_s=1$ (and, hence, diameter $d=2$) were built and studied\footnote{Hereafter, unless otherwise mentioned, all distances are expressed in $r_s$ as the unit of length ($r_s = 1$). Similarly, $r_s^{-1}$ is the unit in reciprocal space.}. They fall into two broad families: random irregular polytetrahedral aggregates (RIPA) and random regular polytetrahedral aggregates (RRPA). All of them were built by adding spheres one by one at their final position, tangentially to three existing ones.

\subsection{Random irregular polytetrahedral aggregates}
Most aggregates were of RIPA type. They were built using algorithms that have been presented in details in \cite{BB15} and are only summarized here. 
A seed of three spheres forming an equilateral triangle is used. Each new sphere $P$ is positioned tangentially to at least three already present spheres (noted $O$, $A$ and $B$). The new sphere can be introduced in a hole whose size is maximized (MAX algorithms) or chosen randomly (RAN algorithms) in the vicinity of the local origin $O$, itself chosen randomly in the aggregate. Moreover, it is possible to insert from 1 up to 9 spheres at once (according to the index $NINS$, for {\em Number of sphere INSerted}) around a given origin $O$ (algorithms MAX 1 to 9 and RAN 1 to 9). Finally, the neighbourhood explored to choose spheres $A$ and $B$ around $O$ is a cube whose edge length $a$ can be varied: it controls strongly the packing fraction of the aggregates. The larger the value of $a$, the higher the packing fraction. It is varied between about 3.4 and $8$ as, for $a<3.4$, no aggregate can be generated and for $a>8$, the maximum packing fraction is reached and no evolution of the generated aggregates is noted for higher values of $a$ (see figure 2 in \cite{BB15}).
More than 300 aggregates of $10^6$ spheres were built, by varying $a$ for each family of these algorithms. 

An additional modification with regard to the aggregates studied in \cite{BB15} was to choose the origin $O$ as close as possible from the center (0,0,0) of the growing aggregate (RMIN algorithms), instead of a purely random fashion. This change has the effect of increasing the maximum packing fraction, reached for (MAX-1, $a>3.5$), of about 1~\%, from 0.586 to 0.593. All other aggregates generated by RMIN-MAX algorithms have a slightly higher packing fraction than their MAX counterparts (i.e. same values of $a$ and $NINS$) with no significant changes concerning the structural results presented in \cite{BB15}. This modification also allows for a more homogeneous growth of the aggregate. Only the results obtained for RMIN-MAX-1 will be used hereafter\footnote{Yet another family of algorithms has been developed, for which the newly added sphere is positioned by controlling the value of the building tetrahedra distortion. However, this strategy leads to the exact same $P(r)$ as the one obtained for RA built with the positioning in function of the hole size and has the same maximum packing fraction: these aggregates will not be further studied here.}.

Finally, as it turned out that the previous families of aggregates had some inhomogeneities of their packing fraction close to (0,0,0) (see hereafter), some aggregates were generated by using as seed a set of $N$ spheres taken in previously built aggregates, instead of an equilateral triangle (N-RMIN-MAX-1 algorithm). The positions of the spheres composing the seed are taken from an aggregate with the same value of other parameters ($a$ and $NINS$), far from the origin (0,0,0), which has the effect to remove the central area with higher packing fraction. Typically, the number of spheres in the seed is between 30 and 600. However, this modification entails only slight changes of pair distribution functions or structure factors.

\subsection{Random regular polytetrahedral aggregates}
A last algorithm has been used, that produces aggregates with only regular building tetrahedra \cite{B79}. In this case, the newly inserted sphere $P$ forms a regular tetrahedron with the three already contacting spheres $O$, $A$ and $B$, {\em i.e.} $PO = PA = PB = OA = OB = AB = 2$. For these algorithms, $a$ has virtually no impact on packing fraction, but the number of inserted spheres around a given local origin (NINS) does. 

Once again, the local origin $O$ can be chosen at random, or as the closest one from the center of the aggregate. When it is chosen randomly (RRPA), the maximum packing fraction (0.418) is reached for $NINS = 3$ and a minimum of 0.408 is observed for $NINS = 1$ and beyond 4. 
When $O$ is chosen as the closest possible origin from the aggregate center (RMIN-RRPA), the maximum packing fraction is 0.456 and is reached for $NINS>4$. 
Similar aggregates, so-called saturated polytetrahedra, have been studied by Medvedev and Pilyugina \cite{MP08}. They found a packing fraction of 0.435 for aggregates consisting of about 576~000 spheres. This value falls in between the maximum ones obtained for aggregates choosing $O$ randomly (lower bound) and those taking $O$ as the closest sphere from the origin (higher bound).

\vspace{5mm}

It should be noted that RIPA and RRPA distinguishes aggregates based on their building algorithms, not their structure, which will be discussed in details below.


The isotropy and the randomness of all aggregates has been systematically checked through the distribution of $i-j$ bonds and the nematic tensor formalism, and turns out to be satisfactory. More details on that latter point are provided in supplementary informations.

\section{Packing fraction}
\subsection{Basic relations}
The radius $R_m$ of a large spherical aggregate centered in (0,0,0) and made of $N$ spheres centered in $\vec R_i$ is given to a good approximation by \cite{B79}:
\begin{equation}
\Rmoy = \sqrt{\Rquad^2}
\end{equation}
where $\Rquad$ is the average quadratic radius of all spheres in the aggregate:
\begin{equation}
\Rquad^2 = \frac{5}{3}\frac{1}{N}\sum_{i=1}^N R_i^2
\end{equation}
However, finite aggregates are not fully spherical and exhibit local order oscillations. Therefore, their packing fraction varies as a function of the radius $\Rvoi$ of the sphere cut into the aggregate bulk and deserves special attention. 

The volume $V_s(r)$ shared by a sphere of radius $r_s=1$ whose center is at a distance $r$ from the origin ($0,0,0$) with another sphere $\GS$, of radius $\Rvoi$, centered in ($0,0,0$) is \cite{sphsph}:
\begin{align}
\label{eqvolSs}
V_s(r,\Rvoi) &= \frac{4}{3}\pi r_s^3                                                & r\leq\Rvoi-r_s\nonumber \\
V_s(r,\Rvoi) &= \pi\frac{(\Rvoi+r_s-r)^2(r^2+2rr_s-3r_s^2+2r\Rvoi+6\Rvoi r_s-3\Rvoi^2)}{12r}& \Rvoi-r_s < r < \Rvoi+r_s\\
V_s(r,\Rvoi) &= 0                                                             & r\geq\Rvoi+r_s\nonumber
\end{align}
These relationships can also be used for any sphere $\GS$ centered in $x$, $y$ and $z$ by a mere change of reference frame. 

Hence, the packing fraction of the sphere $\GS$ can be directly determined for any radius $\Rvoi$, as:
\begin{equation}
\gamma (\Rvoi) = \frac{\sum_{i=1}^N V_{s}(r_i,\Rvoi)}{\frac{4}{3}\pi\Rvoi^3}
\end{equation}
where $i$ accounts for all spheres in the aggregate.

Finally, it is possible to determine the packing fraction of shells of arbitrary thickness $w = R_o-R_i$, where $R_i$ is the inner radius and $R_o$, the outer radius of the shell, simply by removing the portion of spheres outside of the shell, according to the volume complementary of relation \ref{eqvolSs}.

\subsection{Packing fraction of spheres inscribed in the aggregate}
\begin{figure}[htbp]
\subfigure[]{\includegraphics[width=0.48\textwidth]{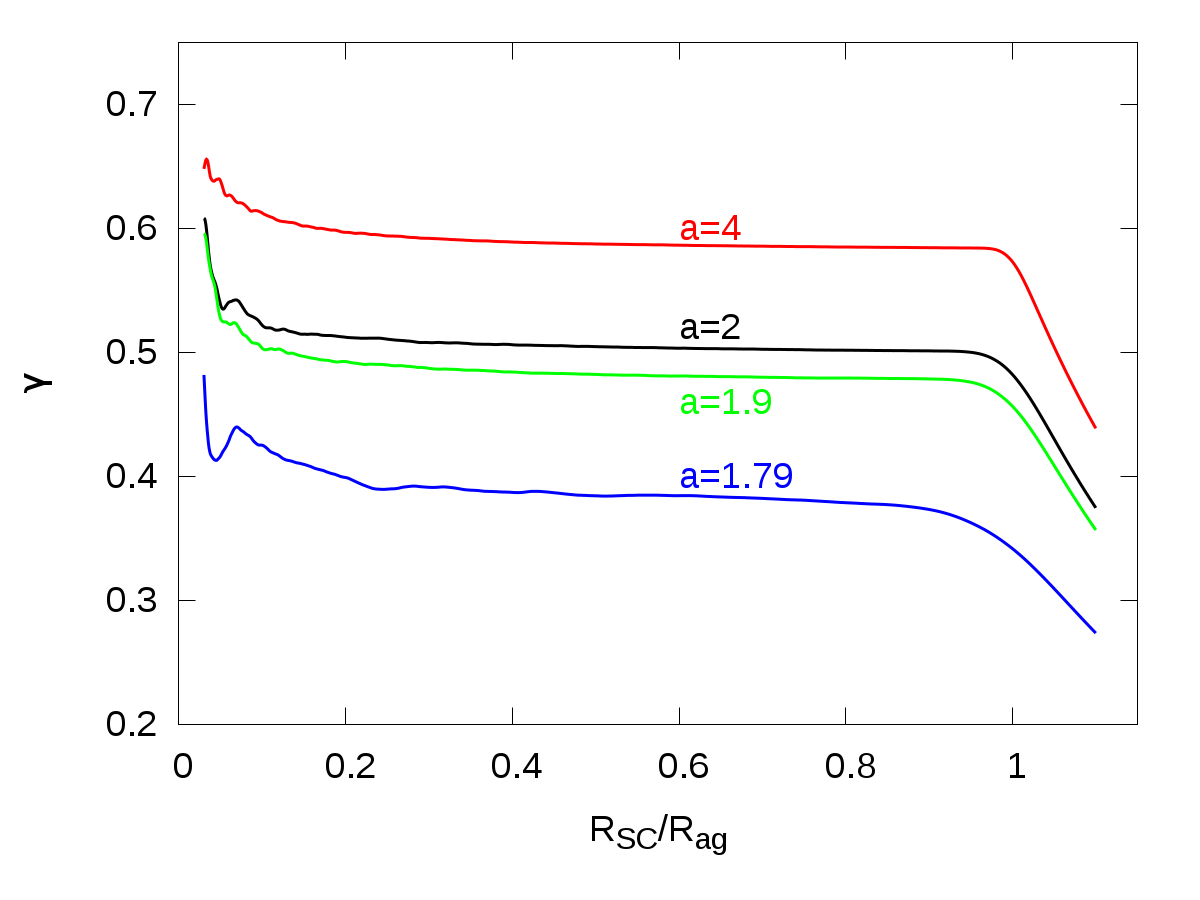}}
\subfigure[]{\includegraphics[width=0.48\textwidth]{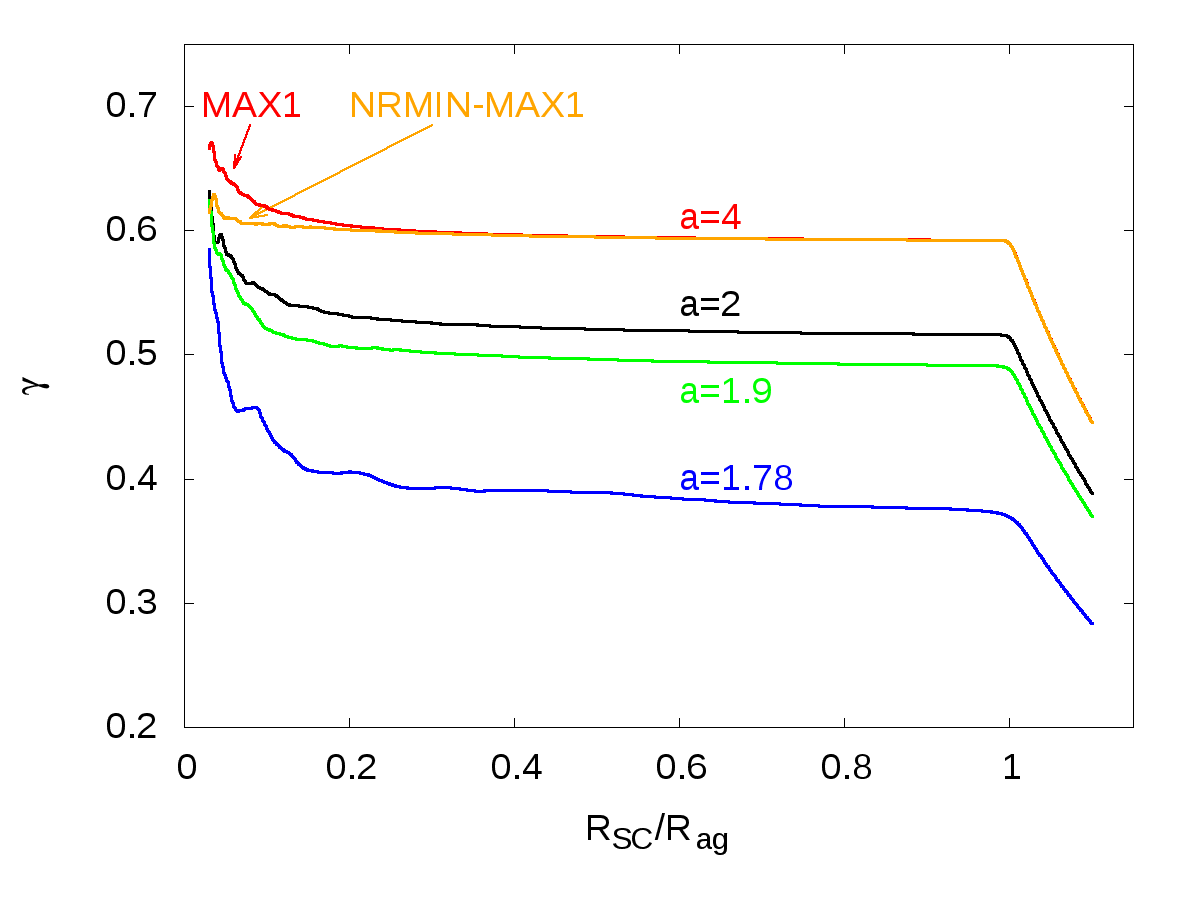}}
\caption{ $\gamma(\Rvoi/\Rmoy)$ for several a) MAX-1 aggregates and b) RMIN-MAX-1 aggregates built with various values of $a$ and a single N-RMIN-MAX-1 aggregate (seed composed of 600 spheres), for the sake of comparison, with $\GS$ centered in $0,0,0$,.\label{DfRvoi}}
\end{figure}

The evolution of the packing fraction of spheres inscribed in the aggregate as a function of their radius $\gamma = f(\Rvoi/\Rmoy)$ (figure \ref{DfRvoi}) shows that, for all aggregates, whatever their building algorithm, the packing fraction decreases slightly when $\Rvoi$ increases.

A seed effect appears as for MAX-1 and RMIN-MAX-1 aggregates (3 spheres forming an equilateral triangle as seed), a first regime of fast decrease is observed for $\Rvoi < 0.2\Rmoy$, i.e. for a number of spheres below approximately 8000, then the packing fraction tends to plateau whereas for N-RMIN-MAX-1 aggregates (seed consisting of spheres taken far from the origin in a previously generated aggregate), this initial decrease is much faster. Nevertheless, the exact range of effect of the seed can only be asserted by the packing fraction of shells studied hereafter.

This effect probably stems from the fact that contacting equilateral-triangle configurations are extremely rare in high packing fraction aggregates. As a matter of fact, this seed dependency disappears for lower packing fraction aggregates, in which such configurations are rather frequent.

On the other hand, N-RMIN-MAX-1 and RMIN-MAX-1 aggregates show exactly the same behaviour for larger values of $\Rvoi$.

For every aggregate, a second regime is observed when $\Rvoi\to\Rmoy$, i.e. when $\GS$ reaches the limit of the aggregate: logically, the packing fraction decreases faster. For a perfectly spherical aggregate, when $\Rvoi>\Rmoy+r_s$, then the "packing fraction" of $\GS$ should decrease as $\Rvoi^{-3}$. When this decrease is at first more progressive, it shows that the aggregate has an imperfect shape and either has protuberances on its surface or is not overall perfectly spherical. 

For the same value of $a$, RMIN-MAX-1 aggregates tend to have a higher packing fraction than the corresponding MAX-1 aggregates, as well as a sharper decrease of $\gamma$ when $\Rvoi \to \Rmoy$, which shows that RMIN-MAX-1 aggregates have a more regular surface than MAX-1 aggregates. For the latter, the thickness of "imperfect aggregate" is about $1.5d$ for $a=3.5$ ($\gamma = 0.586$) and about $5d$ for $a=1.79$ ($\gamma = 0.370$). For RMIN-MAX-1 aggregates, the thickness of imperfect aggregate is roughly $d/4$ for $a=3.5$ ($\gamma = 0.593$) and about $1.5d$ for $a=1.78$ ($\gamma = 0.378$).

\subsection{Packing fraction of shells}
\begin{figure}[htbp]
\subfigure[]{\includegraphics[width=0.48\textwidth]{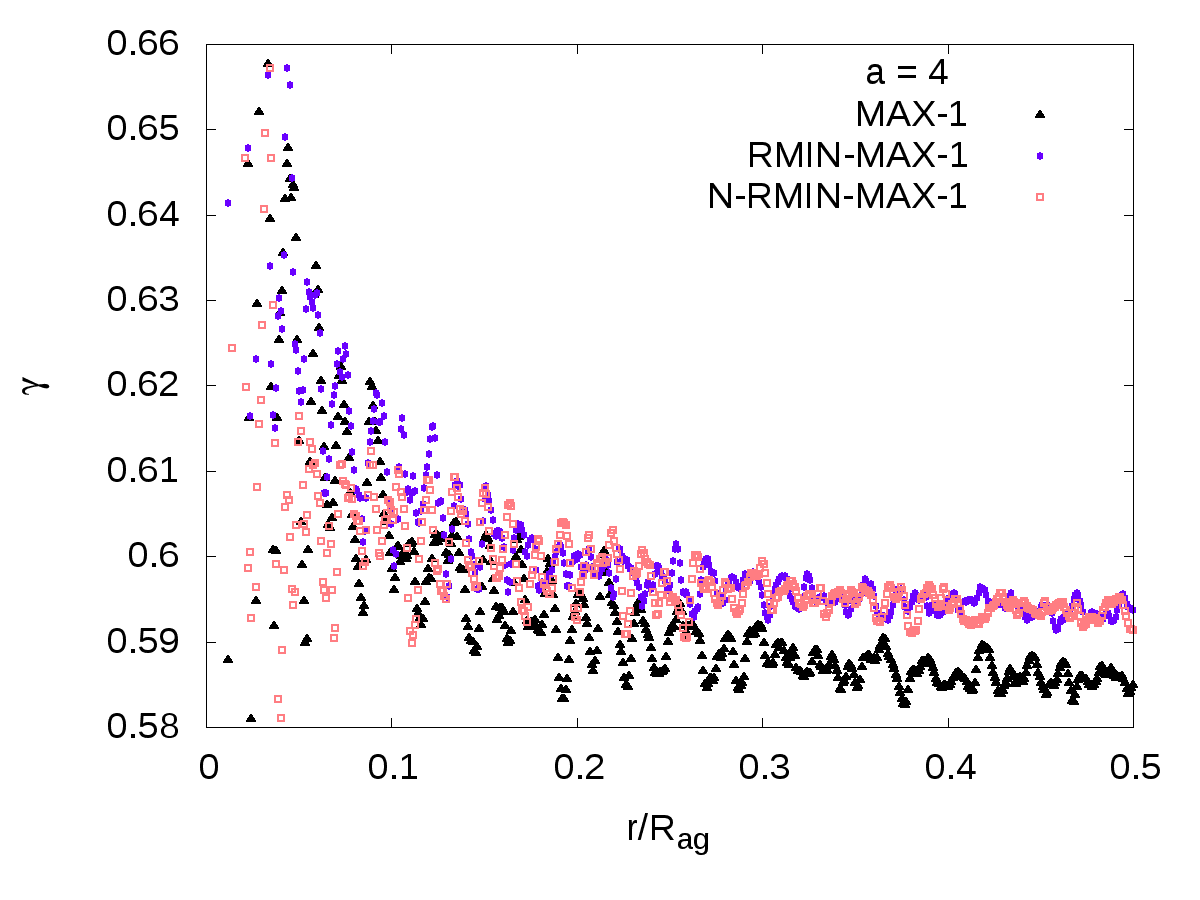}}
\subfigure[]{\includegraphics[width=0.48\textwidth]{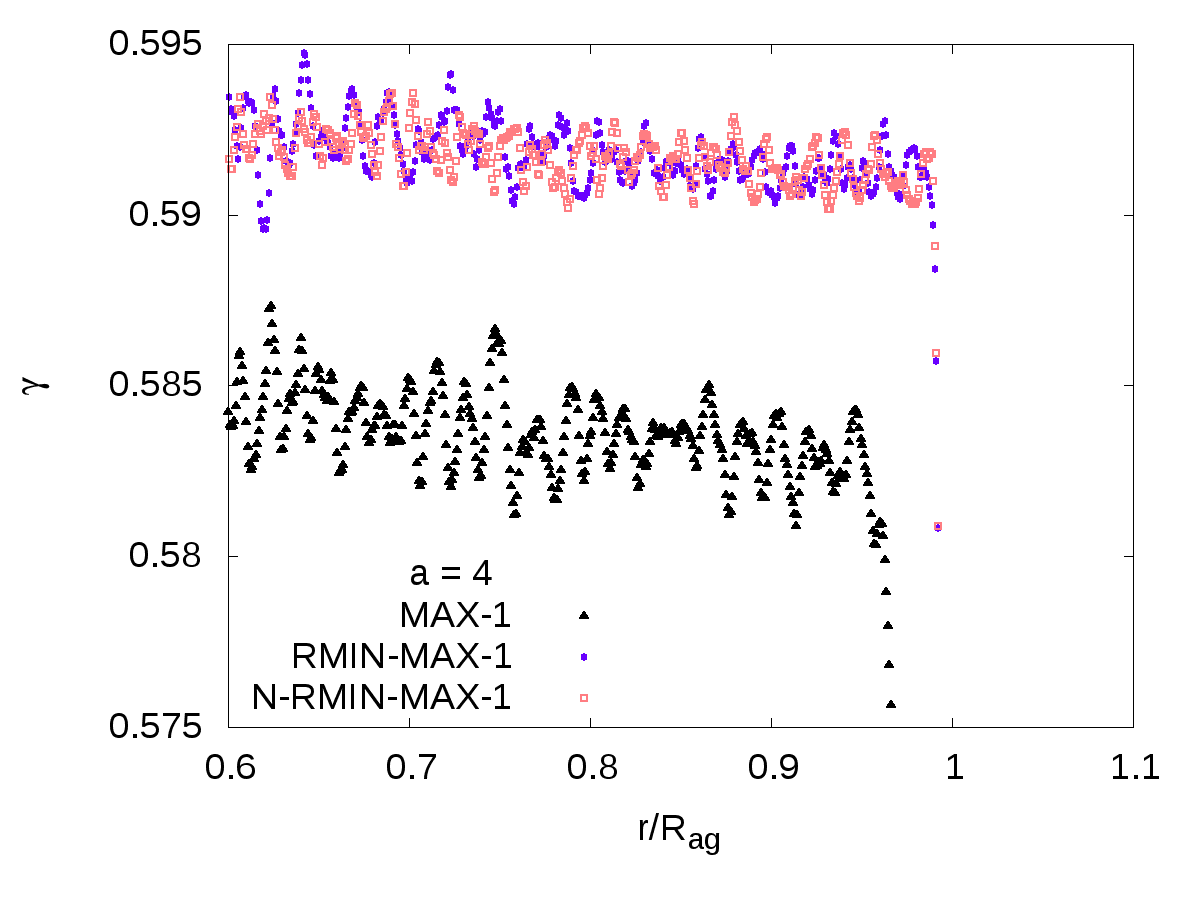}}
\caption{Packing fraction of shells with thickness $w=0.1 r_s$ on the densest aggregates built with three strategies: MAX-1 (3 spheres seed), RMIN-MAX-1 (3 spheres seed, RMIN) and N-RMIN-MAX-1 with a spherical seed of radius $10$ extracted from a previously generated aggregate with the same value for $a$ and $NINS$ a) for relatively small $r$ ($r/\Rmoy \in[0;0.5]$) and b) for large $r$ ($r/\Rmoy \in[0.5;1]$).\label{DShell_r}}
\end{figure}

\begin{figure}[htbp]
\subfigure[]{\includegraphics[width=0.48\textwidth]{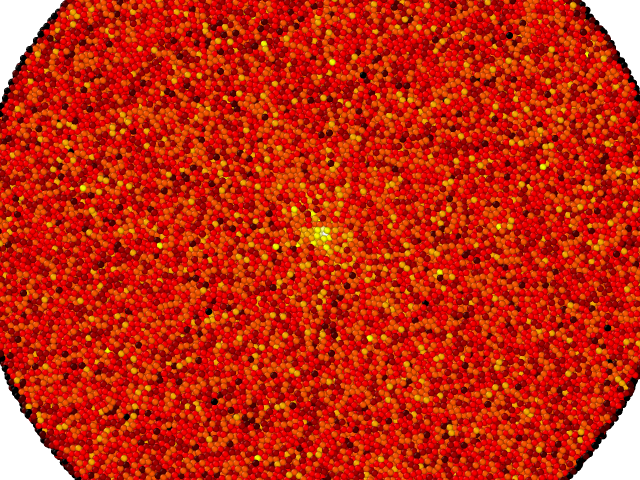}}
\subfigure[]{\includegraphics[width=0.48\textwidth]{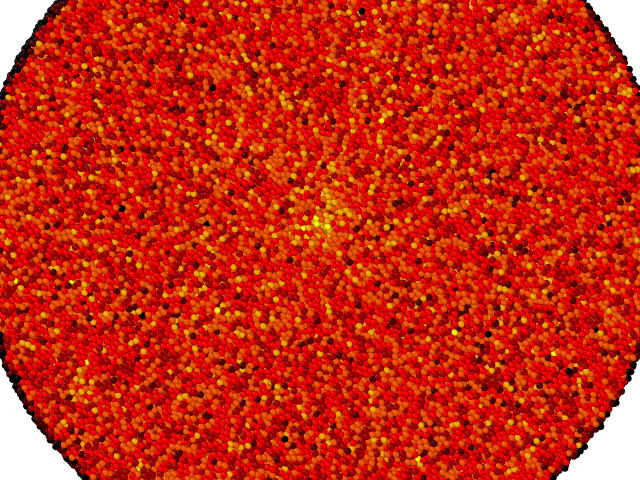}}
\caption{Slices of aggregates generated by two algorithms: a) RMIN-MAX-1 with $a=4$, b) N-RMIN-MAX-1 with $a=4$ and a seed consisting of $N=400$ spheres. Colors (gray scale) correspond to contact coordination number: brighter spheres have a higher CCN. (These figures were generated with Ovito software \cite{S10}.)\label{ovitseed}}
\end{figure}

Figures \ref{DShell_r}.a and b represent the variation of packing fraction in shells with thickness $w=0.1$ for the densest aggregates produced by algorithms MAX-1, RMIN-MAX-1 and N-RMIN-MAX-1. Globally, the packing fraction of such shells oscillates with $r$. Its average value decreases from a higher value near the seed, to a smoother behaviour when $r$ increases. N-RMIN-MAX-1 aggregate presents virtually no effect of the seed: the packing fraction of shells reaches the average behaviour for very small values of $r$, which seems logical as for these aggregates the seed consists of a set of spheres with the average structure. RMIN-MAX-1 and N-RMIN-MAX-1 converge for $r\in [0.1\Rmoy,0.15\Rmoy]$: the effect of the initial equilateral-triangle seed of RMIN-MAX-1 aggregates seems then to act on about $10^3$ spheres in the whole aggregate, consisting of $10^6$ spheres, i.e. significantly less than suggested above by the comparison, in figure \ref{DfRvoi}.b, of $\gamma = f(\Rvoi/\Rmoy)$ for the two same aggregates. Figure \ref{ovitseed} compares two aggregates built by RMIN-MAX-1 and N-RMIN-MAX-1 algorithms, where spheres are colored based on their CCN. It appears that the former aggregate has a brighter atypical central area in the region of the seed, denoting unusual structural properties with respect to the rest of the aggregate, whereas the latter displays a seed area much more similar to the rest of the aggregate.

Moreover, for aggregates with lower packing fraction (i.e. for aggregates built using lower values of $a$ and in which regular polytetrahedra appear), the range of aggregate affected by the seed decreases and completely disappears for the lowest packing fraction aggregates. In that case, indeed, the structure contains a significant amount of equilateral triangles and the initial seed ceases to be special in comparison with the rest of the structure. 
At large values of $r$ (figure \ref{DShell_r}.b), oscillations can still be detected in the packing fraction of shells, however with a much smaller amplitude. A slight decrease of the average value is noticeable: the farther a shell is from the center of the aggregate, the lower its packing fraction, on average. The origin of this phenomenon is not obvious to us. For $r>\Rmoy$, the packing fraction of shells falls rapidly to 0.

\section{Tetrahedral structure}
The structure of random packings of spheres is commonly assessed via the tetrahedra connecting sphere centers, forming the so-called Delaunay tesselation \cite{D24} (these tetrahedra are noted by the subscript $_D$ in what follows). For the present study, Delaunay tessellations were built using the cgal library \cite{CGAL,cgal:pt-t3-16b}. 

In \cite{BB15}, another type of tetrahedra was studied, called building tetrahedra (noted by the subscript $_{BT}$ hereafter). A building tetrahedron is formed by spheres $O$, $A$, $B$ and $P$ when adding the new sphere $P$ tangentially to the three other ones, $O$, $A$ and $B$. 
It should be noted that such tetrahedra may or may not belong to Delaunay's tessellation.
The distortion of building tetrahedra was shown to be a very significant structural parameter, allowing the correlation of various structural traits of the aggregates. In this section, we focus on two distortion parameters of Delaunay tetrahedra.

\subsection{Distortion parameters}
The first tetrahedral distortion parameter has been defined for the characterization of building tetrahedra \cite{BB15} by relation:
\begin{equation}
\kappa_{BT} = \frac{3d^2 + OA^2 + OB^2 + AB^2}{6d^2}
\end{equation}
where $O$, $A$ and $B$ are the three sphere centers used to add the new sphere $P$ and the term $3d^2$ corresponds to the three necessary sphere contacts $PO,\ PA,\ PB$ imposed to building tetrahedra by the algorithm. The maximum value of $\kappa_{BT}$ is 2 and is obtained for a centered equilateral triangle with side $d\sqrt{3}$ of three spheres, with the additional sphere $P$ at its barycenter, while $\kappa_{BT}$ minimum value, 1, corresponds to a regular tetrahedron.

The definition of the parameter $\kappa_{BT}$ is immediately extended to Delaunay tetrahedra by relation:
\begin{equation}
\etaD = \sum_i\sum_{j>i}\frac{d_{ij}^2}{6d^2}
\end{equation}
where $i$ and $j$ are the vertices of the tetrahedron and $d_{ij}$ the vertices length. The smallest distance possible between sphere centers is $d_{ij}=d=2$, hence the smallest possible value is obtained for a regular tetrahedron and is $\etaD = 1$.

The last distortion parameter to be studied hereafter is the longest edge length ($\Lmax$) of the considered tetrahedron. The smallest possible value of $\Lmax$, is $d$, which is found in the case of a regular tetrahedron. The behaviour of $\Lmax$ has already been studied, along with others, notably by Anikeenko et al \cite{AMA08} on aggregates built using Jodrey-Tory (JT) algorithm \cite{JT85} and Lubachevsky-Stillinger (LS) algorithm \cite{LS90,SDST06}. Anikeenko et al \cite{AMA08} have found that $\Lmax$, in spite of its simplicity, shows a great consistency when compared with two other parameters, namely the edge differences and the procrustean distance.

\subsection{Distributions of distortion parameters}

\begin{figure}[htbp]
\subfigure[]{\includegraphics[width=0.48\textwidth]{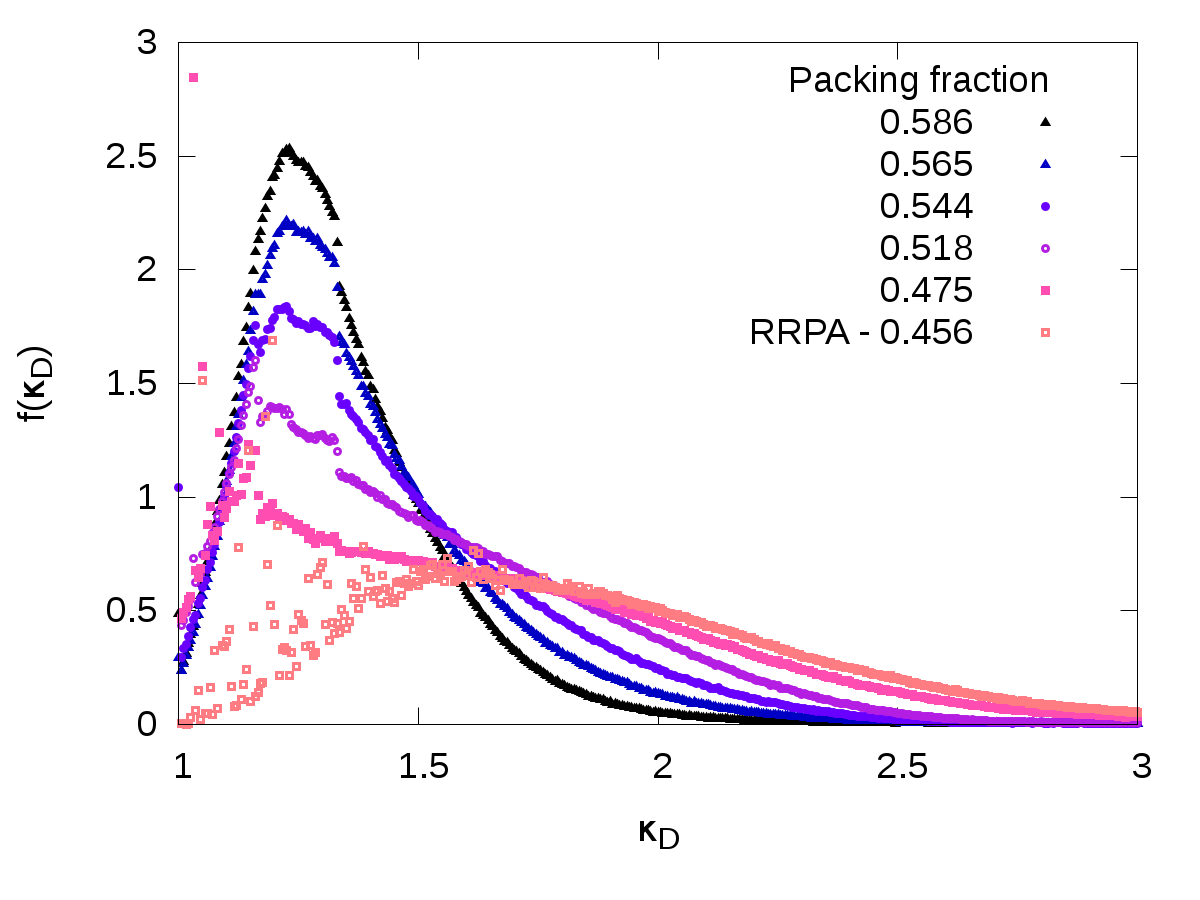}}
\subfigure[]{\includegraphics[width=0.48\textwidth]{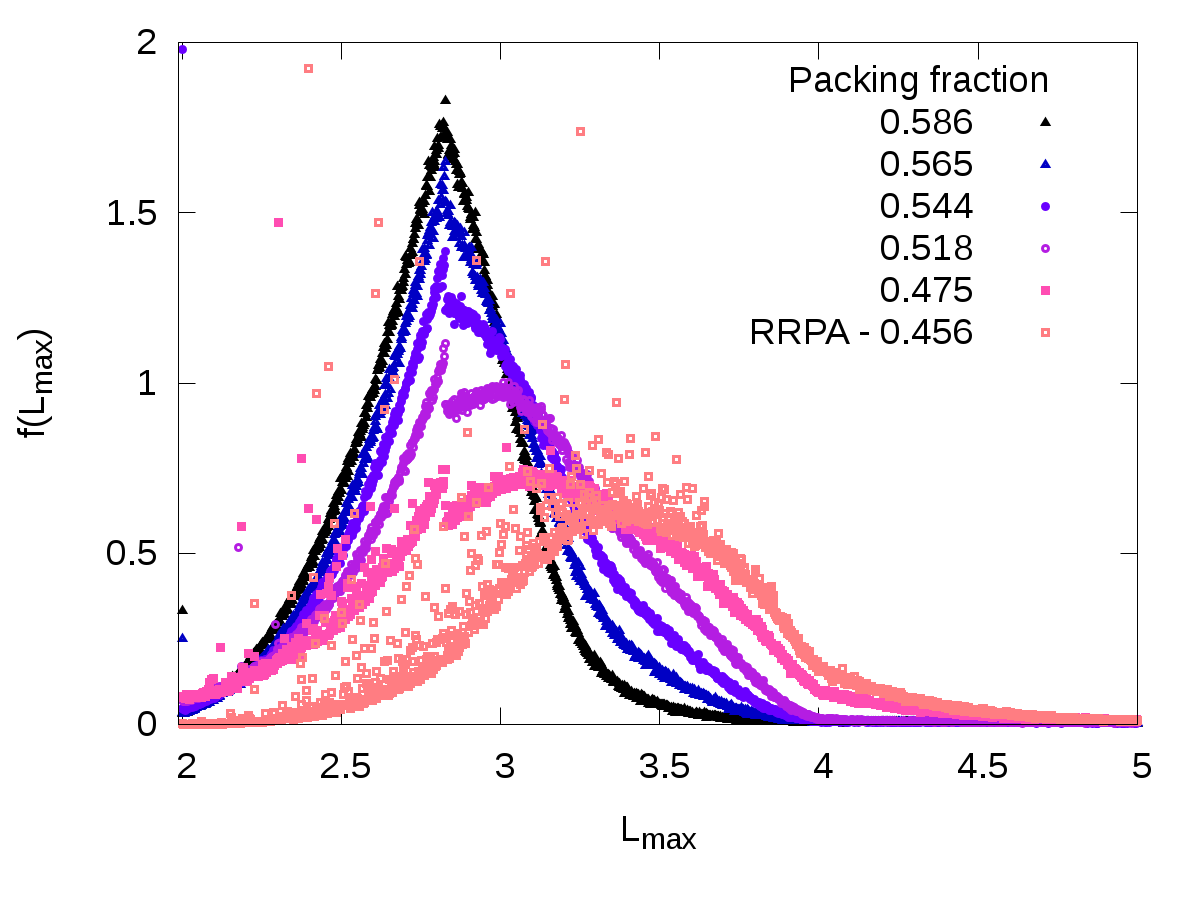}}
\caption{Normalized distributions of a) $\etaD$ and b) $\Lmax$ for various packing fractions. Aggregates were generated with MAX-1 algorithm, with the exception of the RRPA. Error bars are smaller than point size.\label{distperf}}
\end{figure}

\begin{figure}[htbp]
\subfigure[]{\includegraphics[width=0.48\textwidth]{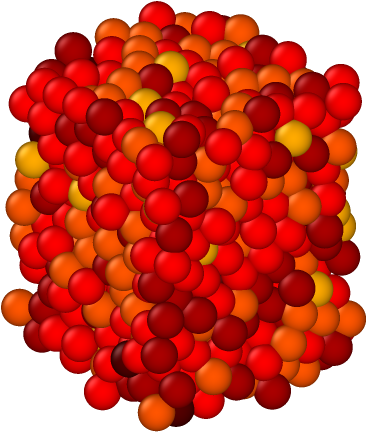}}
\subfigure[]{\includegraphics[width=0.48\textwidth]{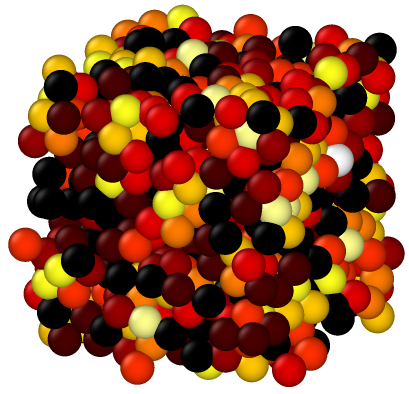}}
\caption{Slices of aggregates generated by two algorithms: a) RMIN-MAX-1 with $a=4$, b) RRPA. Colors (gray scale) correspond to contact coordination number: brighter spheres have a higher CCN. (These figures were generated with Ovito software \cite{S10}.)\label{ovitloc}}
\end{figure}

Globally, the distribution of distortion parameters suggest that two limiting aggregates exist, the densest one, produced by (RMIN)-MAX-1 algorithms (here for $\gamma = 0.586$) and the RRPA. The first limiting aggregate will be called fully random (FR) component, and the second one, regular polytetrahedra (RP) component. The notation FR and RP components refer to structural traits of the aggregates studied here. As a matter of fact, RRPA aggregates are fully RP, whereas RIPAs may share features of these two basic components in a variable proportion, depending on their packing fraction and building algorithm.
Figure \ref{ovitloc} shows local structures of (a) the RIPA aggregate with the smallest proportion of RP component and (b) a RRPA aggregate. The RRPA aggregate presents larger holes in the structure, however the polytetrahedral nature of the latter is quite difficult to visualize. On the other hand, it is clear that the RIPA with high packing fraction presents a much more homogeneous structure than the RRPA.

Figure \ref{distperf}.a presents various distributions of $\etaD$ obtained for aggregates built using MAX-1 algorithm (RIPA) and one RRPA, and figure \ref{distperf}.b represents the distribution of $\Lmax$ for the same aggregates. Concerning MAX-1, these distributions present very similar behaviours: their maximum decreases with packing fraction while their full width at half maximum (FWHM) increases when packing fraction decreases. A bimodal component appears for the lowest packing fraction aggregates on both distributions, respectively centered around $\etaD \approx 2$ and $\Lmax \approx 3.5$. 
These two values may be related in this way: assuming that tetrahedra with $\Lmax \approx 3.5$ have their 5 other edge lengths regularly distributed in the range $[2 ; 3.5]$, leads to a $\etaD$ value of approximately $1.96$, i.e. close to 2, suggesting that these two modes are indeed associated.
$\delta$ peaks appear for the lowest packing fraction, which is consistent with the existence of well defined distances observed on pair distribution function for the same aggregates, noting the existence of regular polytetrahedra and recurrent configurations of spheres in the structure (see \cite{BB15}). 
The RRPA, on the other hand, appears as a limiting case. Indeed, the first peak observed for $\etaD$ and $\Lmax$ distributions in the case of MAX aggregates completely disappears and is replaced by a series of $\delta$ peaks (some out of scale) and a slow evolution with respect to the second mode of the distributions of MAX aggregates. 

Moreover, the distributions of $\etaD$ present a discontinuity at $\etaD \approx 1.33 $, and the distributions of $\Lmax$ have one at $\Lmax=2.827$, which is likely associated. 
The distributions of the lowest packing-fraction aggregates have a change of slope for $\Lmax=4$.
Additional work is needed to analyse precisely the configurations corresponding to these discontinuities.

\subsection{Average values of distortion parameters}

\begin{figure}[htbp]
\begin{center}
\subfigure[]{\includegraphics[width=0.48\textwidth]{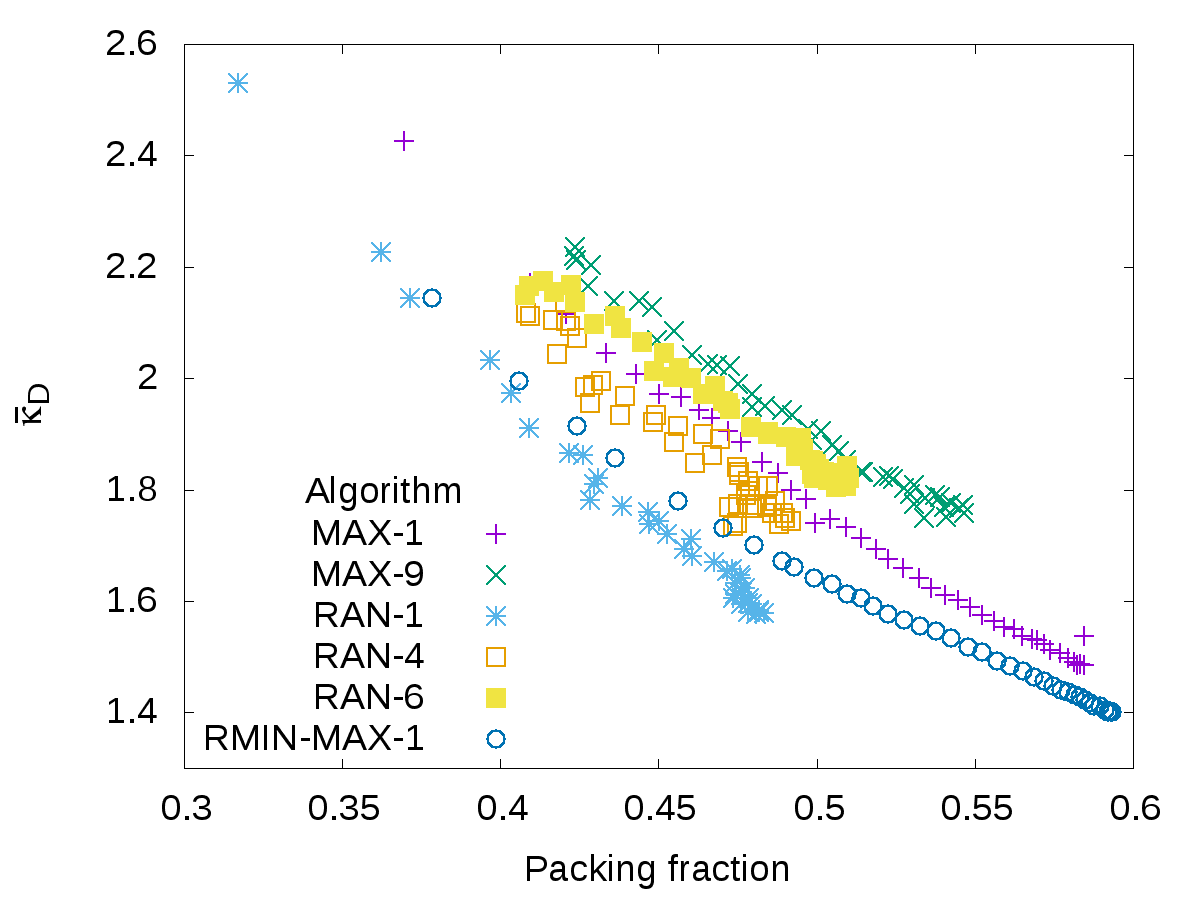}}
\subfigure[]{\includegraphics[width=0.48\textwidth]{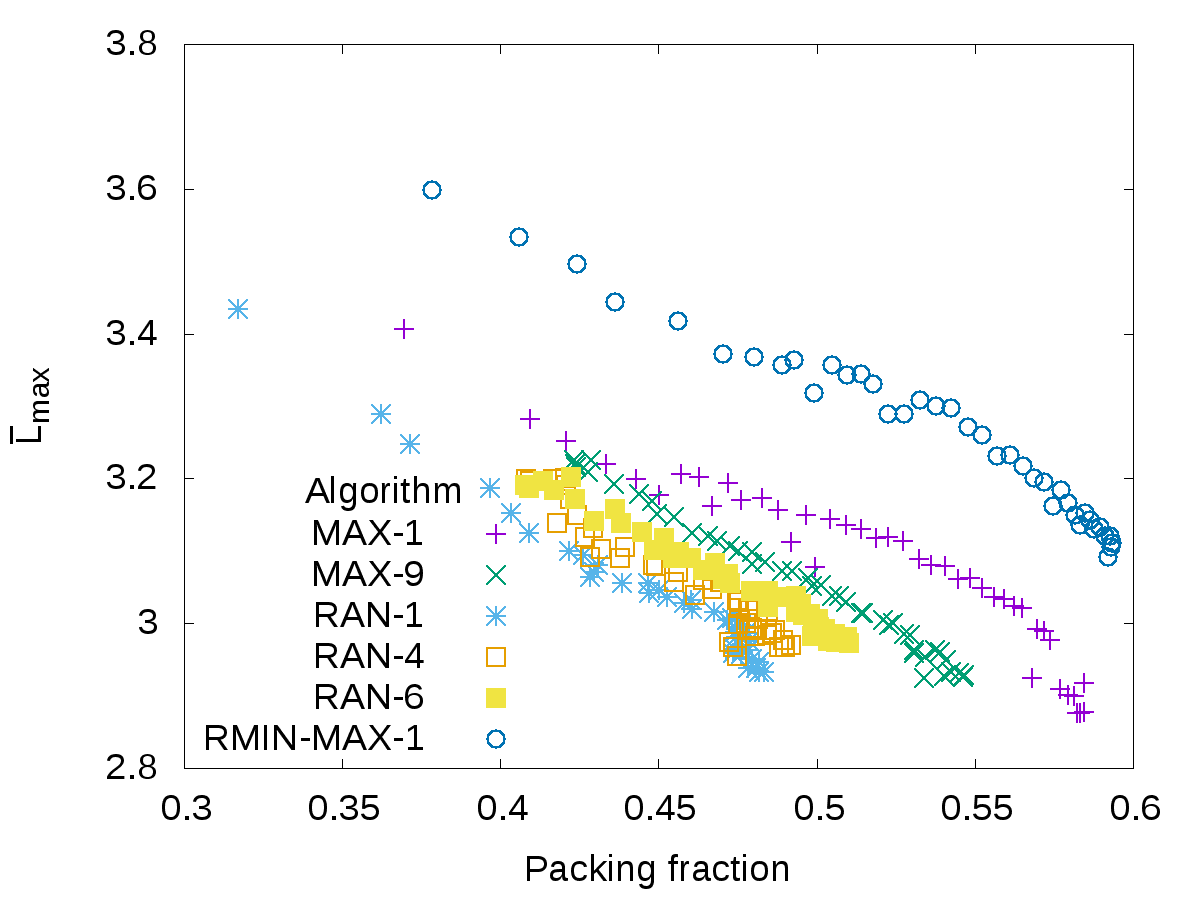}}
\caption{\label{compaperf1} a) Dependency of $\betaD$ with packing fraction b) dependency of $\bLmax$ with packing fraction. Error bars are smaller than point size.}
\end{center}
\end{figure}

Although average distortion parameters are basic structural parameters, they do not determine the packing fraction of the aggregate because they only involve the short range order of spheres (through the tetrahedral description) but do not take into account longer range order. It is therefore worthwhile studying the variation of packing fraction with distortion parameters for all aggregates.
Overall, on average, the higher the packing fraction, the less distorted the tetrahedra, which can appear as slightly counterintuitive as the lowest packing fraction aggregates have a high RP component in their structure and as RP means a regular tetrahedral basis.

Figure \ref{compaperf1}.a presents the dependency of $\betaD$ with packing fraction, which decreases when $\gamma$ increases and turns out to be linear for each family of algorithm.
More specifically, the various algorithms are roughly distinguished as, for MAX-$i$ aggregates, the average distortion increases with $i$ for a given packing fraction. RMIN-MAX-1 stands on its own, with a slightly different slope. The same is observed for RAN-$i$ aggregates, which span a larger interval of $\betaD$ for a given packing fraction (approximately 3 times as wide as that of MAX aggregates) and a certain overlap is observed as MAX-1 to 4 are between RAN-4 and RAN-6 for $\gamma<0.5$, which is the maximum packing fraction that RAN aggregates can reach. 

The behaviour of $\bLmax=f(\gamma)$ is globally the same (figure \ref{compaperf1}.b) as it also decreases when packing fraction increases. However, its behaviour deviates more from linearity: in the case of MAX-1 and RMIN-MAX-1 aggregates a change of slope is observed between 0.47 and 0.5 and they behave very differently from the rest of the other aggregates, with a higher value of $\bLmax$ than any other aggregate for a given packing fraction. Furthermore, there is no overlap between RAN and MAX aggregates. RAN aggregates appear more dispersed than MAX aggregates and, for the latter, they more or less converge on the same curve (that of MAX-9) with the noticeable exception of MAX-1 aggregates.

Hence, surprisingly, the average distortions measured by both indicators do not agree as, for example, in the case of $\betaD$, RMIN-MAX-1 aggregates appear as the least distorted and as the most distorted according to $\bLmax$.

\subsection{Proportion of regular and quasi regular Delaunay tetrahedra}
Using either distortion parameter, it is possible to evaluate the volume fraction of regular Delaunay tetrahedra ($\Phi_V$, with $\etaD=1$ or $\Lmax = 2$), which globally decreases when packing fraction increases, as the RP component of aggregates decreases also. 

RRPA algorithms, in particular, give the highest $\Phi_V$, ranging from 0.101 for $\gamma = 0.452$ to 0.073 for $\gamma = 0.415$. These values remain rather low and prompt the question: what is the geometrical upper bound of the volume fraction of regular tetrahedra in random aggregates? 

\begin{figure}[htbp]
\begin{center}
\subfigure[]{\includegraphics[width=0.48\textwidth]{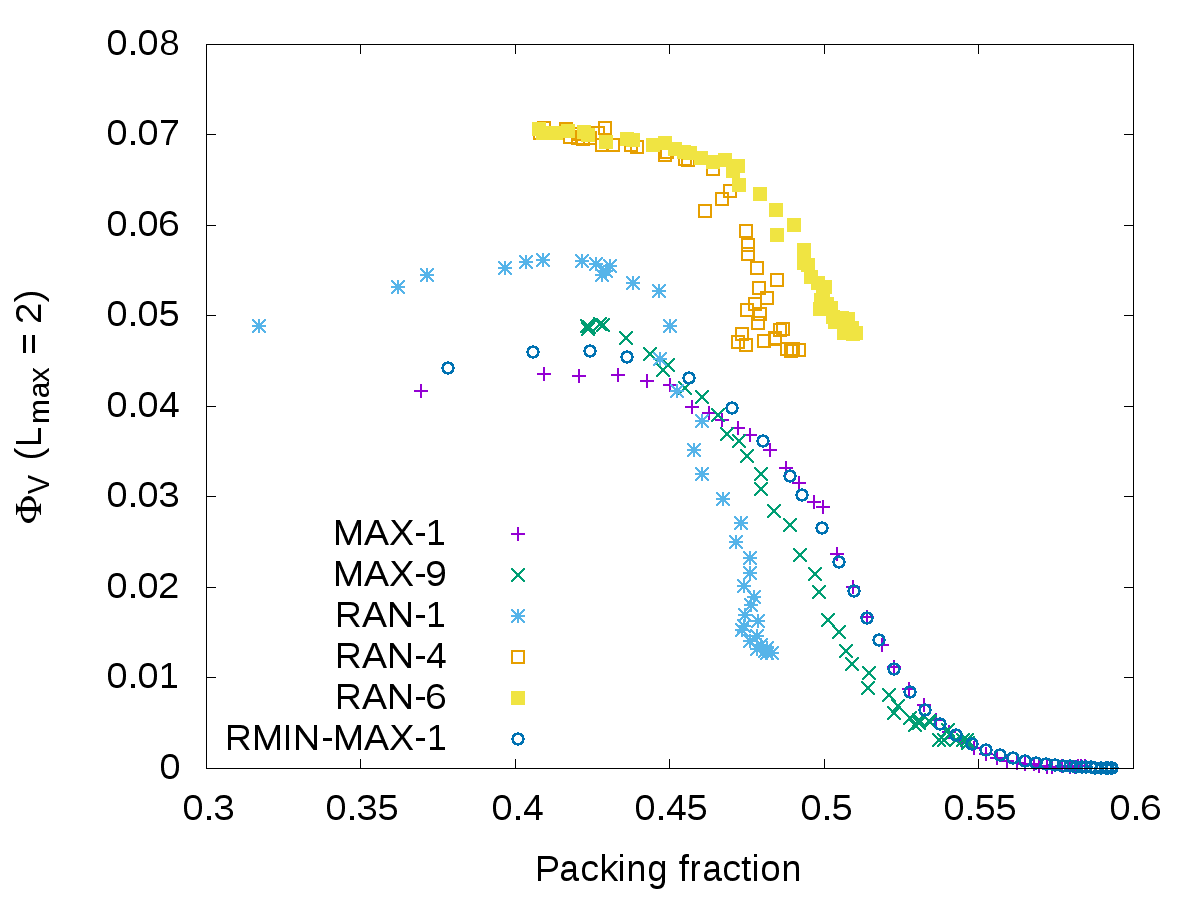}}
\subfigure[]{\includegraphics[width=0.48\textwidth]{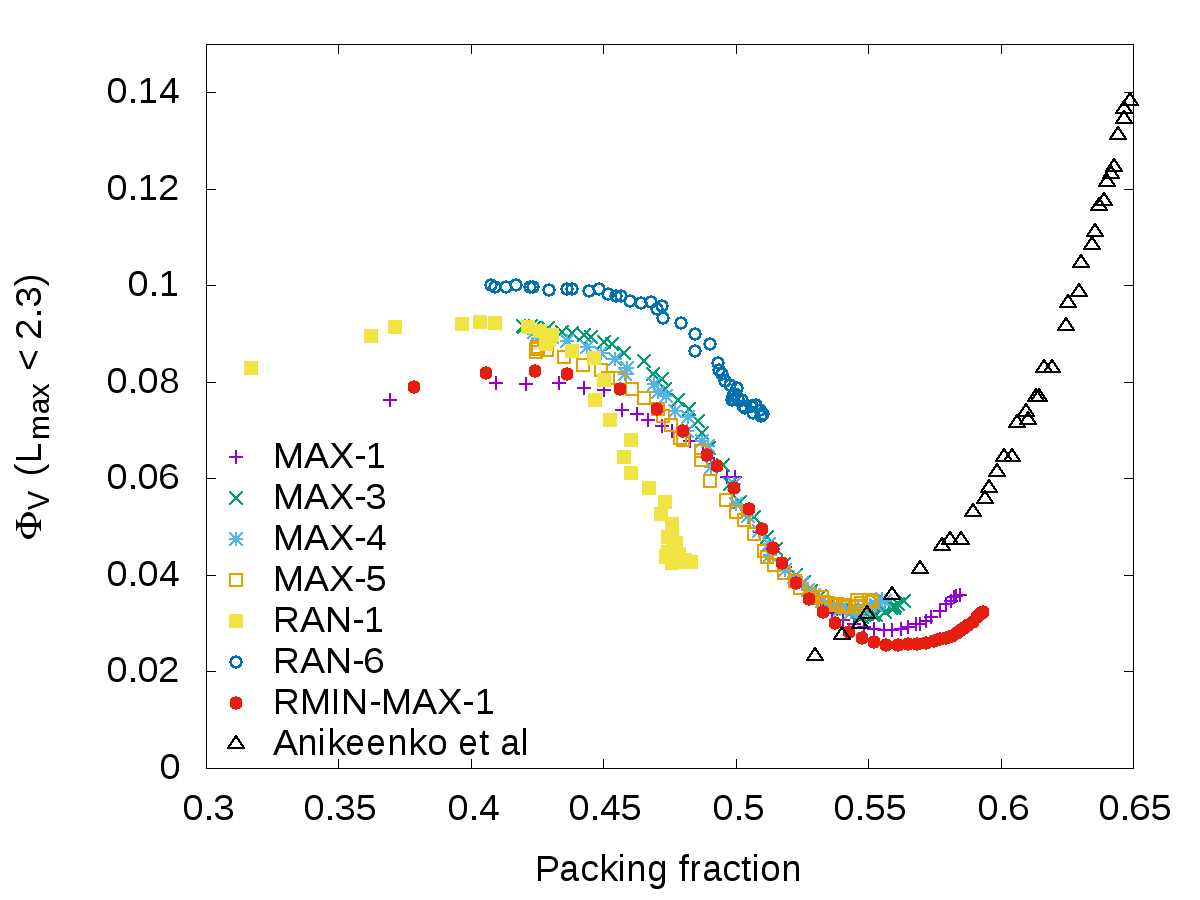}}
\caption{\label{compaperf} a) Dependency of the volume fraction, $\Phi_V$, of perfect tetrahedra ($\Lmax = 2$) with packing fraction. b) Dependency of volume fraction, $\Phi_V$, of tetrahedra with $\Lmax < 2.3$ for the aggregates of the present study and results obtained by Anikeenko and Medvedev in \cite{AM07}.}
\end{center}
\end{figure}

The results for random packings of RIPA type are presented in figure \ref{compaperf}.a, which represents the variation of $\Phi_V$ with packing fraction. The highest proportion is obtained for RAN-6 aggregates ($\Phi_V$ = 0.141, $\gamma = 0.422$). RAN-1 turns out to be the family of aggregates with the smallest fraction of perfect tetrahedra for $\gamma\in [0.45;0.48]$ and outside of this interval, RMIN-MAX-1 outside of this interval. The proportion of perfect tetrahedra goes to 0 at the highest packing fractions.

Anikeenko and Medvedev \cite{AM07} have studied the volume fraction of quasi regular Delaunay tetrahedra (PQRT, i.e. with $\Lmax <2.3$) within aggregates generated with JT and LS algorithms (see figure \ref{compaperf}.b of the present article). They have found that this volume proportion increases with packing fraction in their studied interval of packing fraction, i.e. roughly between 0.53 and 0.71. Figure \ref{compaperf}.b superimposes their results with the ones found in the present study. 
Interestingly, the curves for MAX-4 and MAX-3 match over a rather narrow packing fraction interval, around $\gamma=0.55$, and all MAX-$i$ and RMIN-MAX aggregates show a similar increase of the volume fraction of tetrahedra with $\Lmax <2.3$ for $\gamma\in [0.56;0.59]$, however with lower satisfying quantitative agreement. 
The highest value of the volume fraction of quasi regular tetrahedra found here is 0.127 and is obtained for RMIN-RRPA aggregates.

This comparison shows that regular and quasi regular tetrahedra are two distinct populations: $\Phi_V$ decreases when packing fraction increases and goes to 0 beyond a threshold that depends on the algorithm but is roughly $\gamma = 0.57$, whereas PQRT goes to a minimum for a given packing fraction that depends on the algorithm (between $\gamma = 0.55$ and $\gamma = 0.57$) and then increases with packing fraction beyond this value of $\gamma$.

\section{Effect of sphere coordination}

The contact coordination number (CCN) of each sphere is determined when the aggregate is built: when a sphere is added in contact with another one, both their CCN are increased by 1. Hence, by the end of the building process, each sphere is associated to its CCN.

\subsection{Contact coordination number}
\subsubsection{Partial distributions of contact coordination numbers} 
\begin{figure}[htbp]
    \centering
    \subfigure[]{
        \includegraphics[width=0.48\textwidth]{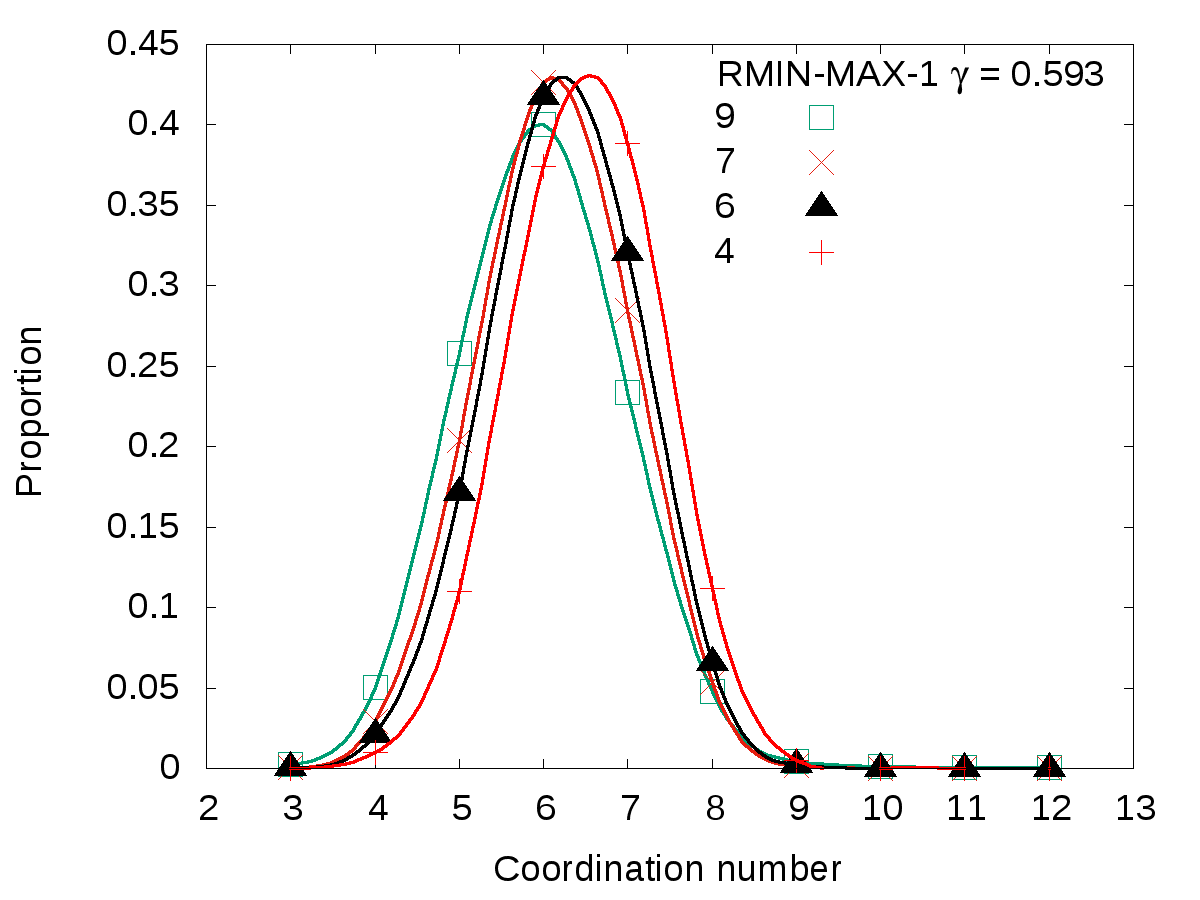}
        \label{perfp333:a}
    }
    \subfigure[]{
        \includegraphics[width=0.48\textwidth]{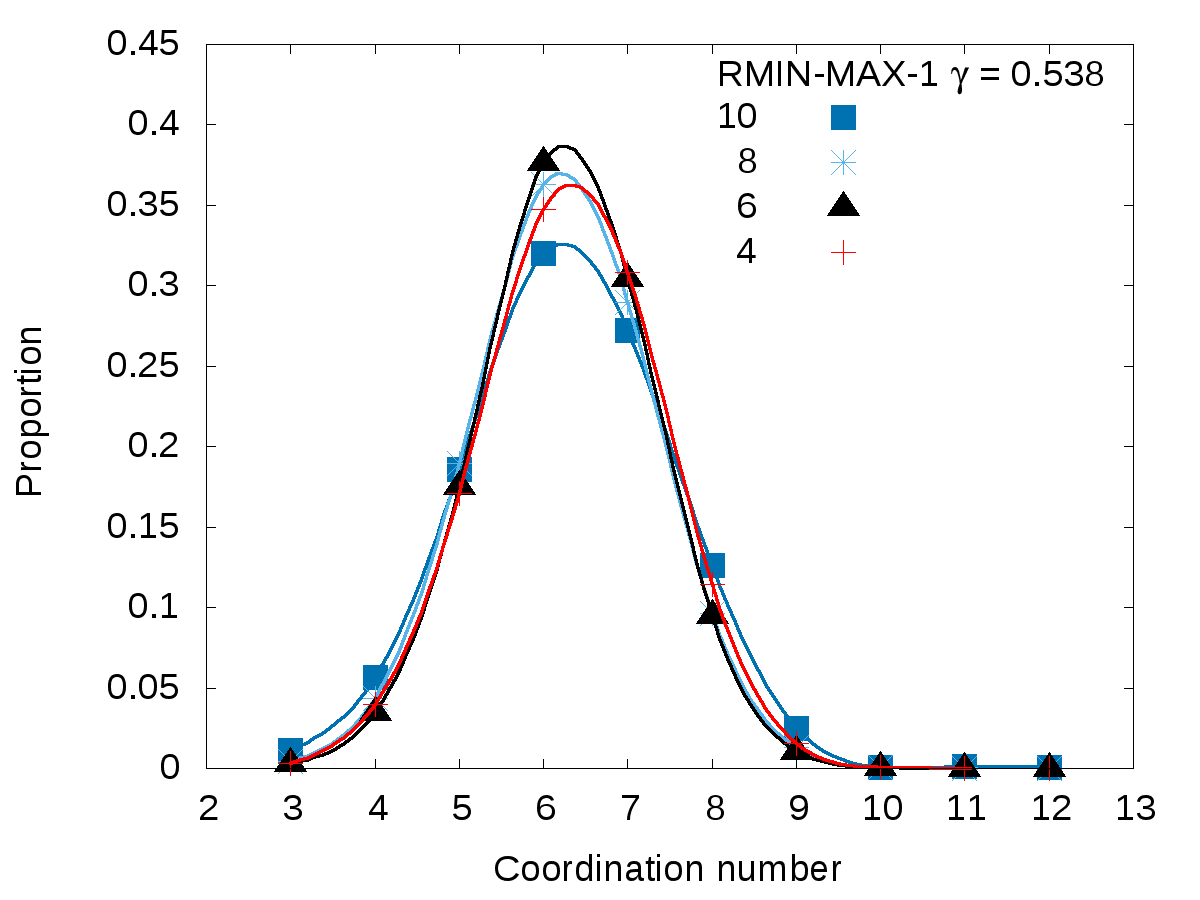}
        \label{perfp333:b}
    }\\
    \subfigure[]{
        \includegraphics[width=0.48\textwidth]{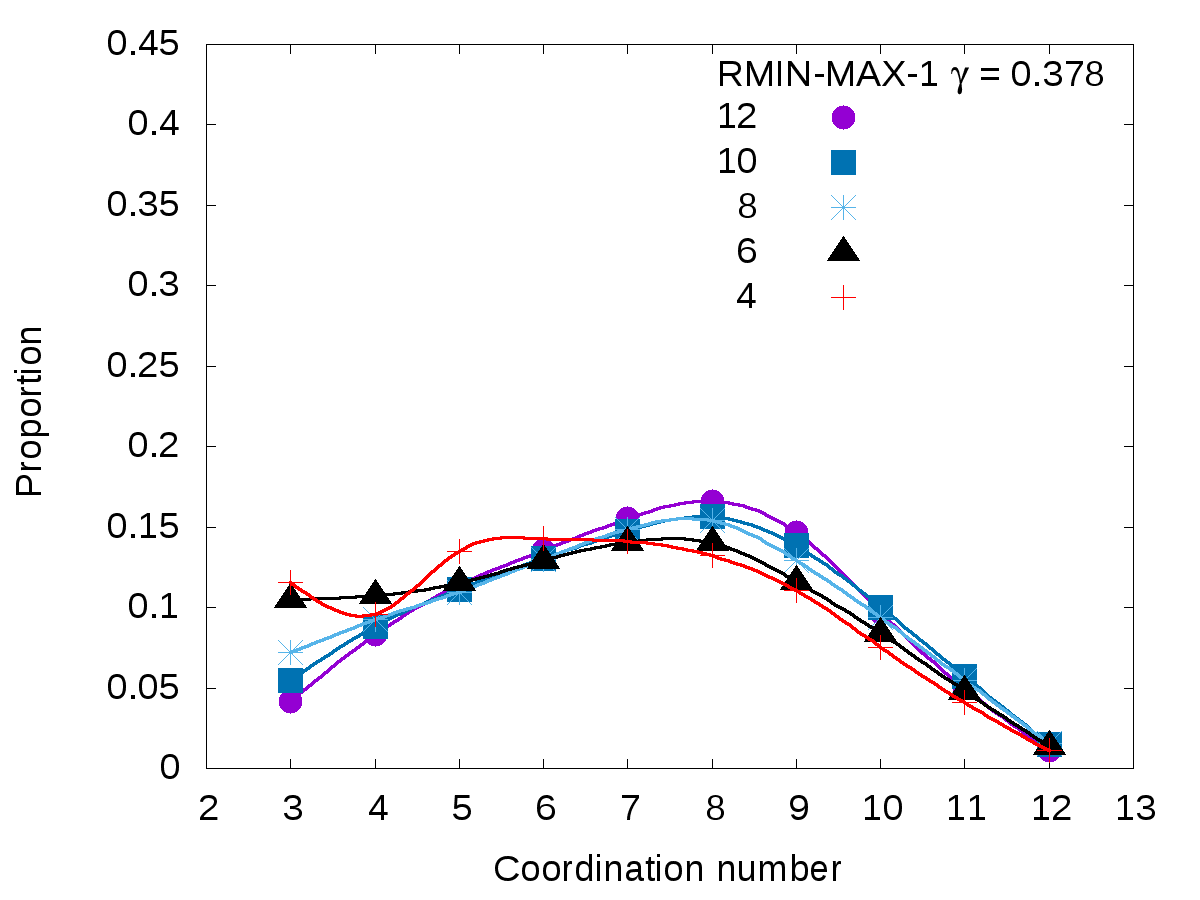}
        \label{perfp333:d}
    }
    \subfigure[]{
        \includegraphics[width=0.48\textwidth]{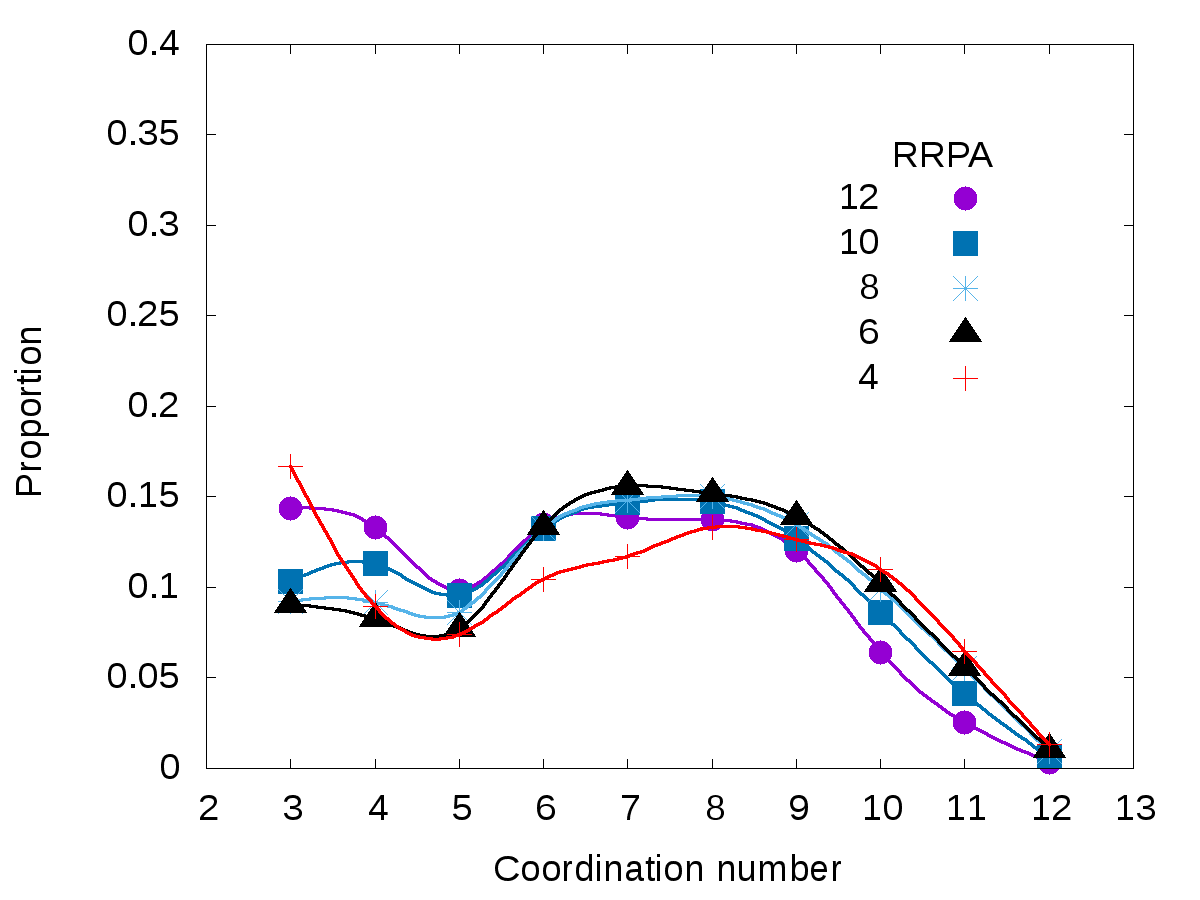}
        \label{perfp333:c}
    }
    \caption{
        Normalized distributions of $\eta_{ij}$ (lines are mere guides for the eye) for aggregates produced by RMIN-MAX-1 algorithm with a) $\gamma = 0.593$, b) $\gamma = 0.538$ and c) $\gamma = 0.378$ and by d) RRPA with $\gamma = 0.456$. 
        \label{distetaij}}
\end{figure}

Let $\eta_{ij}$ be the number of contacts between spheres with CCN $i$ and $j$ respectively. 
The (normalized) distributions of $\eta_{ij}$ can be easily determined from the sphere positions for all values of $i$ and $j$. 
Figure \ref{distetaij} introduces distributions of $\eta_{ij}$ for three aggregates built by RMIN-MAX-1 algorithms, from the highest to the lowest packing fraction (fig. \ref{distetaij}.a to c) and a RRPA aggregate (with the highest packing fraction among RRPA, fig. \ref{distetaij}.d). 

These distributions show a progressive shift of the maxima from high to low packing fraction (fig. \ref{distetaij}.a to c). At high packing fraction, spheres with low CCN tend to be surrounded by spheres with higher CCN thus reducing local fluctuations of the CCN. Then, as the packing fraction decreases, all $\eta_{ij}$ curves more or less collapse, meaning that in this regime, all spheres--whatever their contact coordination number $\eta_i$--have the same $\eta_{ij}$ distribution, centered on the average contact coordination number, hence a very similar environment in terms of contact neighbours. Finally, for even lower packing fraction, an inversion is observed and high CCN spheres are preferentially surrounded by high CCN spheres, which corresponds to a contact segregation effect. 
At the same time, the FWHM of the distributions widen as the packing fraction decreases and they become less symmetrical.
For the lowest packing fraction (fig. \ref{distetaij}.c) $\eta_{ij}$ distributions are very highly spread, with still a higher proportion of high CCN spheres in the vicinity of other high CCN spheres.

The RRPA presents rather similar $\eta_{ij}$ distributions (fig. \ref{distetaij}.d) as low packing fraction RIPAs, going through a minimum for $\eta_{i5}$. However, the order of the various $\eta_{ij}$ distributions appears inverted in the case of the RRPA as, on the high $j$ end, the spheres with the highest proportion of contact with high CCN spheres are spheres with lower CCN ($i$), i.e. $\eta_{3,11}>\eta_{4,11}>...>\eta_{12,11}$, whereas for RIPA with low packing fraction, the opposite situation is observed, i.e. $\eta_{12,11}>\eta_{11,11}>...>\eta_{3,11}$ (fig. \ref{distetaij}.c).

\subsubsection{Evolution of $\beij$ around $i$ sphere}
\begin{figure}[htbp]
    \centering
        \subfigure[]{
        \includegraphics[width=0.48\textwidth]{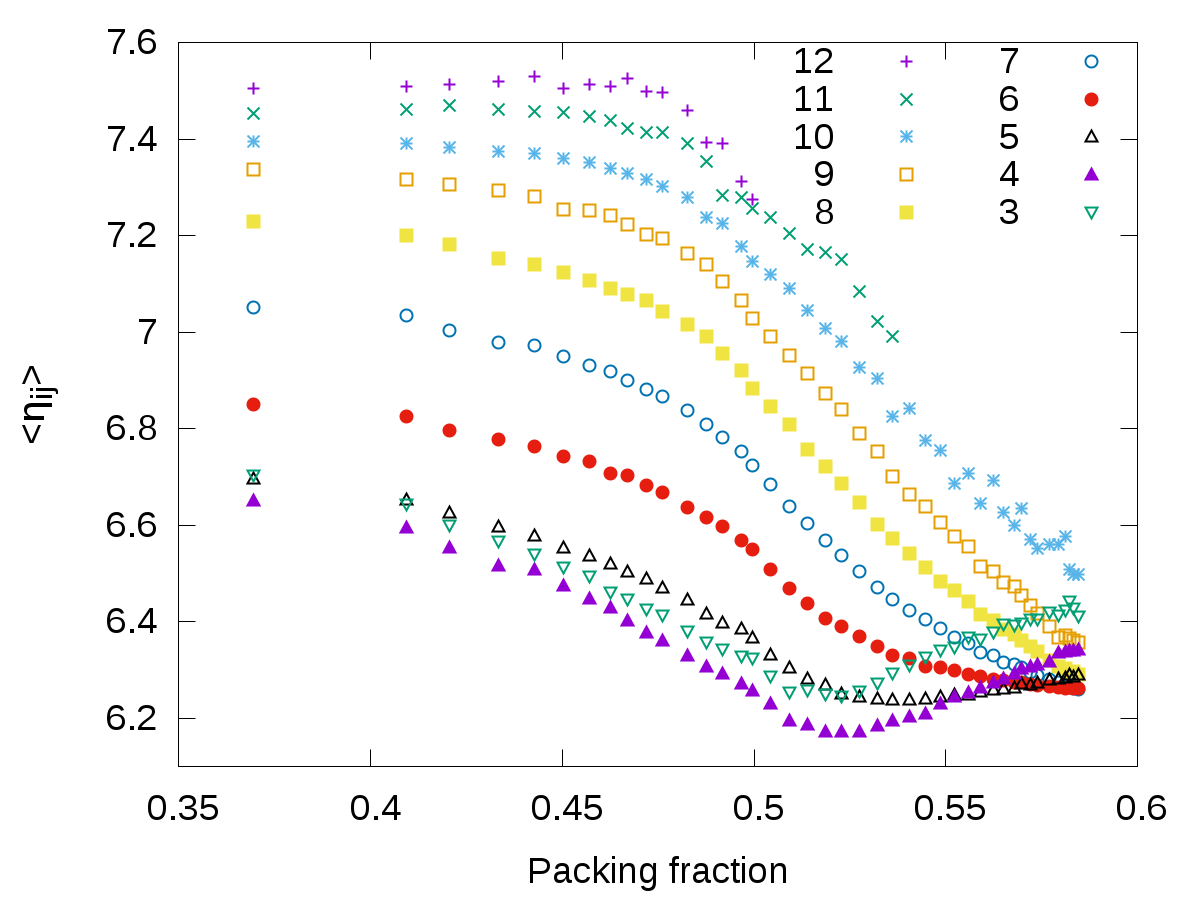}
        \label{}
    }
    \subfigure[]{
        \includegraphics[width=0.48\textwidth]{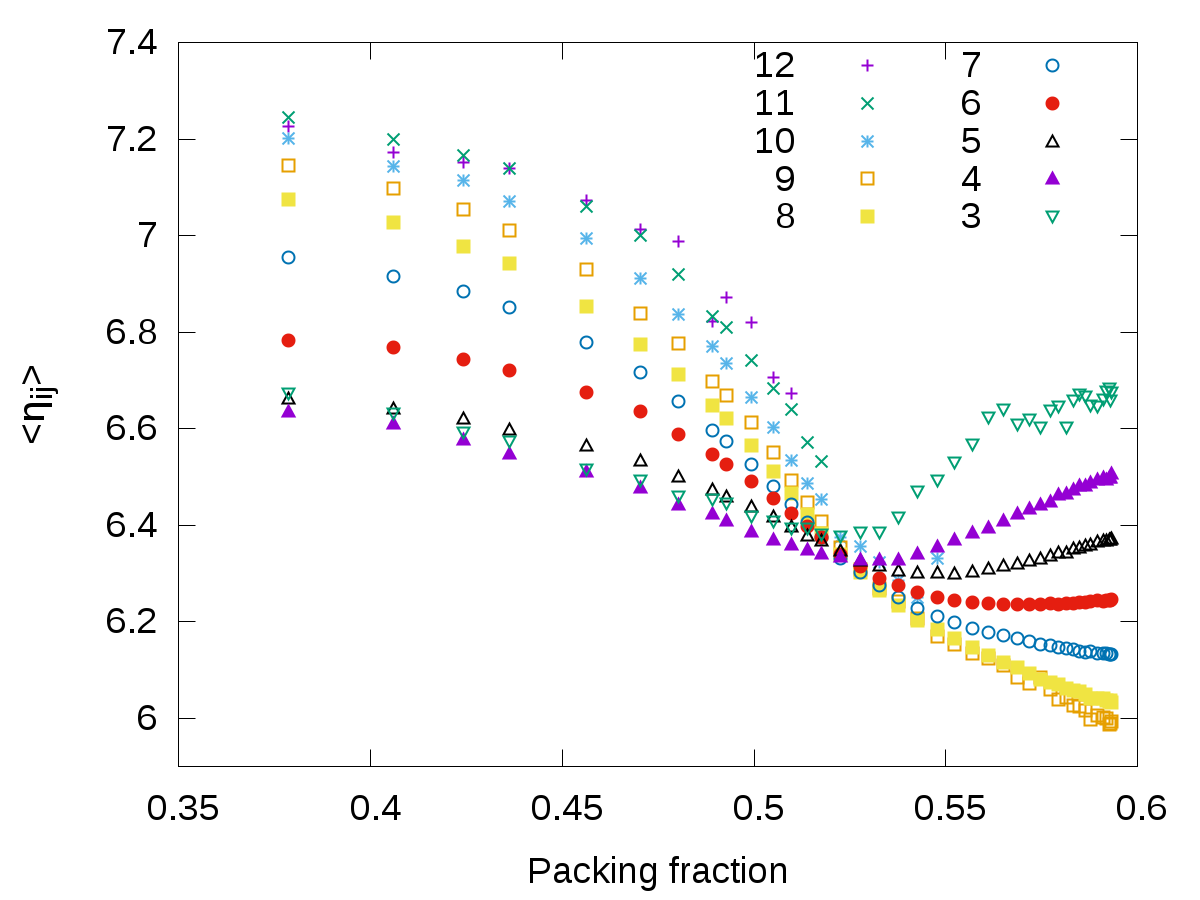}
        \label{}
    }
    \caption{$\beij$ for various values of $i$ in the case of a) MAX-1 and b) RMIN-MAX-1 algorithms.\label{avercont}}
\end{figure}

$\beij$, the average value of $\eta_{ij}$ around spheres with CCN $i$, can easily be determined from the distributions presented above. Figure \ref{avercont} presents the evolution of $\beij$ for all values of $i$ as a function of packing fraction for MAX-1 and RMIN-MAX-1 algorithms. The inversion suggested by the shift of the maximum of distributions in the previous section appears clearly for RMIN-MAX-1 aggregates (figure \ref{avercont}.b) as they form a crossover for $\gamma\approx 0.52$ which separates a low and a high packing fraction regimes. In the low packing fraction regime, high coordination spheres tend to be surrounded by spheres with higher CCN than the spheres surrounding low CCN sphere i.e. $\langle\eta_{12,j}\rangle\ \ >\ \ \langle\eta_{11,j}\rangle\ \ > ...\ >\ \  \langle\eta_{4,j}\rangle$, with the exception of the limit case $\langle\eta_{3,j}\rangle$, corresponding to the segregation effect seen when discussing $\eta_{ij}$ distributions. In the high packing fraction regime, this situation is inverted: spheres with low CCN are surrounded -- on average -- by spheres with higher CCN, thus reducing density fluctuations.

In the case of aggregates generated by MAX-1 algorithm, this crossover is not captured but might take place at higher values of packing fraction -- unaccessible by this algorithm -- as all curves begin to collapse for $\gamma>0.55$. In this case, only the low packing fraction regime is observed.

\subsubsection{Radial dependency of the average contact coordination number}
\label{sec:radDepAvCCN}
\begin{figure}[htbp]
    \centering
        \subfigure[]{
        \includegraphics[width=0.48\textwidth]{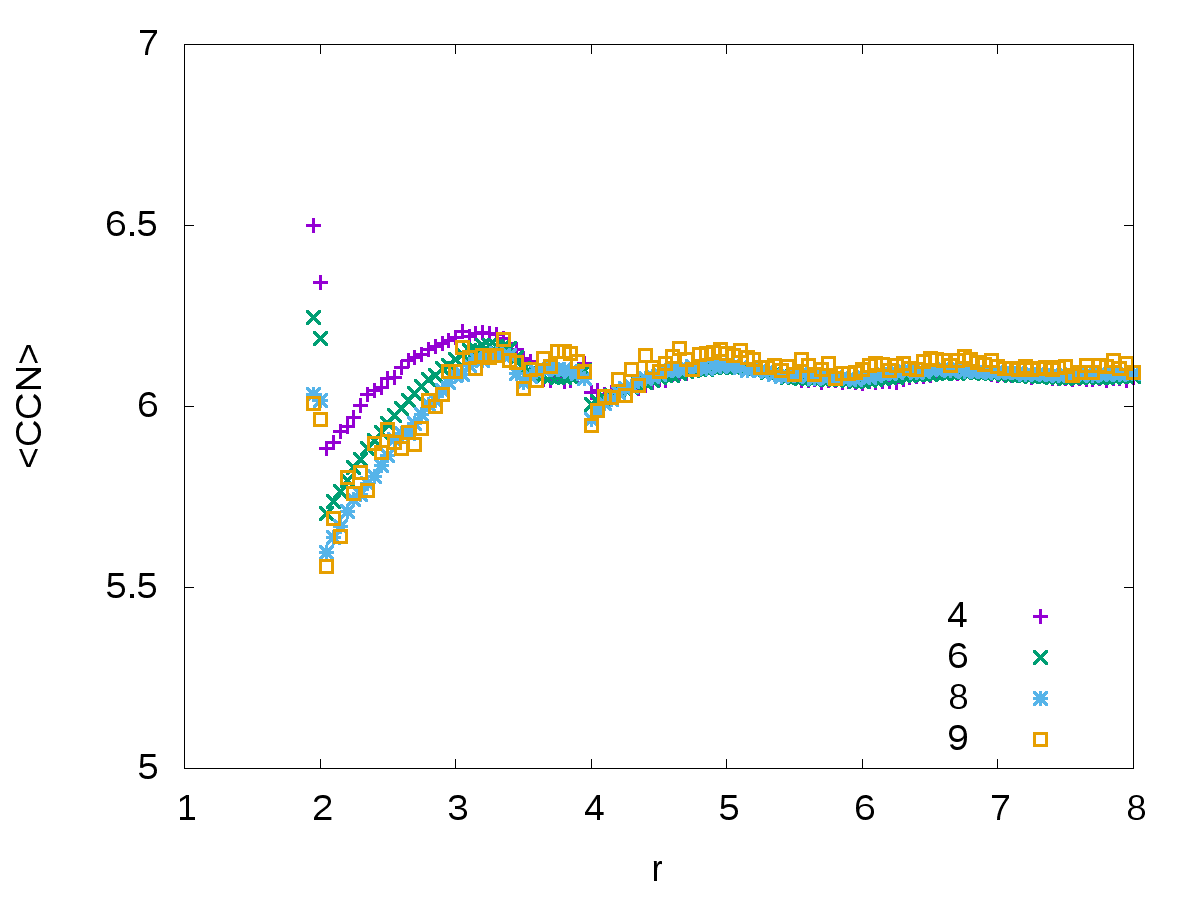}
        \label{}
    }
    \subfigure[]{
        \includegraphics[width=0.48\textwidth]{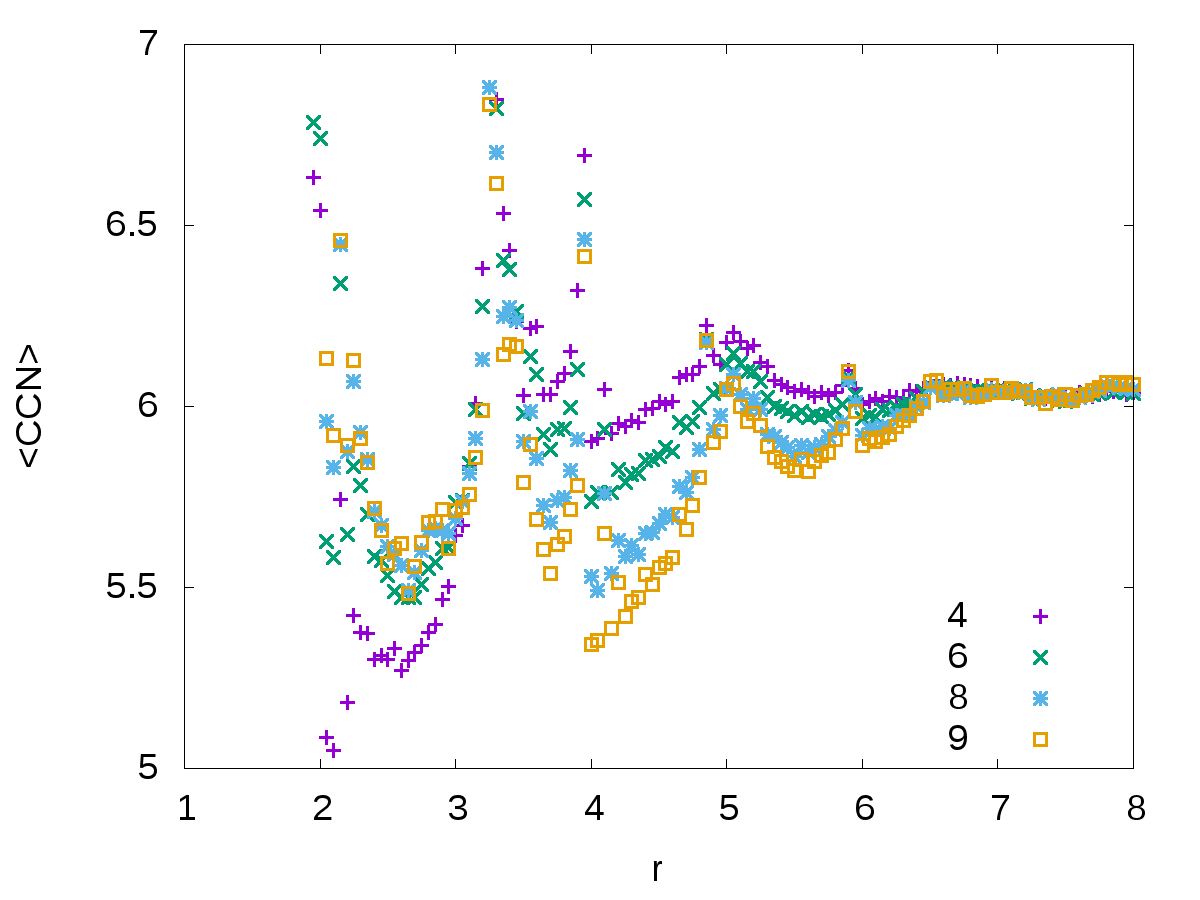}
        \label{}
    }
    \caption{Radial value of the average CCN around $i$ coordinated spheres in RMIN-MAX-1 aggregates. a) RMIN-MAX-1, $\gamma = 0.593$, b) RMIN-MAX-1, $\gamma = 0.378$.\label{radialavercont}}
\end{figure}

The radial dependency of the average contact coordination number $\langle CCN \rangle$ of spheres within $[r;r+dr]$ from an $i$ coordinated sphere has been determined for all aggregates and all values of $i$. Figure \ref{radialavercont} introduces two examples obtained for the RMIN-MAX-1 aggregates with the highest and lowest packing fractions (figure \ref{radialavercont}.a and b).

At high packing fraction (see figure \ref{radialavercont}.a), the average value of the CCN of spheres at a distance $r<3.5$ from a low coordinated sphere is higher than that of spheres surrounding a sphere with higher CCN. Then, as the packing fraction decreases, the relative positions of the various curves for $r<3.5$ are progressively inverted: they superimpose for $\gamma \approx 0.54$, which incidentally matches the packing fraction of the crossover of the various $\beij$ curves in figure \ref{avercont}.b for the same aggregates. At even lower packing fraction, the inversion is complete as it is exemplified in figure \ref{radialavercont}.b: low coordinated sphere are, on average, surrounded by quasi-first-neighbour spheres with low CCN and vice versa. This regime corresponds to a segregation effect in the range of quasi first neighbours instead of contact neighbours. At even larger values of $r$ ($r>4$), this segregation effect gets inverted.

All curves end up superimposing one another beyond some value $r=r_e$, however $r_e$ increases when packing fraction decreases (from $r_e\approx 4$ for $\gamma=0.593$ up to $r_e\approx 6.5$ for $\gamma=0.378$), suggesting that structural inhomogeneities extend over larger and larger scales as $\gamma$ decreases.

Discontinuities are observed at $r=d\sqrt{3}$ and $r=2d$, which matches discontinuities of the various PDF of the same aggregates (cf. section \ref{sec:PPDFRA} below and \cite{BB15}). The amplitude of these discontinuities increases with the CCN of the central sphere.

At low packing fraction (fig \ref{avercont}.b), strong local increases of $\langle CCN \rangle$ are noticeable for values of $r$ which matches $\delta$-peaks on the PDF, hence corresponding to distances characteristics of the presence of regular polytetrahedra in the aggregate (cf. section \ref{sec:PPDFRA} below). 

\subsection{Bond angle distributions}
Two spheres $i$ and $j$ form a bond when they are contact neighbours. The bond angles $\alpha$ around a sphere $i$ are defined as the angles formed between all possible vectors $\vec R_{ij}$ between contacting neighbours. Hence, for a sphere $i$ with $n$ contact neighbours there are $n(n-1)/2$ bond angles.

\subsubsection{Global bond angle distributions}
Bond angle distributions $\alpha$ have been calculated. The smallest possible bond angle is between three contacting spheres, i.e. $\alpha = \pi/3$, and the largest is, of course, $\pi$.

\begin{figure}[htbp]
\begin{center}
\subfigure[]{\includegraphics[width=0.48\textwidth]{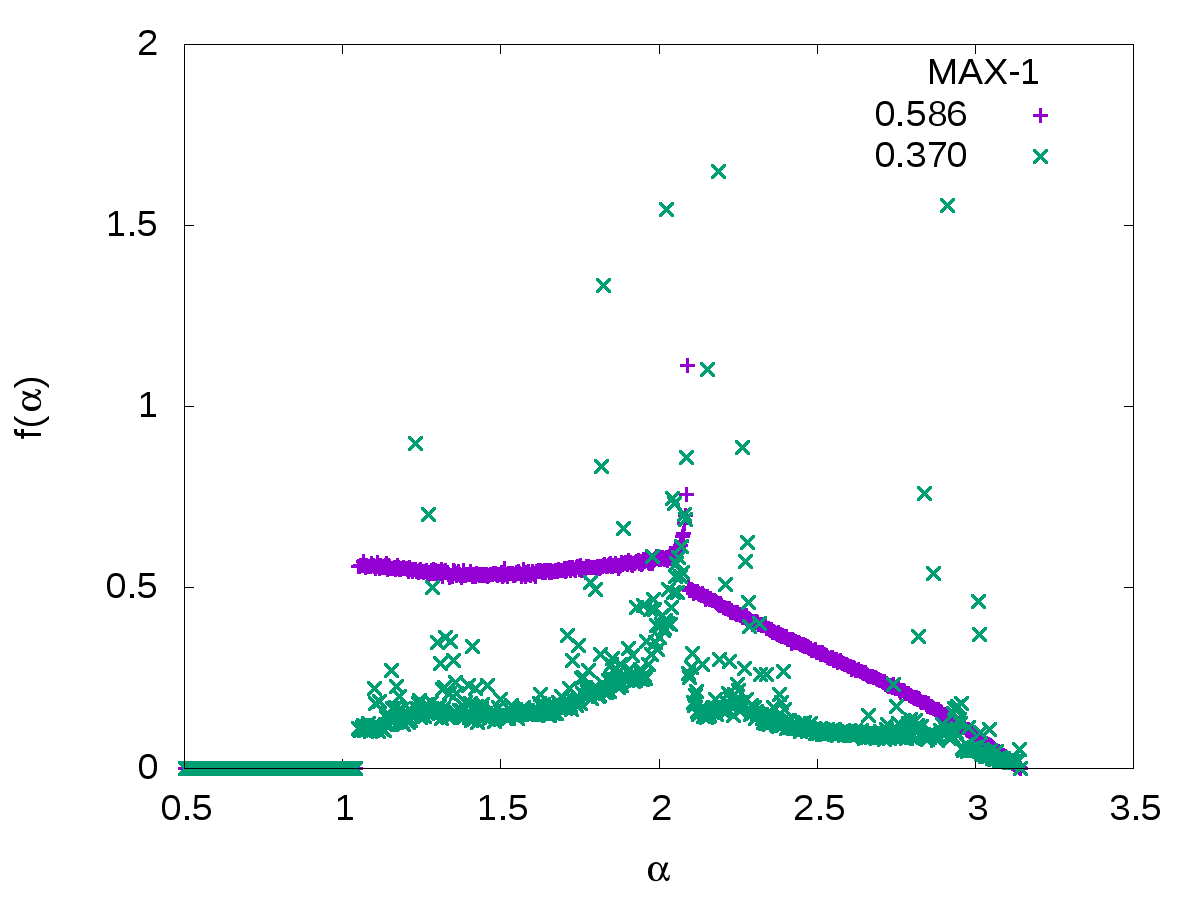}}
\subfigure[]{\includegraphics[width=0.48\textwidth]{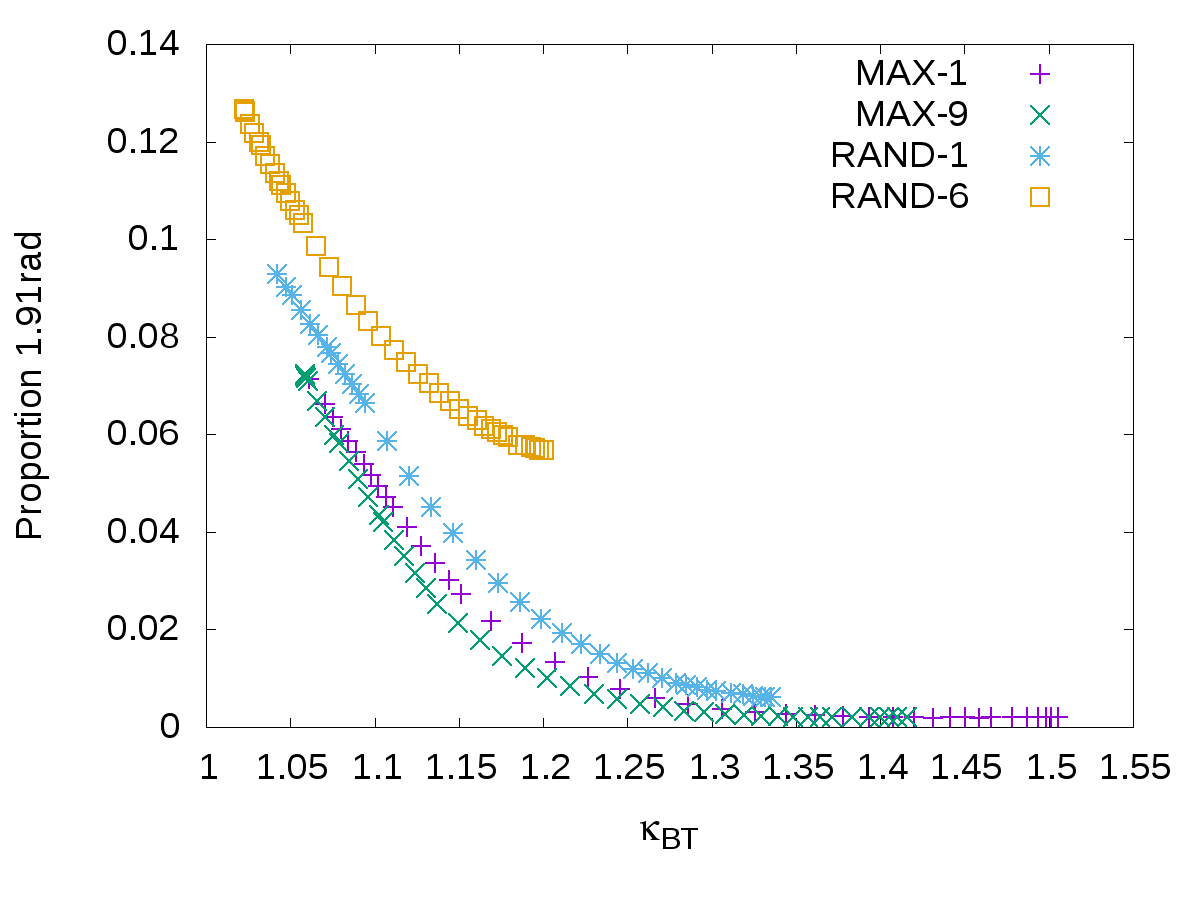}}
 \caption{a) Bond angle distributions for high and low density aggregates (values for $\alpha = \pi/3$ fall out of range of the figure). b) Dependency of the 1.9106~rad $\delta$-peak in the bond angle distribution with the irregularity index of building tetrahedra $\bar\kappa_{BT}$.\label{bondangdist}}
\end{center}
\end{figure}

Figure \ref{bondangdist}.a presents bond angle distributions for the most and least dense aggregates obtained by MAX-1 algorithm. It shows that the distribution of the densest aggregate is mostly smooth, marked by two discontinuities: the first one, the minimum value, is $\alpha = \pi/3$: it corresponds to the configuration of three spheres in contact with each other. The value here is so high that it is outside the scale of the figure. The second discontinuity occurs for $\alpha = 2.093\pm 0.002$, close to $2\pi/3$, which corresponds to the situation where four spheres form two coplanar equilateral triangles sharing a common side.
In \cite{KFL09} Karayiannis et al obtained a very similar bond distribution in their structures of chains of joined monodisperse hard spheres in the MRJ (maximally random jammed) state. The main difference is the broader shape of the peaks in the distributions of ref \cite{KFL09}, which we interpret as a finite size broadening since the maximum number of spheres considered in \cite{KFL09} is 54000 instead of $10^6$ used here.

For the aggregate with the smallest density, many singularities appear, in the same way as $\delta$-peaks appear on its pair distribution function. On PDF, these $\delta$-peaks are characteristic of the presence of regular polytetrahedra in the disordered structure (see \cite{BB15}). The addition of another sphere on top of three contacting ones does not introduce a new bond-angle. The addition of a fifth sphere forms then a trigonal bi-pyramid, with $\alpha \approx 1.9106~\textrm{rad}$. This angle should correlate with the $\delta$-peak at $r=d\sqrt{8/3}$ observed on the pair distribution function of low density aggregates (see \cite{BB15}). It shows the same smooth dependency with the irregularity index of building tetrahedra (see figure \ref{bondangdist}.b) as $P(r=d\sqrt{8/3})$.

\subsubsection{Partial bond angle distributions}

\begin{figure}[htbp]
\begin{center}
\subfigure[]{\includegraphics[width=0.48\textwidth]{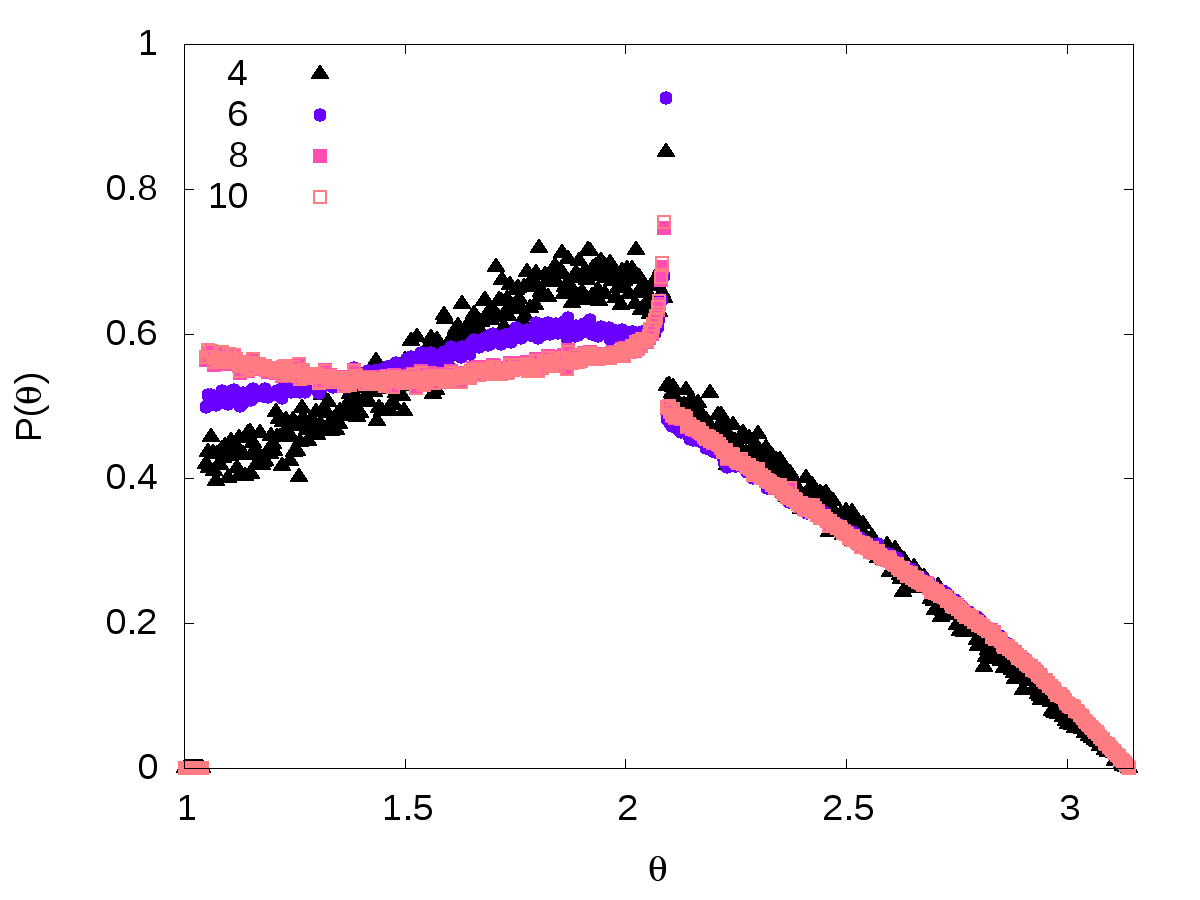}}
\subfigure[]{\includegraphics[width=0.48\textwidth]{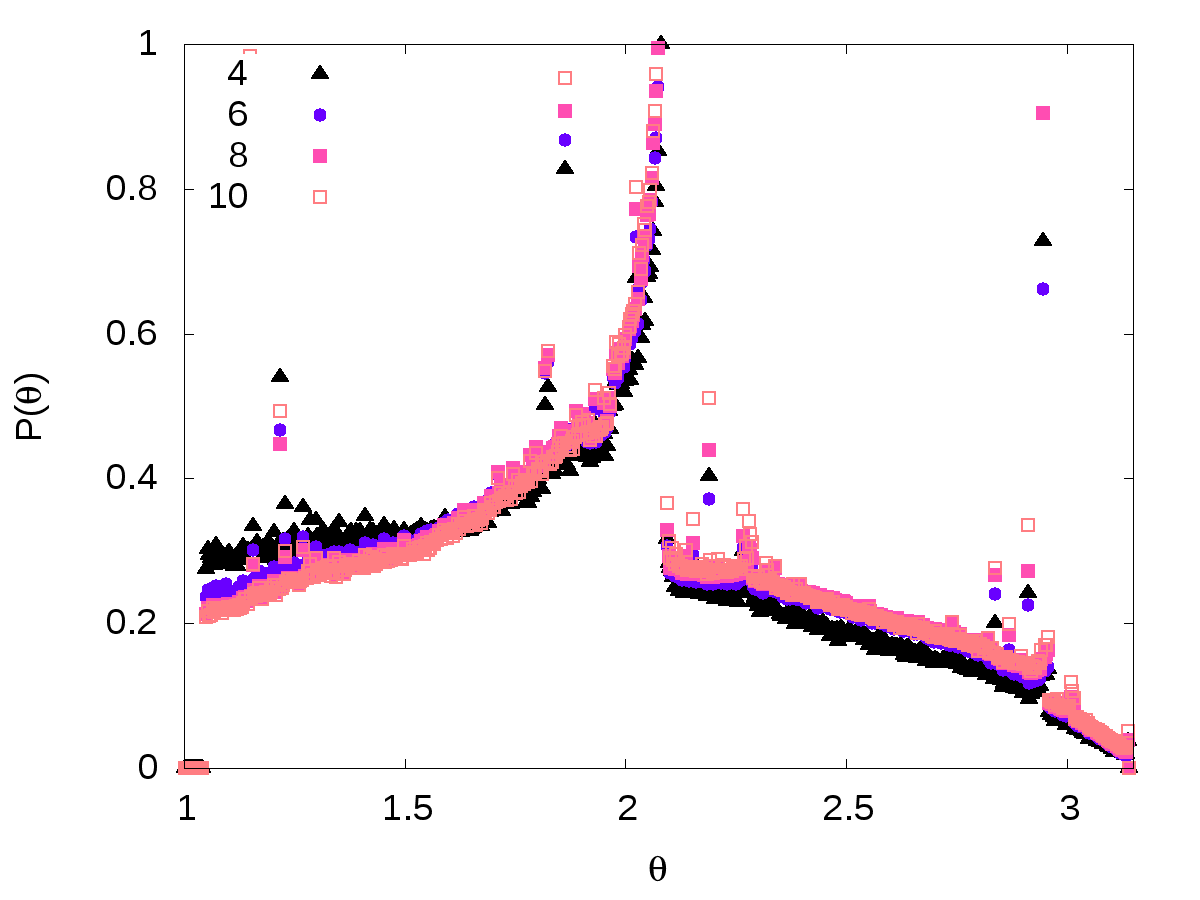}}\\
\subfigure[]{\includegraphics[width=0.48\textwidth]{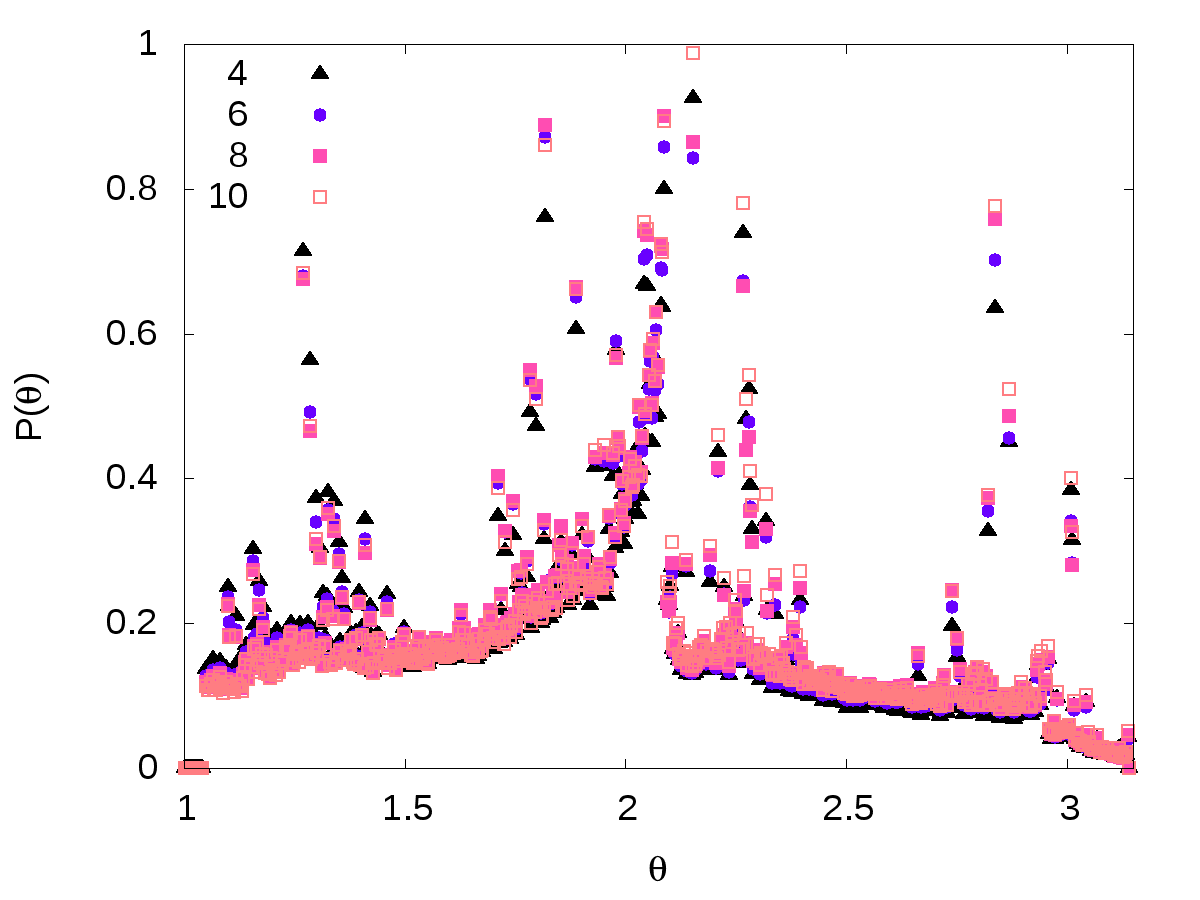}}
\subfigure[]{\includegraphics[width=0.48\textwidth]{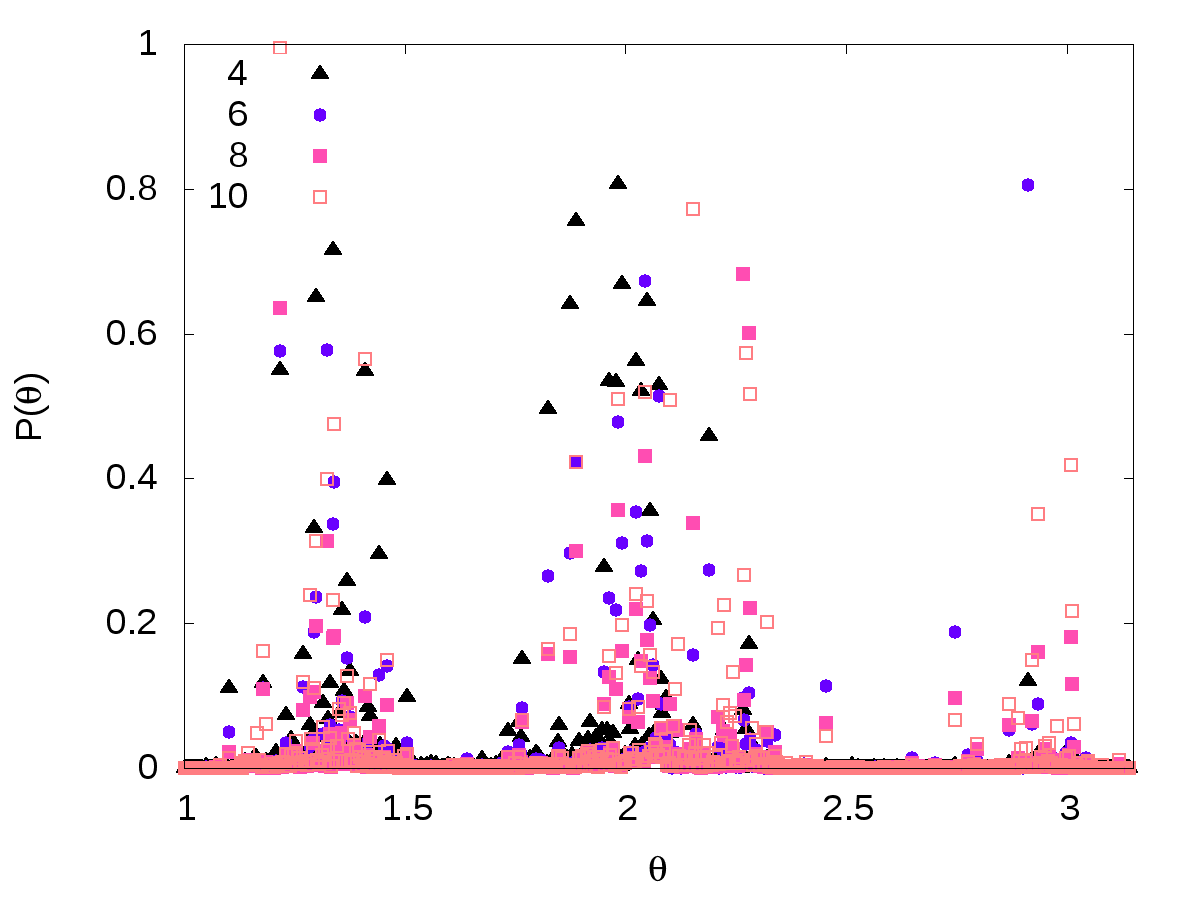}}
 \caption{Partial bond angle distributions for various MAX-1 aggregates a) $\gamma = 0.586$ b) $\gamma = 0.500$ c) $\gamma = 0.370$ and d) one RRPA aggregate, $\gamma = 0.456$. Values for $\alpha = \pi/3$ fall out of range of the figure.\label{bondangdist_i}}
\end{center}
\end{figure}

Bond angle distributions for specific contact coordination numbers are shown in figure \ref{bondangdist_i}. Their global behaviour is similar to global bond angle distributions, but peculiarities can be seen, in function of packing fraction and/or contact coordination number. 

It turns out that for aggregates with high packing fraction, the angular environment depends strongly on the CCN. Low CCN spheres have a stronger asymmetry between $\alpha = \pi/3$ and $\alpha = 2\pi/3$ than spheres with high CCN. For high packing fraction aggregates, distribution of low CCN spheres show a depletion of lower bond angles and an excess of high angles and the distributions get more even when the CCN increases. 
When the packing fraction decreases, low and high CCN spheres tend to have more similar angular environment.

The appearance of regular polytetrahedra, associated with $\delta$ peaks, like in the case of the global bond angle distribution, is logically correlated with a decrease of the continuum component of the distribution. This continuum disappears completely for the RRPA (figure \ref{bondangdist_i}.d).

\subsection{Partial pair distribution functions}
The structure of random sphere packings is usually characterized by the probability per unit volume of finding a sphere center at a distance $r$ from another sphere center, $P(r) \times N/V$, where $N$ is the number of spheres in the aggregate of volume $V$ and the pair distribution function (PDF) $P(r)$ is normalized to 1 when $r\to\infty$.

Distinguishing spheres by their contact coordination number allows a much more detailed structural description by partial pair distribution functions (PPDF), which are defined as the probability $P_{ij}(r)$ of finding a sphere with contact coordination number $j$ at a distance $r$ from another sphere with contact coordination number $i$, normalized to 1 at large $r$.

\subsubsection{Principle for partial PDF $P_{ii}(r)$, $P_{ij}(r)$ and $P_i(r)$}
In practice, coordination numbers range from 3 to 12, the maximum CCN in 3D space. For a spherical aggregate with radius $R$, $P_{ij}(r)$ write:
\begin{equation}
P_{ij}(r) = \frac{V^2}{N_iN_j}\frac{\Delta N_{ij}}{(2-\delta_{ij})S(r)\Delta r}
\end{equation}
where $N_i$ and $N_j$ are the number of spheres with CCN $i$ and $j$, respectively, in the volume $V=4\pi R^3/3$ of the aggregate; 
$\Delta N_{ij}$ is the number of sphere pairs of CCN $i$ and $j$ respectively, lying in the interval $[r : r+\Delta r]$;
$\delta_{ij}$ is the usual Kronecker symbol;
$S(r)$ is the spherical shape factor of the aggregate \cite{F51}:
\begin{equation}
S(r) = \frac{\pi^2}{6}r^2(2R-r)^2(4R+r)
\end{equation}

$P_{ii}(r)$ PPDF describe the arrangement of $i$ coordinated spheres, while $P_{ij}(r)$ PPDF with $i\neq j$ describe the mutual arrangement or "chemical order" between $i$ and $j$ coordinated spheres.

In the case of sticky hard spheres with diameter $d$, the peak of contacting  neighbours in $P_{ij}(r)$ is represented by \cite{B90}:
\begin{equation}
P_{ij}(d) = \bar\eta_{ij} \frac{V}{4\pi N_j d^2}\delta(r-d)\ \mathrm{i.e. numerically}\ \eta_{ij} = 3\frac{N_j}{R^2}d^2 2\sigma P(d)
\end{equation}
where $\bar\eta_{ij}$ is the average number of $j$ coordinated spheres contacting an $i$ coordinated sphere and $\sigma = 0.01$, is the length step used in $P(r)$ calculations. Numerical values of $P_{ij}(d)$ fall out of range of the $P_{ij}(r)$ figures presented below.

One can then define the probability $P_i(r)$, to find a sphere with any coordination number at a distance $r$ from a sphere with coordination number $i$. It writes:
\begin{equation}
P_i(r) = \sum_{j=3}^{12} C_j P_{ij}(r)
\end{equation}
where $C_j = N_j/N$ is the concentration of $j$ coordinated spheres ($N=\sum_{j=3}^{12}N_j$).

These $P_i(r)$ characterize the global arrangement of spheres around a sphere with coordination $i$.

Finally, the global PDF writes:
\begin{equation}
P(r) = \sum_{i=3}^{12}C_i P_i(r)
\end{equation}

\begin{figure}[htbp]
\begin{center}
\includegraphics[width=0.48\textwidth]{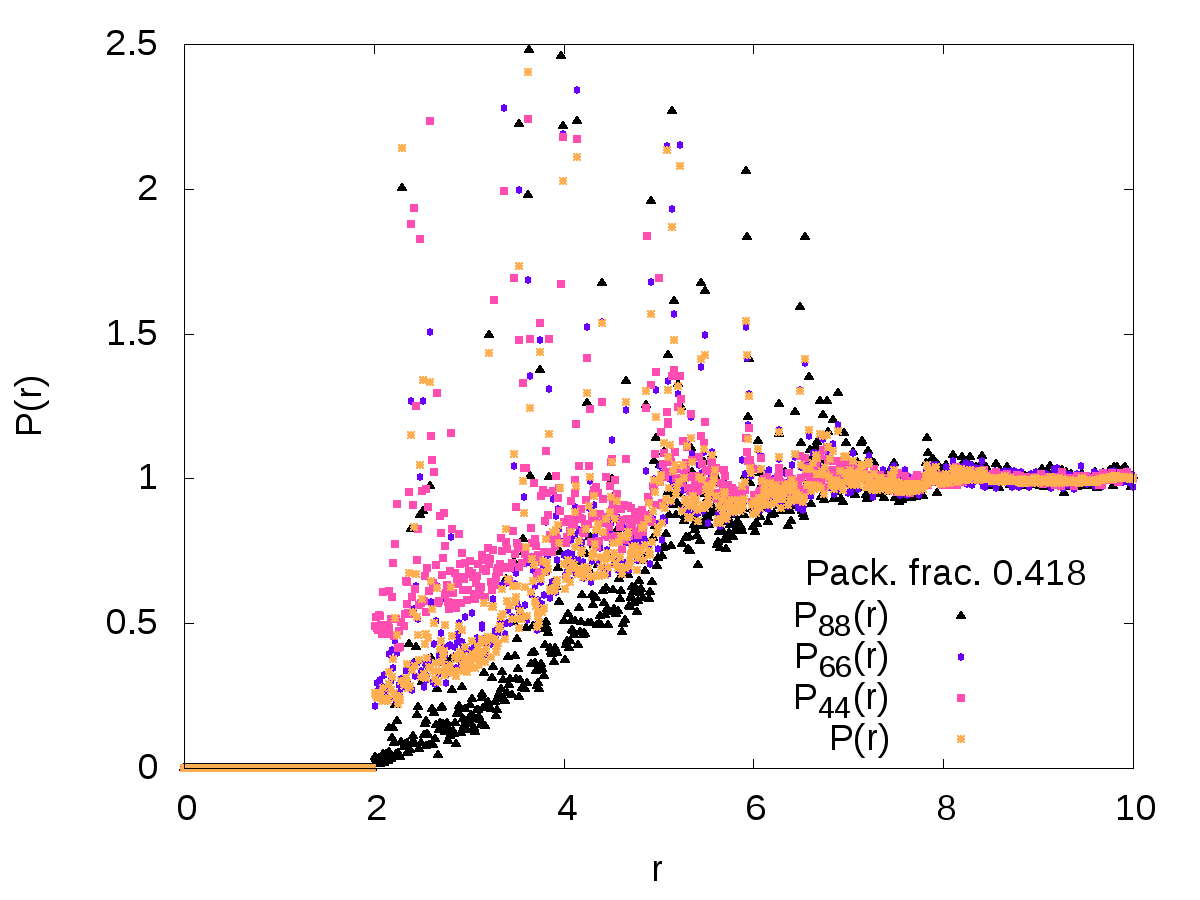}
 \caption{Global and partial pair distribution functions ($P(r)$ and $P_{ii}(r)$ for $i=4,6$ and 8) for a random regular polytetrahedral aggregate.\label{pdr_rrpa}}
\end{center}
\end{figure}

\paragraph{Random regular polytetrahedral aggregates}
Figure \ref{pdr_rrpa} presents $P_{ii}(r)$ curves (with $i=4,\ 6$ and 8) of an RRPA with $\gamma = 0.418$, as well as its global PDF (see \cite{BB15} for the calculation procedure of the latter).

Concerning the global PDF, a striking difference from what was observed for RIPA (see \cite{BB15}), is the disappearance of the topological discontinuities at $r = \sqrt{3}d$ and $r=2d$. The continuous structure of the PDF observed in RIPAs almost disappears and is replaced by a set of polyhedral $\delta$ peaks (\cite{B90,MP08}), which are due to a large (virtually infinite) regular polytetrahedron. The positions of these peaks, which are due to precise configurations of spheres with well defined distances, are identical to those studied by Medvedev et al \cite{MP08}. The continuous structure grows almost linearly as a function of $r$ and goes to 1 for $r\approx 3.5d = 7$.

The partial PDFs ($P_{ii}(r)$) show a systematic tendency, as their continuous regime in the region of the quasi first neighbours (QFN) starts from 0 for spheres with coordination number $i$ equal or superior to 8 and increases like the global $P(r)$. The lower coordination numbers ($i<8$) lead to $P_{ii}(r)$ starting from higher values in the QFN area. This can be qualitatively understood: the more contacting first neighbours a sphere has, the less quasi first neighbours it can accept.
\paragraph{Random irregular polytetrahedral aggregates}
\label{sec:PPDFRA}
The global PDF of RIPA was studied in \cite{BB15}, with the exception of RMIN aggregates. However, the latter bring no qualitative differences to these results. 

\begin{figure}[htbp]
\begin{center}
\subfigure[]{\includegraphics[width=0.48\textwidth]{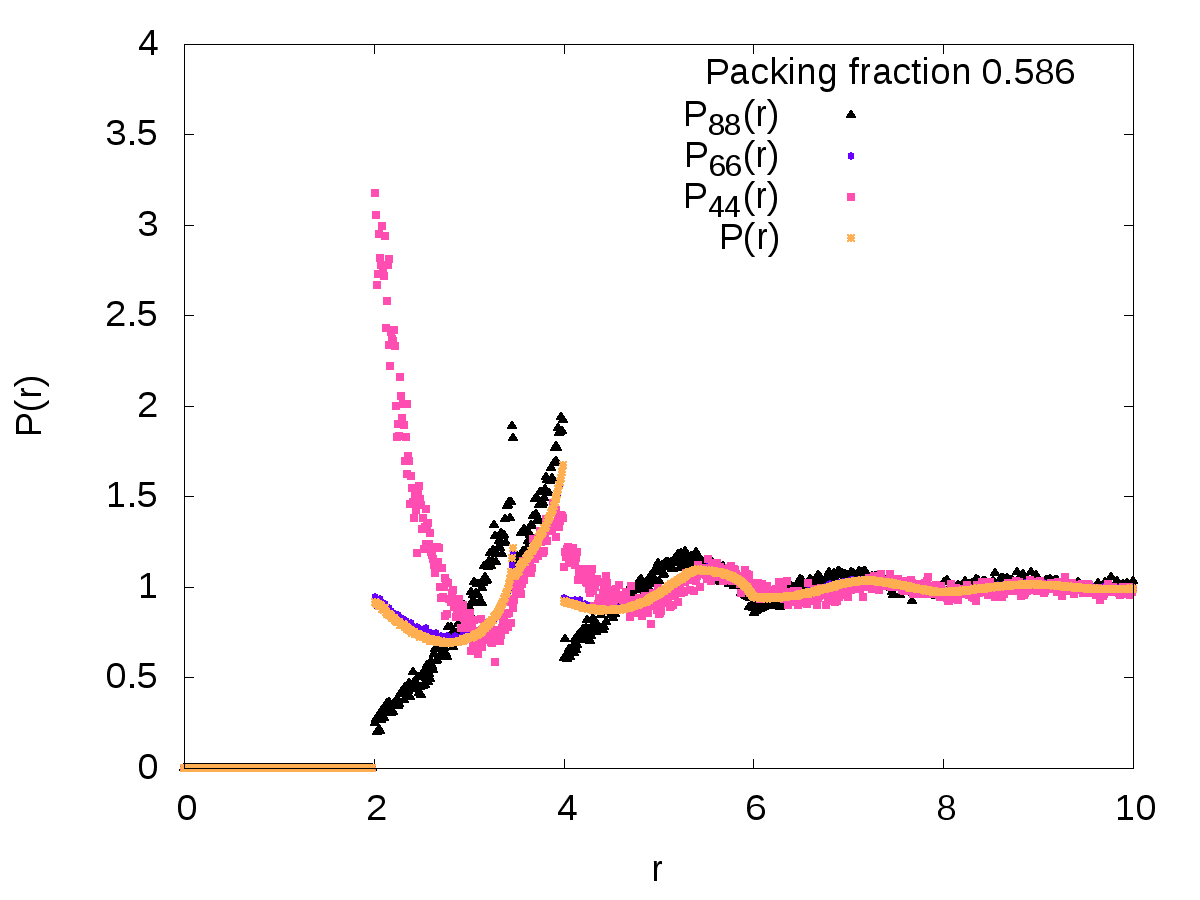}}
\subfigure[]{\includegraphics[width=0.48\textwidth]{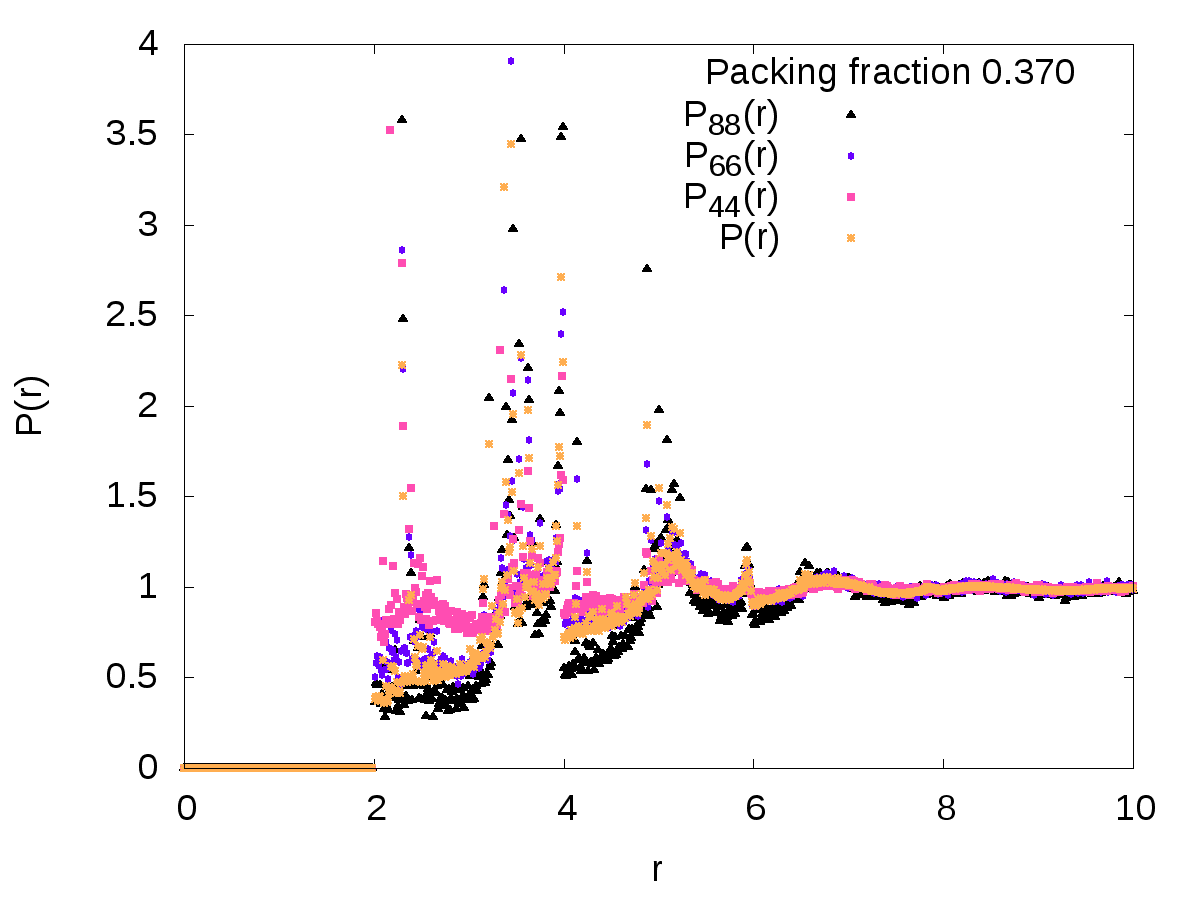}}\\
 \caption{Sample $P(r)$ and $P_{ii}(r)$ obtained for aggregates generated by algorithms a) MAX-1 ($\gamma = 0.586$) b) MAX-1 ($\gamma = 0.370$).\label{pii_de_r_compa}}
\end{center}
\end{figure}

\subparagraph{$P_{ii}(r)$} The PPDF $P_{ii}(r)$ together with the corresponding PDF are presented in figures \ref{pii_de_r_compa}.a and b for MAX-1 aggregates with the two most extreme packing fraction, $\gamma = 0.586$ and $\gamma = 0.370$.
First and foremost, for a given packing fraction, the number of quasi first neighbours (corresponding to values of $r$ close to $d$) decreases when the coordination number ($i$) increases, like in the case of RRPA and for the same reasons: the most contacting neighbours a sphere has, the less quasi first neighbours it can accept. 

On the other hand, the comparison between figure \ref{pii_de_r_compa}.a and b, shows that the number of quasi first neighbours decreases with packing fraction. At low packing fractions, they form plateaus whose level increases when the coordination number decreases, whereas they have a very distinct behaviour for high packing fractions: low coordination number spheres possess a lot of quasi first neighbours while high coordination number ones have a very limited number of QFN.

Both topological discontinuities (at $r = \sqrt{3}d$ and $r=2d$) increase when the coordination number increases and when packing fraction increases.

Finally, (see supplementary informations for figures) "iso-packing-fraction" and "iso-$\betaD$" aggregates obtained by different building algorithms show significant differences in their global and partial PDF, confirming that parameter $\gamma$ and $\betaD$ are greatly insufficient for a full structural description of random aggregates. Aggregates sharing similar values of these two parameters ($\gamma$ and $\betaD$) still present significant differences in structural properties. Hence, even when used together, $\gamma$ and $\betaD$ are not satisfying predictors of structural properties of disordered systems. Besides, it has been impossible to find aggregates with similar $\bLmax$ and $\gamma$ to compare them in a similar fashion leaving the question open for a possible combination of these two parameters as good predictor of random aggregates global structure. Conversely, it turns out that PPDFs allow a more sensitive distinction between aggregates built by different algorithms than the corresponding global PDF and are thus interesting structural descriptors.

\subparagraph{$P_{ij}(r)$ with $i\neq j$} 
A sampling of $P_{ij}(r)$ curves is shown in supplementary information. They show that QFN are favoured by higher packing fractions and lower coordination numbers $i$ and $j$. The implicit "chemical" ordering in these curves will be studied in more details hereafter.

\begin{figure}[htbp]
\begin{center}
\subfigure[]{\includegraphics[width=0.48\textwidth]{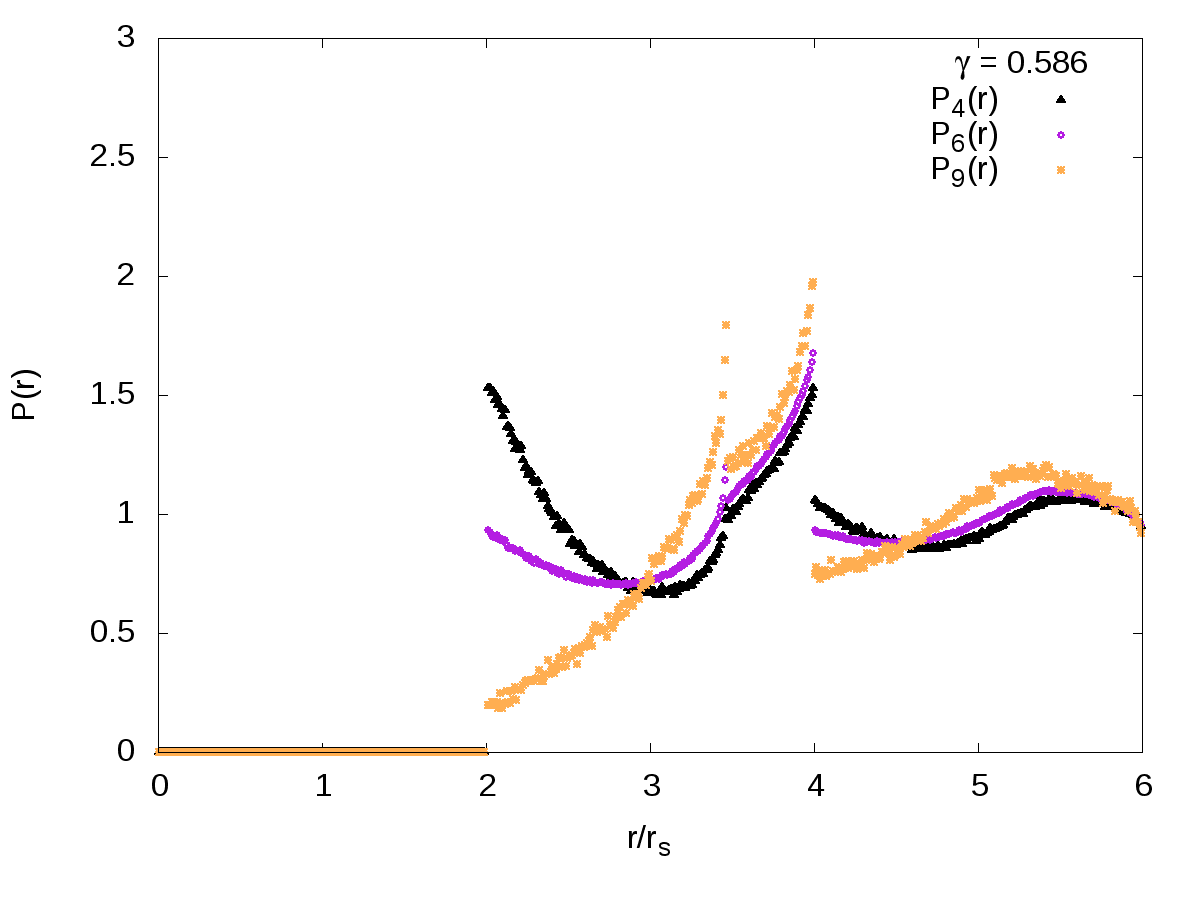}}
\subfigure[]{\includegraphics[width=0.48\textwidth]{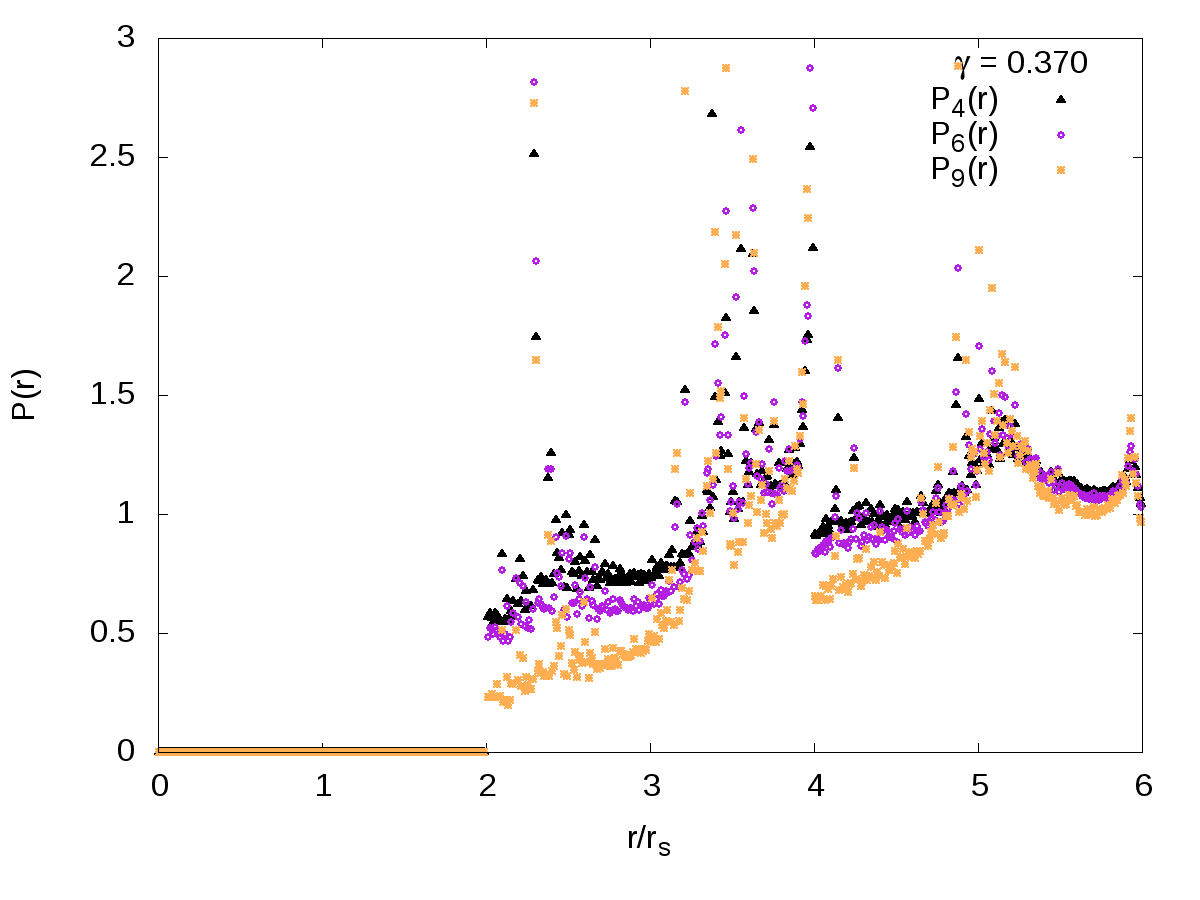}}
 \caption{$P_{i}(r)$ for two packing fractions, algorithm MAX-1.\label{pi_de_r_compa}}
\end{center}
\end{figure}

\subparagraph{$P_i(r)$}
$P_i(r)$ (for $i=4,6$ and 9) obtained for aggregates built by MAX-1 algorithm with packing fraction 0.370 and 0.586 are displayed in figure \ref{pi_de_r_compa}.
These PPDF appear to depend strongly on $i$ and packing fraction. On the one hand, for each packing fraction:
\begin{itemize}
\item quasi first neighbours increase when the coordination number decreases;
\item topological peaks at $r=d\sqrt{3}$ and $2d$ are more intense for high coordination numbers;
\end{itemize}

On the other hand, the main differences between low and high packing fraction aggregates are that:
\begin{itemize}
\item polytetrahedral $\delta$ peaks are noticeable on aggregates with low packing fractions;
\item the transition from the high to the low packing fraction is mostly due to a decrease of quasi first neighbours around low coordination spheres;
\item in the case of the densest aggregate, $P_i$ oscillations are increasingly shifted toward smaller $r$ values as the coordination number $i$ increases, while such a shift is not observed for the lowest packing fraction aggregate.
\end{itemize}

\subsection{Radial evolution of local packing fraction from pair distribution functions}
\subsubsection{Principle}
Knowing pair distribution functions, it becomes possible to study the variation of the local packing fraction around an average sphere as a function of $r$.
Using relation \ref{eqvolSs}, the packing fraction of a sphere $\GS$ with arbitrary diameter $\Rvoi>d$ situated within the aggregate writes:
\begin{equation}
\gamma (\Rvoi) = \frac{4\pi r_s^3/3 + \rho \int_0^{\Rvoi+r_s} P(r)4\pi r^2 V_s(r,\Rvoi) dr}{4\pi\Rvoi^3/3 }
\end{equation}
with $\rho=N/V$, the number of spheres per unit volume in the aggregate;
the $4\pi r_s^3/3$ term corresponds to the sphere in $r=0$; $V_s(r)$  is defined by equation \ref{eqvolSs}; $P(r)$ can be a total or partial pair distribution function. It also comes $\gamma (r\leq r_s)=1$.

It is also possible to remove the contribution of the central sphere and its contact first neighbours to packing fraction, which writes: 
\begin{equation}\label{eq1}
\gamma_{CN}(\Rvoi) = \frac{4\pi r_s^3/3 + iV_s(r=d,\Rvoi)}{4\pi \Rvoi^3/3}
\end{equation}
where $i$ is the number of contacting neighbours of the central sphere. 
It then comes:
\begin{equation}\label{eq:gWCN}
\gamma_{WCN}(\Rvoi) = \gamma (\Rvoi) - \gamma_{CN}(\Rvoi)
\end{equation}
$\gamma_{WCN}$ is the contribution to packing fraction of quasi-first and further neighbours around an average sphere.

\subsubsection{Local packing fraction around average sphere}

\begin{figure}[htbp]
\subfigure[]{\includegraphics[width=0.48\textwidth]{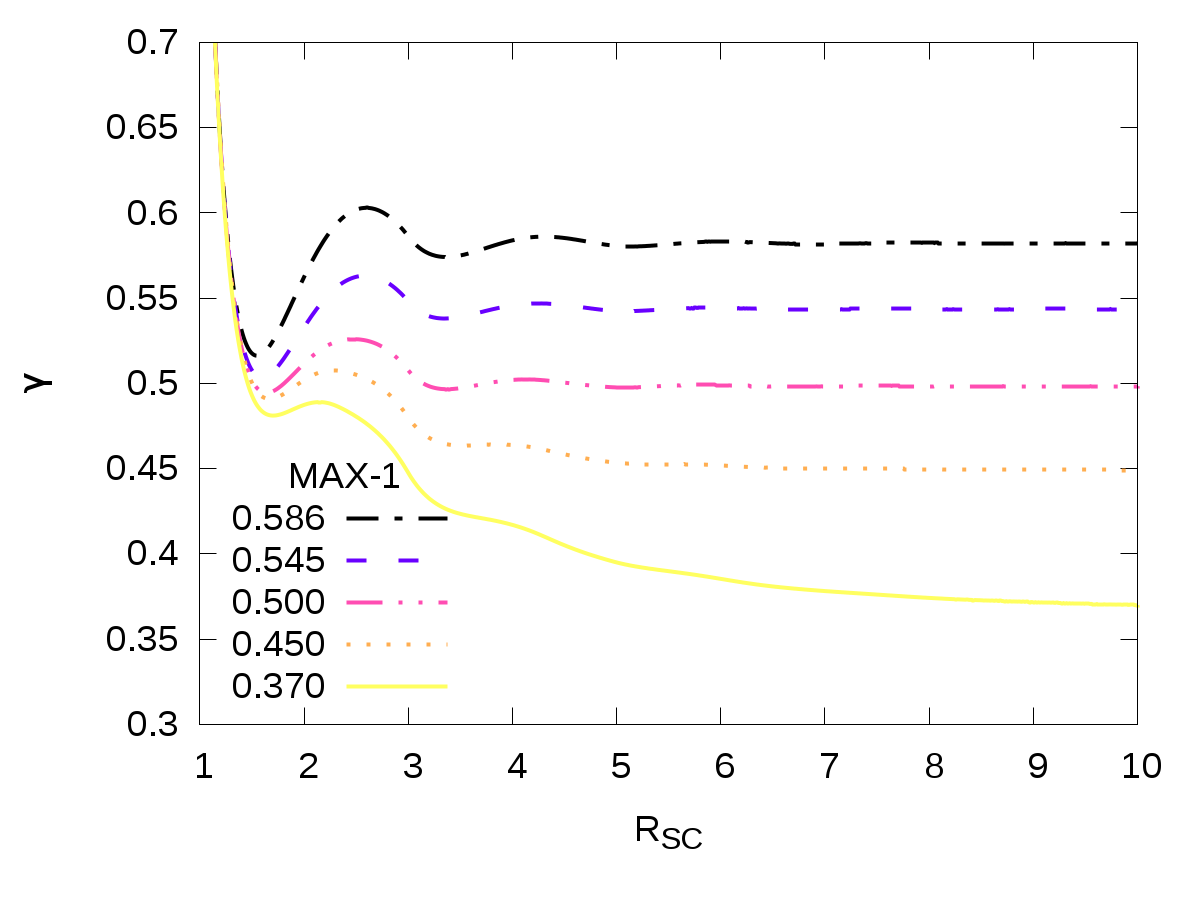}}
\subfigure[]{\includegraphics[width=0.48\textwidth]{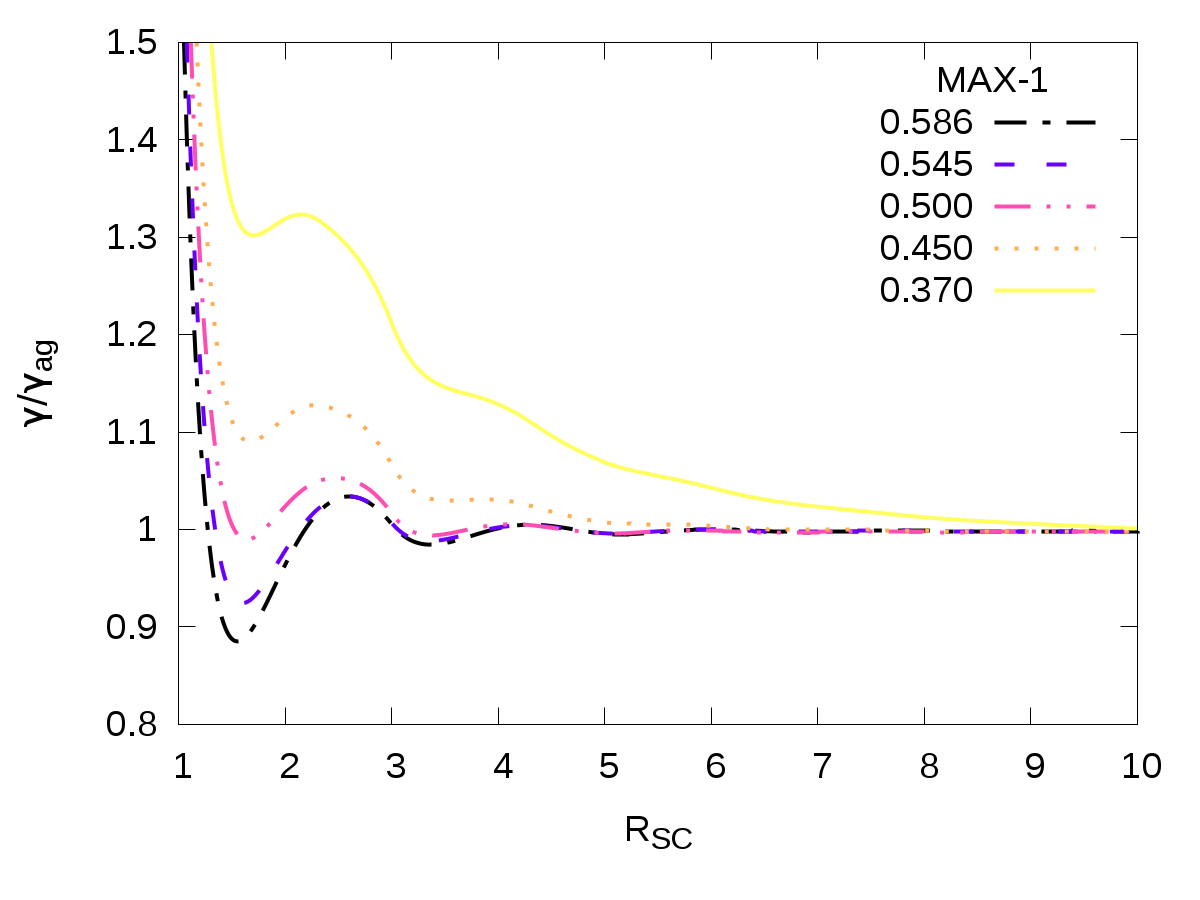} }
\subfigure[]{\includegraphics[width=0.48\textwidth]{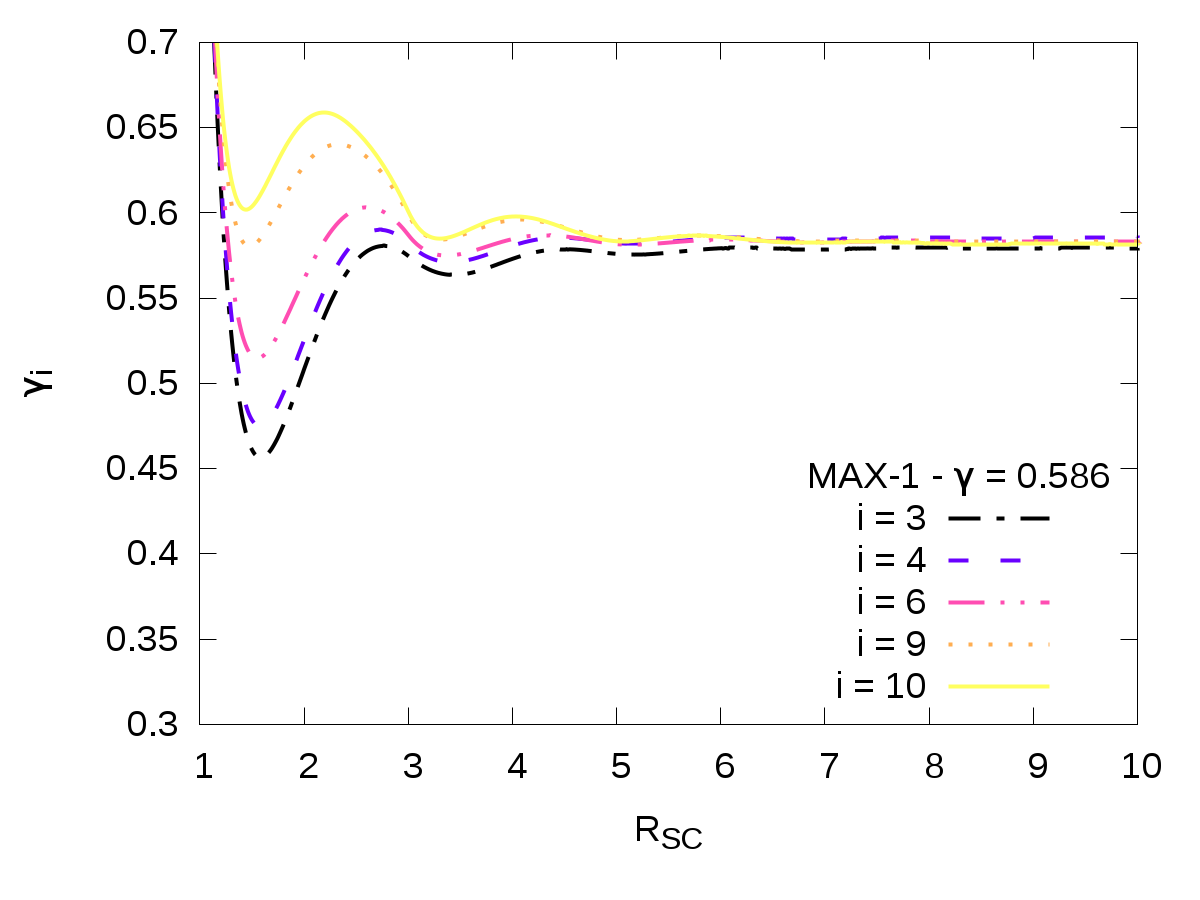}}
\subfigure[]{\includegraphics[width=0.48\textwidth]{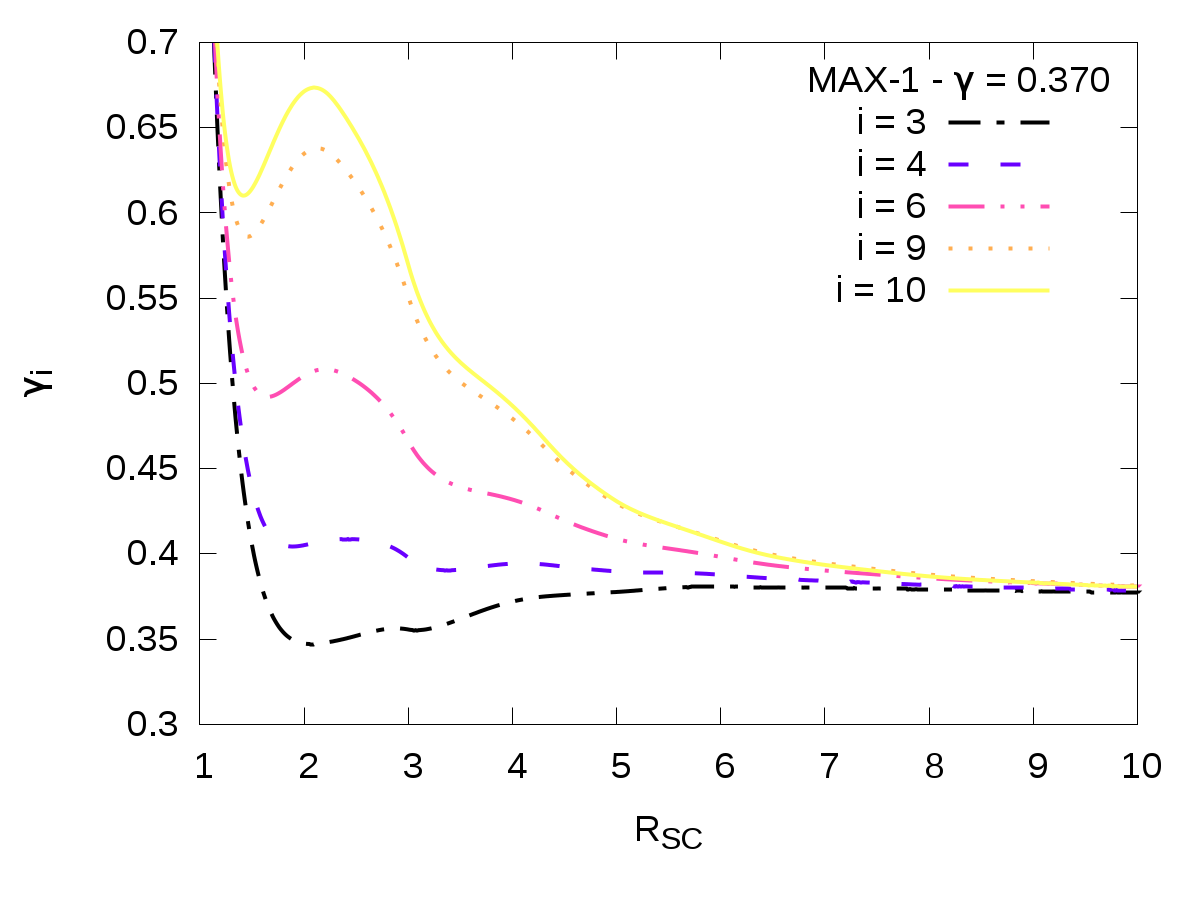}}
\subfigure[]{\includegraphics[width=0.48\textwidth]{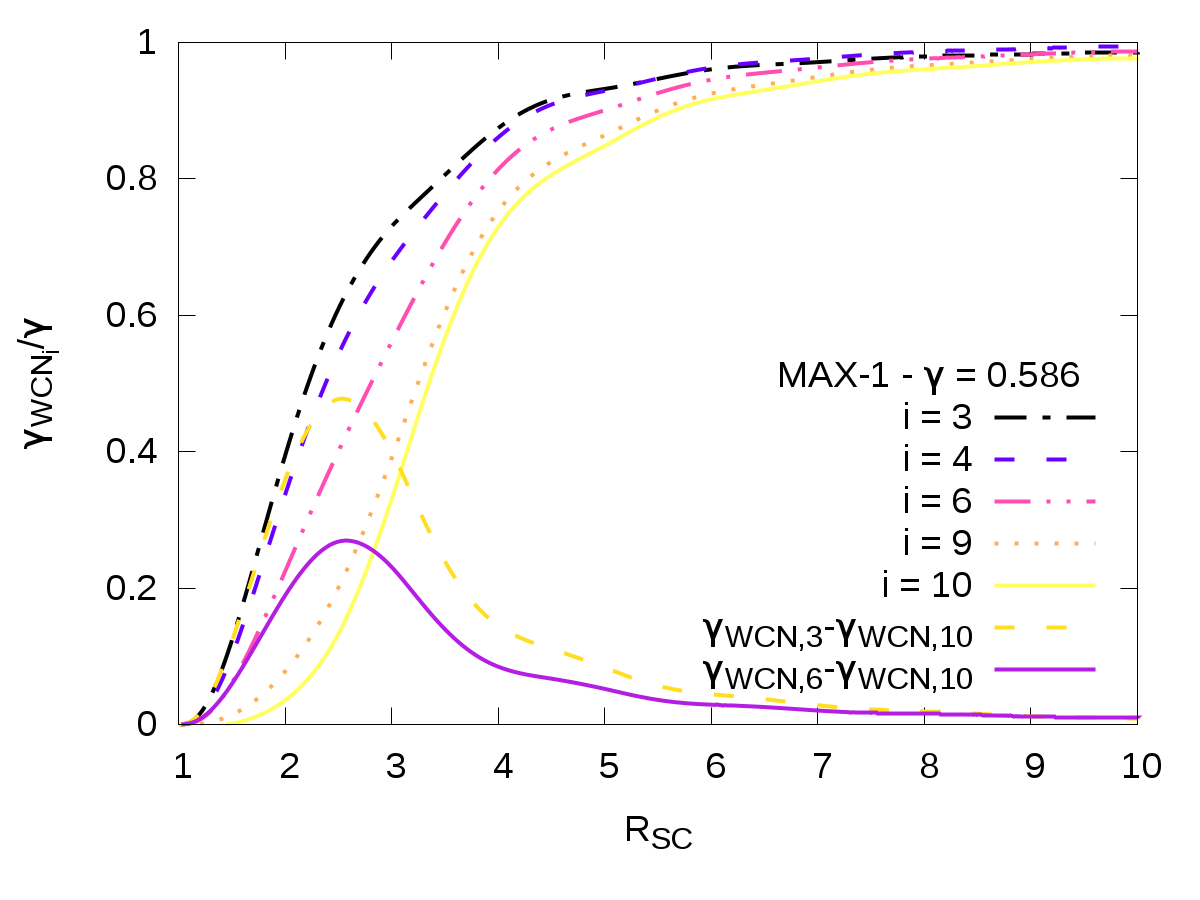}}
\subfigure[]{\includegraphics[width=0.48\textwidth]{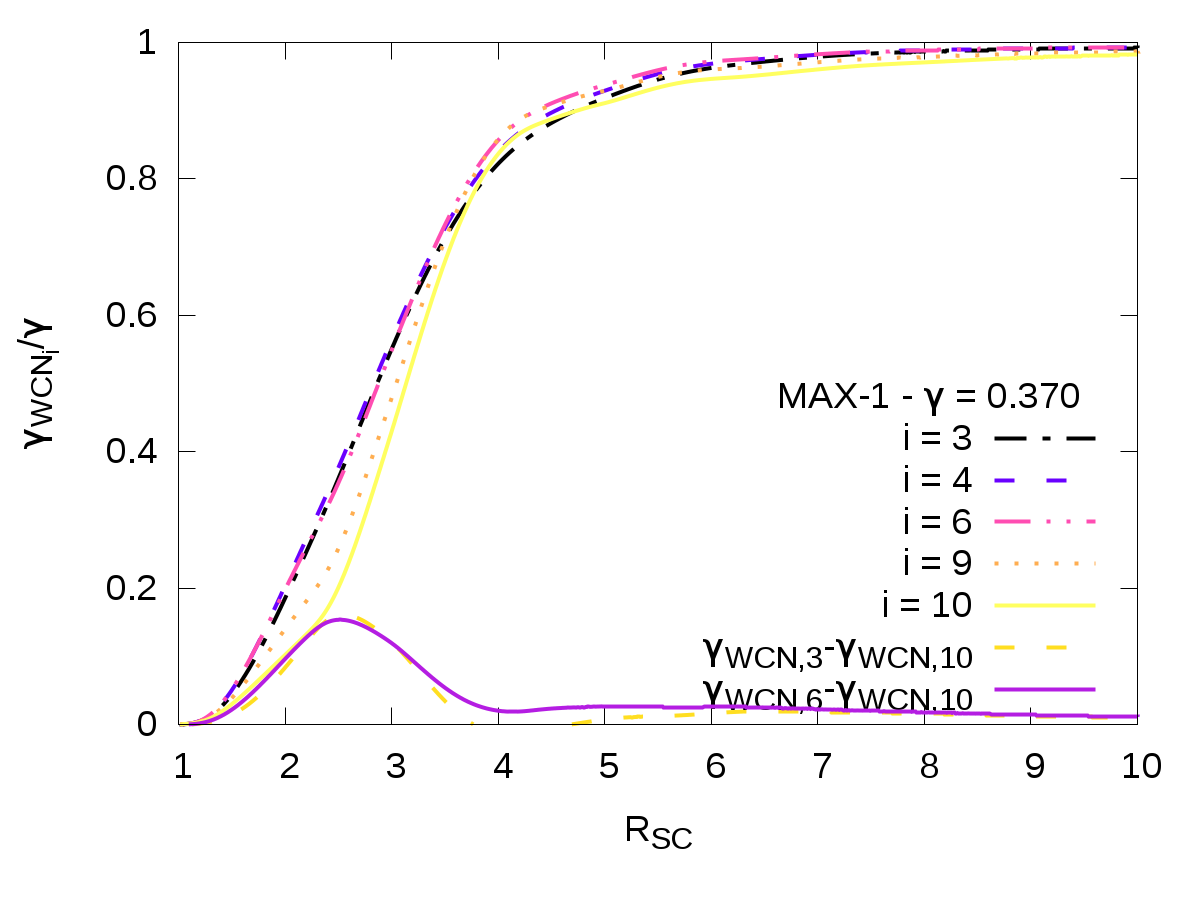}}
 \caption{a) Local packing fraction b) local packing fraction divided by average PF for various MAX-1 aggregates.\label{densitysph}
          Radial evolution of the packing fraction around $i$ coordinated spheres for MAX-1 aggregates with c) $\gamma = 0.586$ and d) $\gamma = 0.370$.
          Radial evolution of packing fraction around $i$ coordinated spheres without contact neighbours (relation \ref{eq:gWCN}) for MAX-1 aggregates with e) $\gamma = 0.586$ (the bottom lines represent the difference between the highest and lowest bound, i.e. $\gamma_{WCN,3}(\Rvoi) - \gamma_{WCN,10}(\Rvoi)$ and intermediate, i.e. $\gamma_{WCN,6}(\Rvoi) - \gamma_{WCN,10}(\Rvoi)$) and f) $\gamma = 0.370$.}
\end{figure}

Using the global PDF, the formalism just introduced gives access to the variation of the local packing fraction as a function of $r$ around an average sphere. Figure \ref{densitysph}.a represents $\gamma(\Rvoi)$ and figure \ref{densitysph}.b represents $\gamma(\Rvoi)/\gamma$, which goes to 1 as $\Rvoi$ goes to $\infty$. All aggregates behave differently depending on their packing fraction. For the densest ones, the packing fraction falls below the average value, before converging more rapidly at large $r$ than the least dense ones. On the other hand, in the case of the least dense ones, packing fraction remains above the average value and converges more slowly. 
Oscillations of local packing fractions are damped at about $r=5d$ for the lowest packing fraction and around $r=3d$ for the highest packing fraction.

\subsubsection{Local packing fraction around $i$ coordinated spheres}\label{sec:locdenssph_i}
Using $P_i(r)$, it is possible to determine the local packing fraction as a function of $r$ for an average sphere with coordination number $i$, $\gamma_i$.
Figures \ref{densitysph}.c and d present the sets of corresponding curves for the two MAX-1 aggregates with the lowest and highest packing fractions. 
The short-range local-density grows with the coordination number and the packing fraction of the aggregate. 
Significant oscillations are observed for the aggregate with the highest packing fraction. Their amplitude is damped for the aggregate with the lowest packing fraction. Figures \ref{densitysph}.e and f show the evolution of packing fraction without the contribution of contact neighbours ($\gamma_{WCN}$) and the central sphere, hence only that of quasi-first and further neighbours. As it could be expected, the lower number of quasi first neighbours (QFN) around spheres with high coordination numbers leads to a significantly lower packing fraction due to QFN than in the case of spheres with low contact coordination number. 

In the case of the aggregate with the highest packing fraction (fig. \ref{densitysph}.e), the various curves converge for $\Rvoi \approx 3.5d $ and the highest difference is observed between $i=10$ and $i=3$ for $\Rvoi \approx 2.53$, $\gamma_{WCN,3}-\gamma_{WCN,10}\approx0.28 = 0.48\gamma_{ag}$.  Between $i=10$ and $i=6$, the maximum is found at $\Rvoi \approx 2.57$, with $\gamma_{WCN,6}-\gamma_{WCN,10}\approx0.16 = 0.27\gamma_{ag}$.

There is a much lower difference in contribution of the QFN between spheres with different contact coordination number in the case of the aggregate with the lowest packing fraction (in terms of absolute as well as relative value). However, the maximum of the difference between $\gamma_{WCN,3}$ and $\gamma_{WCN,10}$ is obtained for approximately the same value $\Rvoi \approx 2.55$ and $\gamma_{WCN,3}-\gamma_{WCN,10}\approx0.06 = 0.16\gamma_{ag}$. Between $i=10$ and $i=6$, the maximum is found at $\Rvoi \approx 2.51$, with $\gamma_{WCN,6}-\gamma_{WCN,10}\approx0.06 = 0.16\gamma_{ag}$, which is remarkably similar to the former. In this case, figure \ref{densitysph}.e shows that $\gamma_{WCN}(\Rvoi)$ for coordination numbers below 7 behave very similarly, suggesting that these types of spheres have more or less the same environment in terms of number of QFN, irrespective of their contact coordination number. 

\subsection{Global and partial structure factors}
The previous structural description  achieved by the PPDF analysis of random packings leads naturally to the study of the corresponding global and partial structure factors which can be compared with the results of diffraction experiments on disordered materials.

\subsubsection{Global and Ashcroft Langreth partial structure factors}
Let us first introduce the global structure factor:
\begin{equation}\label{eq:SQglob}
S(Q) = 1 + \frac{N}{V}\int_0^\infty \Big(P(r)-1\Big) \frac{\sin(Qr)}{Qr}4\pi r^2 dr
\end{equation}
and the partial Ashcroft Langreth (AL) structure factors, defined by:
\begin{equation}\label{eq:AL}
S_{ij}(Q) = \delta_{ij} + \frac{\sqrt{N_iN_j}}{V} \int_0^\infty \Big(P_{ij}(r)-1\Big)\frac{\sin (Qr)}{Qr}4\pi r^2 d r
\end{equation}
where $\delta_{ij}$ is the Kronecker symbol. 

The diagonal terms, $S_{ii}$, represent the structure factors of the partial aggregates formed by spheres with contact coordination $i$ and will be studied first. The non diagonal terms $S_{ij}$ with $i\neq j$ describe the mutual or "chemical" arrangement between $i$ and $j$ coordinated spheres and will be studied later by the more suited Bhatia-Thornton formalism. 

$S(Q)$ and $S_{ij}(Q)$ values for $Q \ll 1/R$ cannot be calculated from relations \ref{eq:SQglob} and \ref{eq:AL} owing to the finite radius $R$ of the aggregates \cite{BB15}. However $S(Q=0)$ values are determined by the statistical density fluctuations according to relation \cite{KB51}:
\begin{equation}\label{eq:flucij}
S_{ij} (0) = \frac{\overline{N_iN_j} - \overline{N_i}\ \overline{N_j}}{\sqrt{\overline{N_i}\ \overline{N_j}}}
\end{equation}
where upper bars represent averages.

Fluctuations were directly derived from the positions of the sphere centers. To that purpose, a large cube with edge $50d$ centered on the aggregates origin has been subdivided into 1000 sub-cubes with edge $5d$, each of them containing $N_i$ and $N_j$ sphere centers with respective coordination $i$ and $j$. Average values of $\overline{N_i},\ \overline{N_j}$ and $\overline{N_iN_j}$ were taken over the 1000 sub-cubes to get the statistical fluctuations involved in relation \ref{eq:flucij}.

\begin{figure}[htbp]
\subfigure[]{\includegraphics[width=0.48\textwidth]{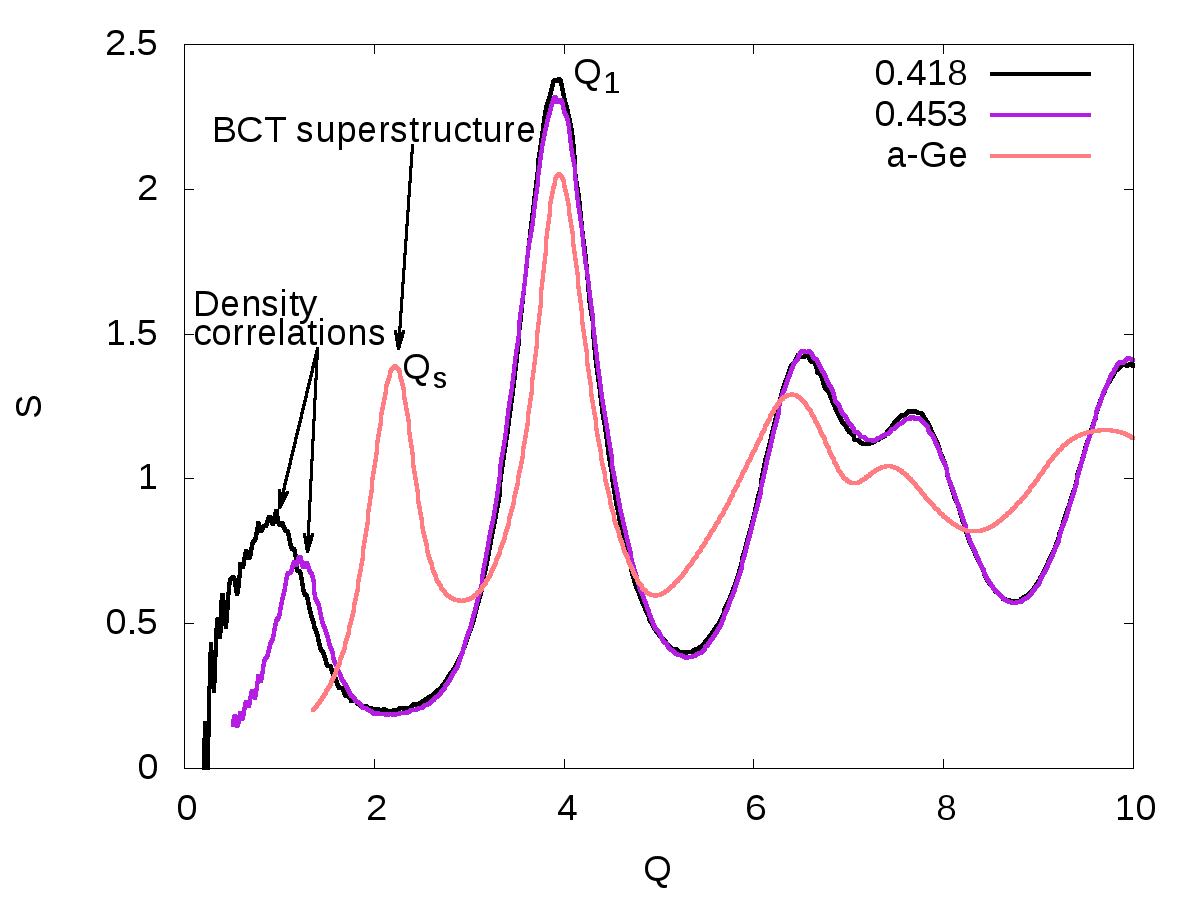}}
\subfigure[]{\includegraphics[width=0.48\textwidth]{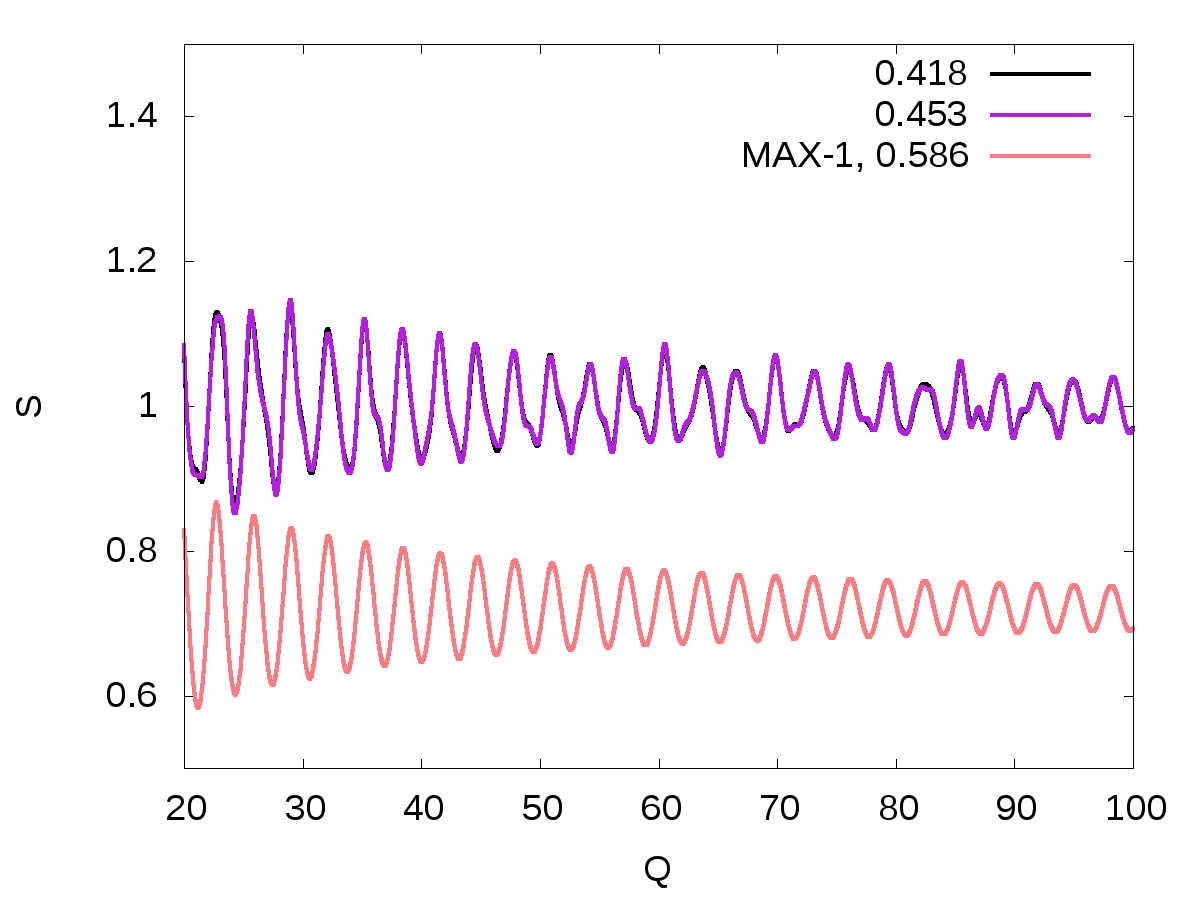}}\\
 \caption{a) Structure factor of two RRPA aggregates and experimental results obtained on amorphous Ge (data obtained in \cite{EWWDCS82}) at small $Q$ b) large $Q$ behaviour of the structure factor of the RRPA aggregates and one (high packing fraction) RA aggregate.\label{SQ_tetraparf}}
\end{figure}

\paragraph{Random regular polytetrahedral packing -- global structure factor}
Figure \ref{SQ_tetraparf} introduces the global structure factor of two RRPA: \ref{SQ_tetraparf}.a (small $Q$) and b (large $Q$ or asymptotic regime).
It turns out that the structure factors of RRPA exhibit a prepeak, whose position depends on the used algorithm or packing fraction. The position and intensity of this pre-peak corresponds to the folding of the polytetrahedra which controls the space correlation of density fluctuations. It accounts for interferences between high density and low density zones of the single polytetrahedron. At high $Q$, $S(Q)$ appears aperiodic due to the infinite series of $\delta(r_p)$ peaks in $P(r)$ corresponding to the distances between vertices in the infinite RRPA and is only weakly affected by small changes in packing fraction (depending on the used algorithm).

\subparagraph{Comparison with the structure of tetravalent elements}
The polytetrahedral structure factor can be usefully compared with the experimentally measured structure factors of pure tetravalent elements (C \cite{GSCM91}, Ge \cite{EWWDCS82}, Si \cite{LKRCRWM99}) or with the tetrahedral structure of oxygen atoms in glassy water \cite{NDL67,NVR76}. 
As a matter of fact the structure of these elements is based on body centered regular tetrahedra (BCT) connected by their vertices \cite{P74}, while RRPA's are made from simple regular tetrahedra connected by their faces. 
As a consequence, a "pseudo superstructure" peak for such hard-sphere based centered-tetrahedra should correspond to $Q_s/Q_1 = d_1/d_s = \sqrt{6}/4 = 0.612$, where $d_1$ is the distance between contacting atoms and $d_s$ is the edge length of the regular centered tetrahedron. 
Experimentally, Etherington finds $Q_s/Q_1=0.572$ \cite{EWWDCS82} for the structure factor of amorphous Ge, Laaziri et al. find $0.583$ for the same ratio in a-Si \cite{LKRCRWM99} and Gaskell et al. find $0.540$ for a-C \cite{GSCM91}. 
The discrepancy from 0.612 for these three values could be due to the error introduced by hard sphere approximation of tetrahedral bond angle between first neighbours of softer potentials, which does not allow reproducing perfectly the behaviour of covalent materials. 

However, this superstructure peak is lacking from the structure factor of RRPA's but another small angle peak at significantly lower value is found around $Q_s/Q_1 \approx 0.25$, varying with packing fraction. The latter can be attributed to the pseudo periodic correlation of density fluctuations present in the RRPA model. These correlations happen over larger distances than those of the BCT, hence the smaller $Q$ values of the corresponding pre-peak. It corresponds roughly to the end of significant oscillations of pair distribution functions.

The reason why this longer correlation-distances pre-peak does not appear for tetravalent elements has probably to do with the fact that tetrahedra forming their structures are connected by vertices, allowing for many more degrees of freedom and preventing longer distances correlations than in the case of RRPA aggregates where tetrahedra are connected by their faces, which is geometrically more constraining. 
On the other hand, the second peak of the structure factor of the RRPA splits into two subpeaks in a completely analogous manner with what is found in amorphous tetravalent elements. 

Finally, for RRPA aggregates, characteristic distances are strictly defined, giving rise to $\delta$ peaks on the PDF resulting in aperiodic large $Q$ behaviour, also absent from the structure factor of tetravalent elements. In contrast, the lack of such well defined distances, because of higher degrees of freedom in the orientations of tetrahedra relatively to one another in the case of tetravalent element, results in a single $\delta$ peak on their PDF, which accounts for their periodic behaviour in the large $Q$ regime \cite{B90}.

\paragraph{Random irregular polytetrahedral aggregates}

\begin{figure}[htbp]
\subfigure[]{\includegraphics[width=0.48\textwidth]{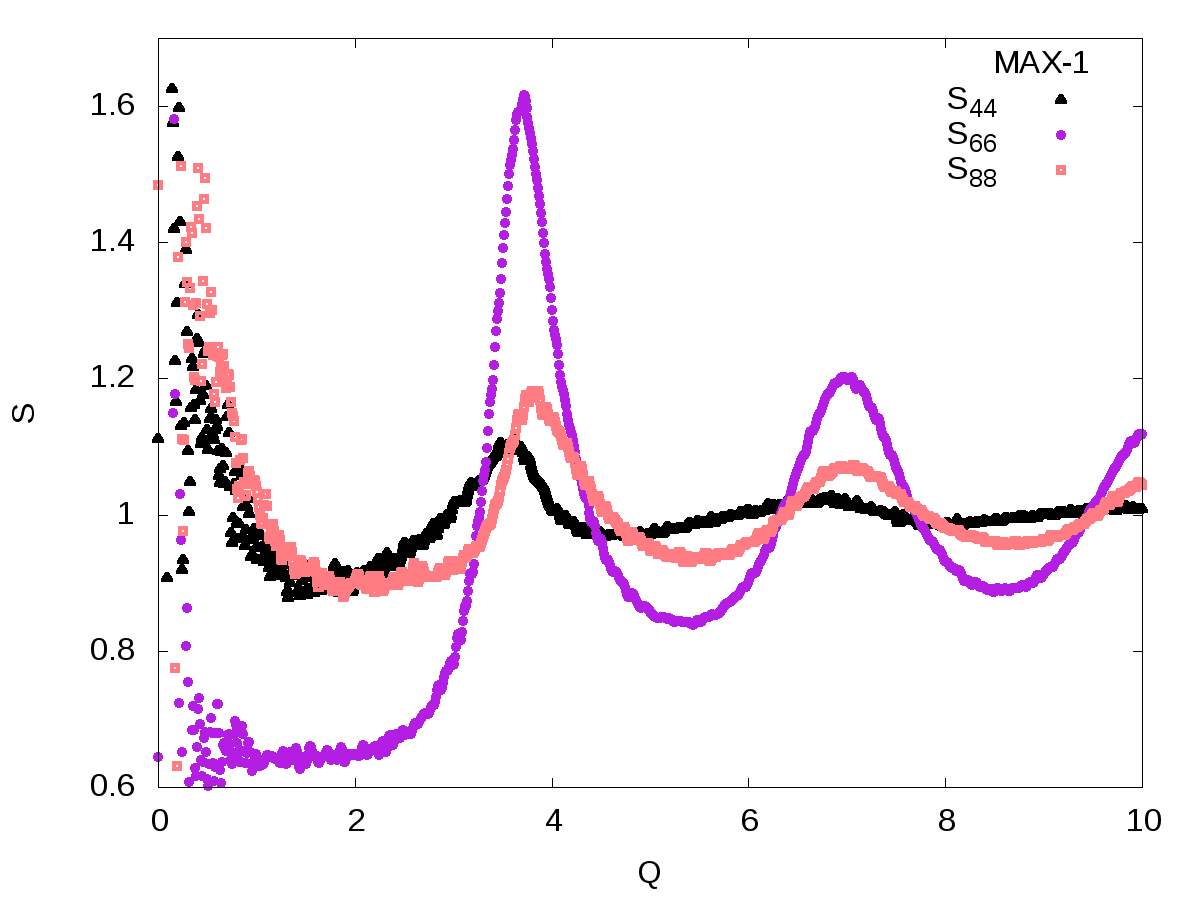}}
\subfigure[]{\includegraphics[width=0.48\textwidth]{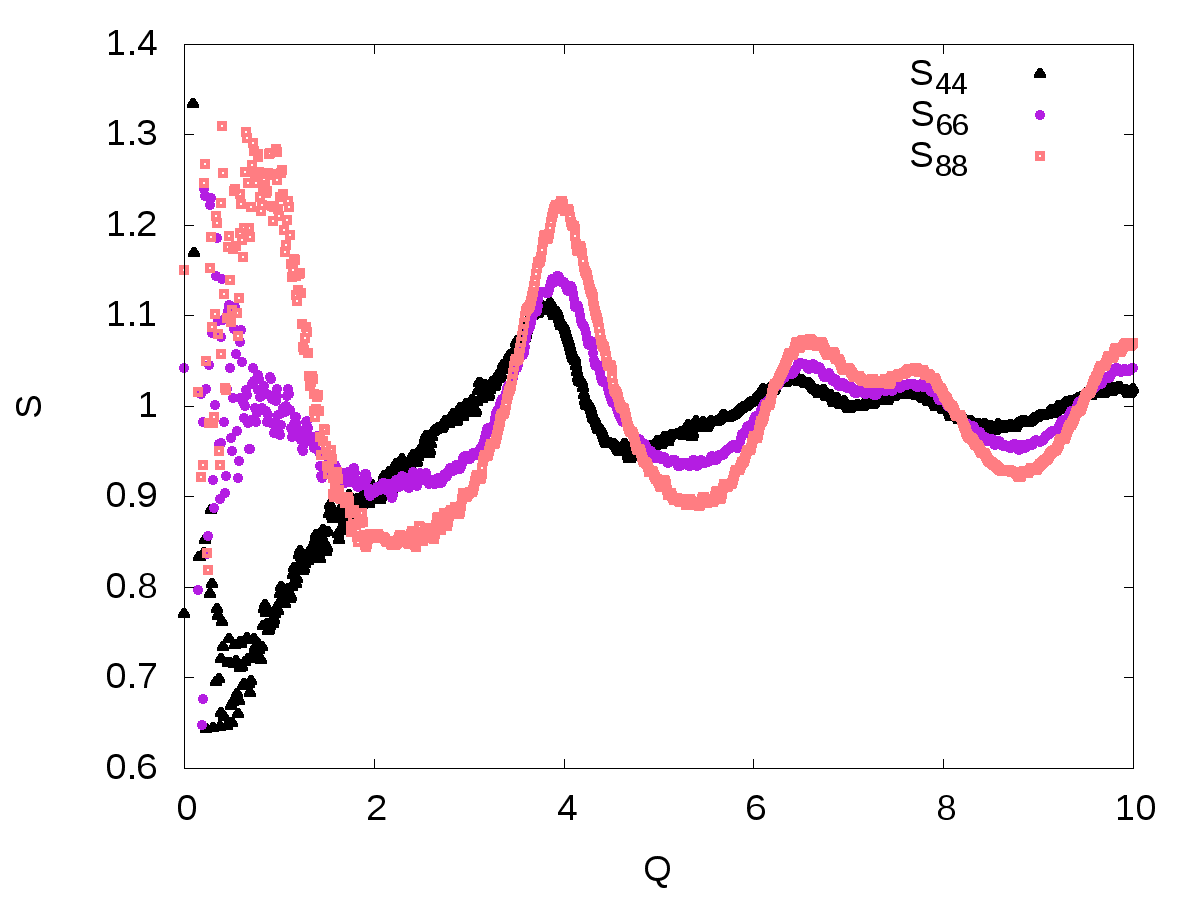}}
 \caption{Ashcroft-Langreth partial structure factor for a) highest packing fraction RIPA, b) RRPA.\label{Ashcroft_petQ}}
\end{figure}

The AL partial structure factors of random irregular polytetrahedral aggregates are presented in figure \ref{Ashcroft_petQ}.a to c, along with the partial structure factors of the RRPA (\ref{Ashcroft_petQ}.d). 
These structure factors all have in common that the global amplitude of their oscillations increases from $i=3$ up to $i=6$ and then decreases as $i$ increases beyond 6.
Moreover, the position of the first peak of $S_{ii}$ goes to small $Q$ when packing fraction increases (whatever $i$ and $\gamma$).
At large $Q$, there is no phase shift, whatever $i$ and $\gamma$, and $S_{ii}$ oscillates like $\sin (Qd)/Qd$ (not shown here).

Some peculiarities are also observed. A shoulder appears on the left of the first peak of $S_{44}$ (around $Q=3.8$). 
For $S_{88}$ and $S_{44}$ an intense pre-peak around $Q=0.8$ is noted, whose intensity increases with the packing fraction for $S_{44}$ and decreases when the packing fraction increases for $S_{88}$. This pre-peak is due to correlated density fluctuations between low or high density regions separated by an average distance of $2\pi/0.8 r_s \approx 8r_s$ lying beyond the main $P_{ii}$ oscillations; they are observed whatever the packing fraction, even if there is no noticeable pre-peak on the global structure factor.
In the case of $S_{66}$, $S_{66}(0)$ decreases when $\gamma$ increases and barely any pre-peak may be observed at all. The first peak remains symmetrical; its position depends little on packing fraction. The first peak of $S_{88}$ becomes strongly asymmetrical for the highest packing fraction. 

Finally, a splitting of the second peak is observed on the AL structure factors with high coordination numbers and low packing fraction. This splitting has been noticed above for the RRPA and it is consistent with the existence of a RP structure embedded in a "continuum" random structure (respectively RP and FR structural components), as was already concluded in \cite{BB15} for these aggregates.

\subsubsection{Bhatia Thornton partial structure factor}
The partial structure factors introduced by Bhatia and Thornton \cite{BT74} are associated with density and concentration correlations in binary alloys. They have been extended to alloys consisting of more than two components \cite{B76} and can then be used by considering random aggregates as alloys of spheres with various coordination numbers.

\paragraph{Formalism}
The partial structure factor corresponding to:
\begin{itemize}
\item density-density correlation function, $S_{NN}$, writes:
\begin{equation}\label{eq:snngen}
S_{NN}(Q) = 1+\rho \int_0^\infty \Big(P(r) - 1\Big) \frac{\sin(Qr)}{Qr} 4\pi r^2 dr
\end{equation}
\item concentration-concentration correlation function, $S_{C_iC_j}$, writes:
\begin{equation}\label{eq:sccgen}
S_{C_iC_j}(Q) = 1+\rho \int_0^\infty \Big(-P(r) + P_i(r) + P_j(r) - P_{ij} (r)\Big) \frac{\sin (Qr)}{Qr}4\pi r^2 d r
\end{equation}
\item density-concentration, $S_{NC_i}$, writes \cite{B90,B79}:
\begin{equation}\label{eq:sncgen}
S_{NC_i}(Q) = \rho \int_0^\infty \Big(P_i(r) - P(r)\Big) \frac{\sin (Qr)}{Qr}4\pi r^2 d r
\end{equation}
\end{itemize}
For the sake of simplicity, we only consider hereafter sphere mixtures made of two components, namely $i$ and $j$ coordinated spheres. The corresponding relations are provided in supplementary informations. 

\paragraph{Results}
Between 15 (for high packing fraction) and 45 sets of Bhatia-Thornton structure factors corresponding to different coordination pairs $i-j$ were calculated for 500 aggregates. In the case of high packing fraction aggregates where "extreme" coordination numbers (i.e. $i=3$ and $i>9$) are scarce, BT sets could only be calculated for coordination numbers lying between 4 and 8, hence the lower number of BT structure factors calculated for them.

A sample of Bhatia Thornton partial structure factors is provided in supplementary informations for three MAX-1 aggregates, with maximum, minimum and intermediate packing fraction along with those of one RRPA aggregate.

Their main characteristics are the following:

\subparagraph{$S_{N_iN_j}$} The $S_{N_iN_j}$ structure factor is related to the overall structure of the two coordinations $i$ and $j$ considered. It looks like the average AL structure factor for small coordination differences ($i-j = 1$ or $2$ at most), while it resembles more to the global structure factor of the aggregate for "large" differences in coordination numbers ($i-j\geq 4$)

\subparagraph{$S_{NC_i}$} structure factors exhibit significant oscillations whose intensity increases with the coordination difference $i-j$. Furthermore, for low coordination differences these oscillations decrease when the coordination numbers increase. All these oscillations come from the fact that the average environment of sphere $i$ and $j$ characterized by $P_i$ and $P_j$ differ from the average global environments $P(r)$ (see relation \ref{eq:sncgen}). This behaviour differs completely from the case of binary "substitution" alloys (with equal atomic diameters) for which $S_{NC}$ does not oscillate at all because $P_i(r)\approx P(r)$, independently of the chemical order of the two alloy components (which only affects $S_{CC}$ \cite{B90}).

\subparagraph{$S_{C_iC_j}$} Most interesting is the study of $S_{C_iC_j}$, which characterizes the "chemical" order between $i$ and $j$ coordinated spheres through its dependence on $P_{ii}-P_{ij}$ (see equation \ref{eq:sccgen}). For small coordination differences (at most 2) and all packing fractions, $S_{C_iC_j}$ oscillates weakly around 1 and there is no "chemical" order effect between $i$ and $j$ coordinated spheres.

\begin{figure}[htbp]
\subfigure[]{\includegraphics[width=0.48\textwidth]{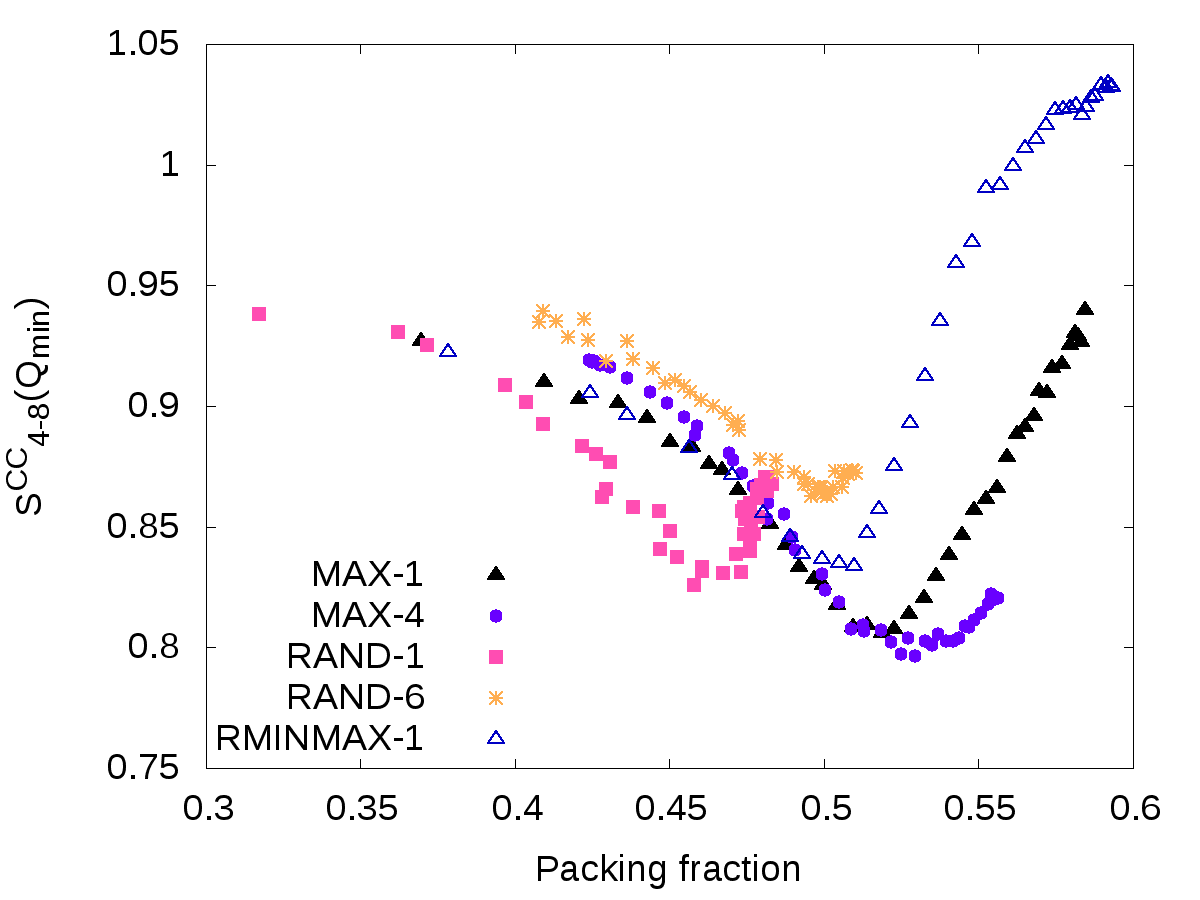}}
\subfigure[]{\includegraphics[width=0.48\textwidth]{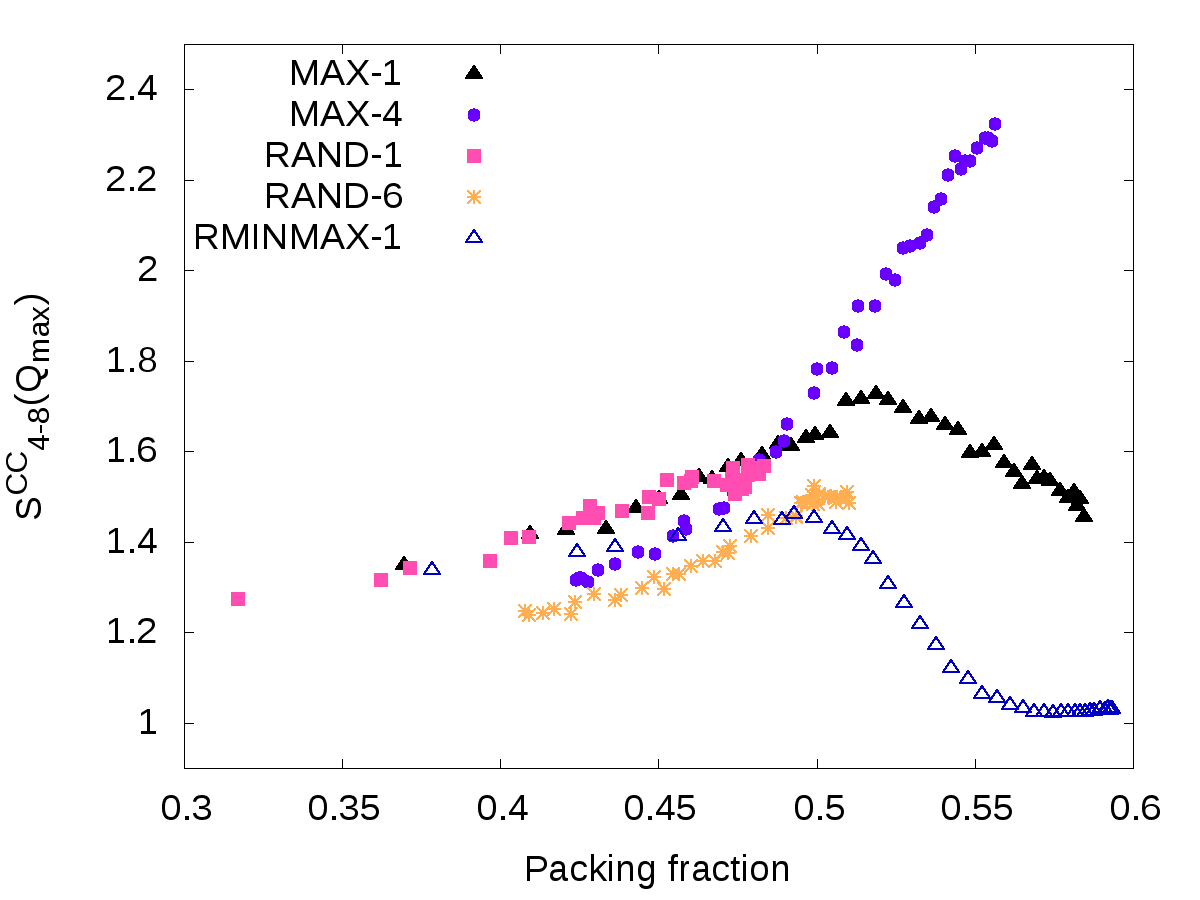}}
\caption{a) Amplitude of $S^{CC}_{4-8}$ minimum ($Q\in [1.7-1.9]$). b) Amplitude of the small $Q$ maxima in Bhatia-Thornton $S^{CC}_{4-8}$ structure factors ($Q<1$).\label{Bhathia_petQ}}
\end{figure}

For large coordination differences ($i-j \geq 4$) a "pre-deep" is observed on $S_{CC}$ around $Q \approx 0.5 Q_1$. 
This pre-deep indicates a segregation effect \cite{B78}, between spheres with large CCN differences. It varies non uniformly with packing fraction (see figure \ref{Bhathia_petQ}.a) and its maximum amplitude lies in the packing fraction interval $\gamma\in[0.5;0.55]$.
This segregation effect occurs over distances larger than contact neighbours, as it was suggested by the radial dependency of $\avCCN$ seen in section \ref{sec:radDepAvCCN}. As a matter of fact, the maximum segregation measured by this pre-deep occurs in a range of packing fraction where there is virtually no contact segregation in the case of RMIN-MAX-1 aggregates (by comparing fig \ref{avercont}.b and fig \ref{avercont}.a), suggesting that there can exist contact and longer-range segregation effects between spheres of various coordination numbers and that these two types of segregation are not necessarily concomitant.

Furthermore, for these high coordination differences a new $S_{CC}$ prepeak appears at very small $Q$ values, around $0.2 Q_1$, which seems to vanish in the aggregates with the highest packing fraction (fig \ref{Bhathia_petQ}.b), with the notable exception of MAX-4. Accordingly, this $S_{CC}$ prepeak is also observed in the RRPA.
This peak should be associated with the existence of segregation effects extending to larger $r$ range, i.e. to the formation of isocoordinated aggregates in an "average matrix". This suggests the existence of regions of low or high coordination in the aggregate matrix. 

\section{Discussion and conclusion}
The so called random packings of adhesive hard spheres cover a wide range of packing fractions, from a lower limit of 0.15 at the three dimensional percolation threshold \cite{Z83} with contact coordination number 2 \cite{B90}, up to a maximum value of about $0.636$ in the close packed random packing (RCP) with coordination number approximately 8.1, as determined experimentally by Scott \cite{S60}.

In this paper we focused on the detailed structural characterization of such packings, using a wide variety of static aggregates-building algorithms. 

The randomness or isotropy of the $10^4$ aggregates (containing $10^6$ spheres each one) was checked first by studying  the angular distribution of pairs of spheres separated by a given distance. The eigenvalues of the corresponding nematic order tensor were shown to be always less than $2.10^{-2}$ and confirmed that aggregates are fully isotropic.

Accurate methods for the determination of the aggregate packing fraction were then introduced. It was thus shown that seed effects at the origin of the aggregates do not extend beyond the fifth neighbour range, i.e. are limited to the first 8~000 spheres, if the seed consists of sphere arrangements that are rare in the aggregate. This seed effect can be totally removed by using as seed a sample with a structure similar to the built aggregate. 

On the other hand it has been shown that the effect of the imperfect spherical surface of the aggregates depends on the building algorithm and packing fraction, and is limited to a thickness of approximately $6r_s$ in the "worst case". Finally, the accuracy on the (average) packing fraction increases with the sphere number of the aggregate which must reach  $10^6$ in order to get a $10^{-3}$ accuracy. 

The detailed structural analysis of random packings could then be undertaken.

As a first step, these structures have been tackled via the Delaunay tessellation of tetrahedra connecting sphere centers. The distributions of these tetrahedra have been characterized by two distortion parameters, $\Lmax$ and $\etaD$. Their distributions show a bimodal character that varies with packing fraction and appear related, they both include discontinuities whose origin remains unclear. Their average values decrease with increasing packing fraction. 
However, they provide complementary characterization of the aggregates produced by the various studied algorithms, suggesting that beyond their similarities, these distributions also present subtle differences that will be the object of a future study. 

Special attention was paid to the populations of regular tetrahedra (formed by 4 mutually contacting spheres, $\Lmax=2$) and quasi regular Delaunay tetrahedra ($\Lmax < 2.3$) which were shown to behave quite differently. As a matter of fact, the volume fraction of regular tetrahedra has been shown to decrease with increasing packing fraction and reaches a maximum value of 0.165 for RPPA aggregates (only built with regular construction tetrahedra). 
This raises a fundamental question: is there a maximum geometrically defined (i.e. irrespective of their building mechanism) proportion of regular tetrahedra in random aggregates?
Conversely, the proportion of quasi regular Delaunay tetrahedra goes to a minimum around $\gamma = 0.56$ and then increases with increasing packing fraction.

New structural characterization methods could then be introduced by taking advantage of the unequivocal definition of sphere contacts and, hence, contact coordination numbers.

First, partial characterization was carried for short distances, i.e. contacting neighbours. The distributions of pairs of spheres with respective CCN $i$ and $j$ ($\eta_{ij}$) were first studied. 
Their FWHM increases when packing fraction decreases and their average values ($\beij$) show distinct behaviours with respect to $\gamma$ for low and high values of $i$ respectively: a non uniform decrease with $\gamma$ is obtained only in the case of low $i$ values.
The evolution of these distributions shows that contact segregation can exist between spheres of various CCN, whereas the evolution of $\avCCN$ for spheres of various coordination shows that another segregation may exist, on the basis of CCN, over a larger range.
Partial distribution of bond angles for spheres with different CCN were also studied. It was shown that high CCN spheres tend to have a smoother distribution, whereas low CCN ones present a depletion of low angle bonds and have a higher proportion of high angle bonds. These differences are particularly distinct in high packing fraction aggregates.

This structural description was then extended to all distances by introducing the partial pair distribution functions $P_{ij}(r)$ i.e. the probability of finding a sphere with CCN $i$ at a distance $r$ from another sphere with CCN $j$ (normalized to 1 for large r) and “local packing fractions” around $i$ coordinated spheres.
The different shapes and discontinuities of the $P_{ij}(r)$ curves were given and this detailed analysis allowed a clear cut distinction between different random packing structures which cannot be distinguished by their packing fraction and/or $\etaD$ parameters. The question remains of whether these aggregates could be characterized by characterizing them through the combination of packing fraction $\gamma$ and distortion parameter $\Lmax$. 

Distinguishing (with Bernal \cite{BM60,B62}) contacting from quasi-contacting spheres, it was shown that high CCN spheres have few quasi contacting neighbours (QCN) while low CCN spheres have a higher number of QCN (as could be expected) and that this local effect extends to larger distances. Moreover this behaviour is amplified when the packing fraction increases. 

Finally, the two sets of partial structure factors respectively introduced by Ashcroft-Langreth ($S_{ij}(Q)$) and Bhatia-Thornton ($S_{NN}$, $S_{NC}$ and $S_{CC}$) were analyzed. They are different Fourier transforms of linear combinations of the $P_{ij}(r)$ and can be directly compared with the results of diffraction experiments on liquid or disordered materials. In particular the diagonal $S_{ii}(Q)$ give the partial structure factors of the partial aggregates made from $i$ coordinated spheres and the variable shapes of these $S_{ii}$, especially their prepeak, first and second $Q$ oscillations, were described. Furthermore the  concentration-concentration partial structure factor of Bhatia gave the mutual or “chemical” order between spheres with different coordinations $i$ and $j$ and put forward hetero-attractions or segregation of spheres within the different aggregates, confirming that several kind of segregation may indeed exist.

From all these results we could conclude that the random irregular polytetrahedral aggregates studied here are made from two basic components, namely fully random component (without regular tetrahedra) and regular polytetrahedral component (only built with regular tetrahedra), whose proportion decreases with increasing packing fraction. This composite nature of the aggregates produces “prepeaks” (at very small $Q$) in the aggregate structure factors and the main features differentiating the RP component from the counterpart FR component are the following:
\begin{itemize}
\item A high proportion of regular Delaunay tetrahedra (as expected)
\item an increase in $\delta$-peaks noticeable in: distributions of Delaunay tetrahedra distortion indexes, angular bond distributions, global or partial pair distribution functions
\item an increase of the second mode in the distributions of Delaunay tetrahedra distortion parameters $\etaD$ and $\Lmax$
\item a spreading of contact coordination numbers distributions $\eta_{ij}$
\item strong variations in the radial dependency of the average contact coordination number of spheres
\item a collapse of the continuum of partial bond angle distributions
\item a more similar quasi-first-neighbours part of partial pair distribution functions of low and high contact coordination number spheres
\item a more similar radial dependency of packing fraction for spheres of any contact coordination number
\item a reduction of topological discontinuities in pair distribution function at $r=2\sqrt{3}$ and 4
\item a splitting of the structure factor's second peak ($Q\approx 7$), similar to what is experimentally found in amorphous tetrahedral materials. 
\end{itemize}

The present results on statically built sticky hard sphere aggregates cover the packing fraction interval 0.370-0.593. It could be interesting to extend this study:

Firstly to the low packing fraction range 0.15 (condensation or percolation limit)-0.370, which involves contact coordination number lower than 3 and cannot be reached by the building algorithms used here.

Secondly, to the high packing fraction range 0.593-0.64 which needs a reduction of the fluctuations of the local packing fraction at all length scales and cannot be reached by our algorithms which only minimize the fluctuations of the local packing fraction up to the second neighbours distances.
In particular, it would be of great interest to study dynamically built aggregates, such as the ones produced by Lubachevsky-Stillinger algorithm, Jodrey-Tory algorithm or molecular dynamics.
The first problem to be resolved would then be to define the distance corresponding to contacting spheres, as, for these aggregates, the first neighbour peaks are spread over a finite length interval. Nonetheless, valuable structural informations might be at hand through the techniques studied in the present work, that might shed new light on various long standing questions concerning random aggregates as well as allowing new comparisons between statically and dynamically built random aggregates.

\bibliographystyle{elsarticle-num}
\bibliography{pap.bib}

\section*{Acknowledgments}
P. C\'en\'ed\`ese is greatly acknowledged for his help in openMP parallelization of some codes. Anonymous referees are thanked for their observations and suggestions that helped in the process of writing the present paper.

Anonymous referees are greatly acknowledged for their critics and suggestions that helped improving the quality of this paper.

\appendix
\section{Supplementary information}

\subsection{Isotropy of the aggregates}

Isotropic disordered aggregates are characterized by an isotropic distribution of $\hat{r}_{ij}$ bonds, where $\hat{r}_{ij}$ are the unit vectors joining the (i,j) sphere centers. This requirement provides a convenient check for the building method and/or the minimum aggregate size beyond which the isotropy is reached. Such a distribution is characterized by a uniform random distribution of azimuthal angles $\varphi$ and a polar angles $\theta$ distribution following a probability density $P_b(\theta)=\sin(\theta)/2$. 

It is worth mentioning that a lattice like distribution of spheres is characterized by a $\theta$--distribution presenting distinguishable peaks; thus the features of the $\theta$ distribution appears to be a convenient tool to control both the aggregates randomness and isotropy. 

In order to go beyond a simple qualitative characterization, we can quantify the deviation from isotropy by looking if a favoured direction emerges among the $\{\hat{r}_{ij}\}$ distribution. For this, we follow the usual method of the framework of the nematic liquid statistics \cite{MR77} according to which, from the diagonalization of the two rank tensor
\begin{equation}
\label{q_1}
  \bar{Q} = \frac{1}{N}\sum_{ij} \frac{1}{2}(3\hat{r}_{ij} \hat{r}_{ij} - \bar{I})
\end{equation}
where $\bar{I}$ is the identity tensor, a nematic order parameter, say $S_1$ is obtained as the largest eigenvalue $\lambda_{max}$ of $\bar{Q}$.
In equation (\ref{q_1}), the sum running over all the $\{i,j\}$ pairs, can be limited to the sum over the bonds $\hat{r}_{ij}$, which can be replaced by a sum over the unit vectors carried by the $\vec{r}_i$, with respect to a fixed reference $\vec{r}_o$, which avoids the surface effects when both $i$ and $j$ are located at the aggregate surface.

\begin{figure}[htbp]
\begin{center}
\subfigure[]{\includegraphics[width=0.48\textwidth]{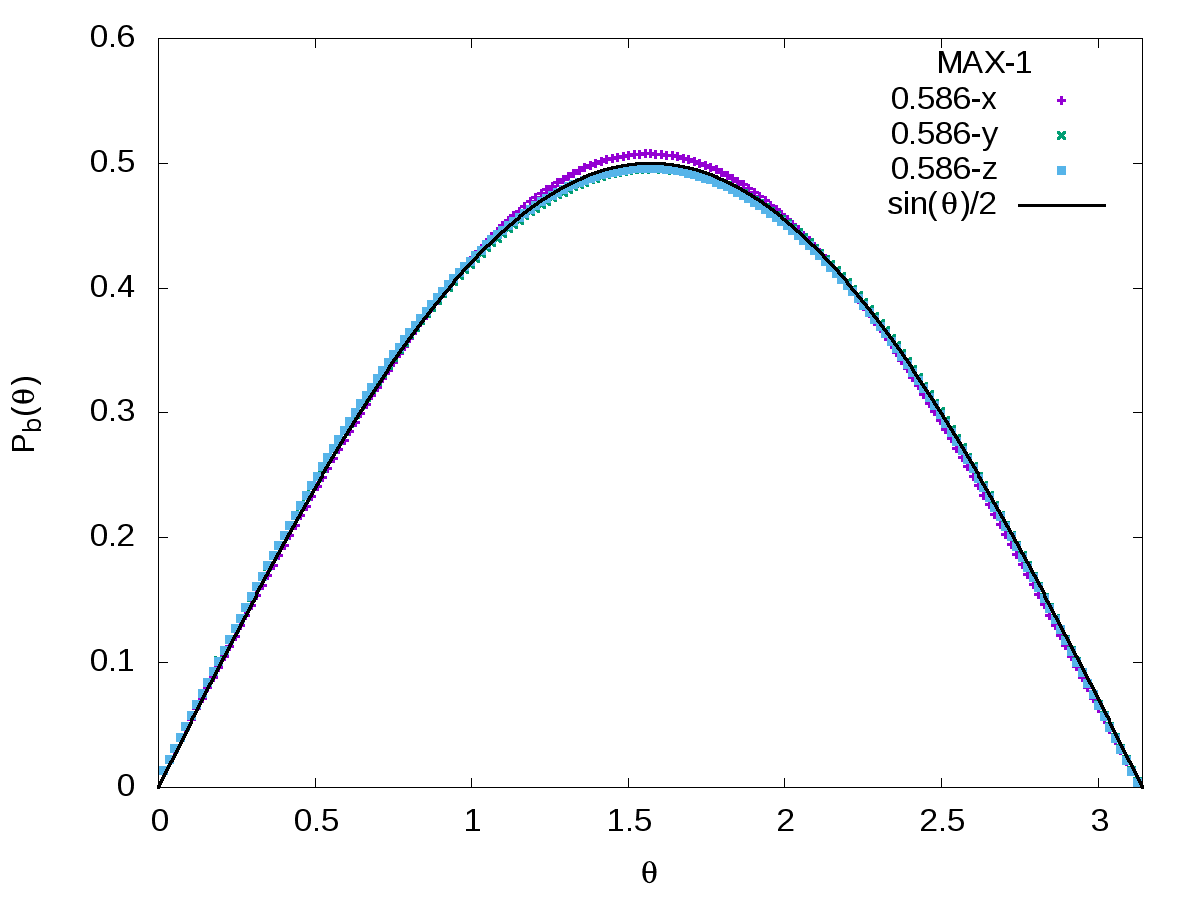}}
\subfigure[]{\includegraphics[width=0.48\textwidth]{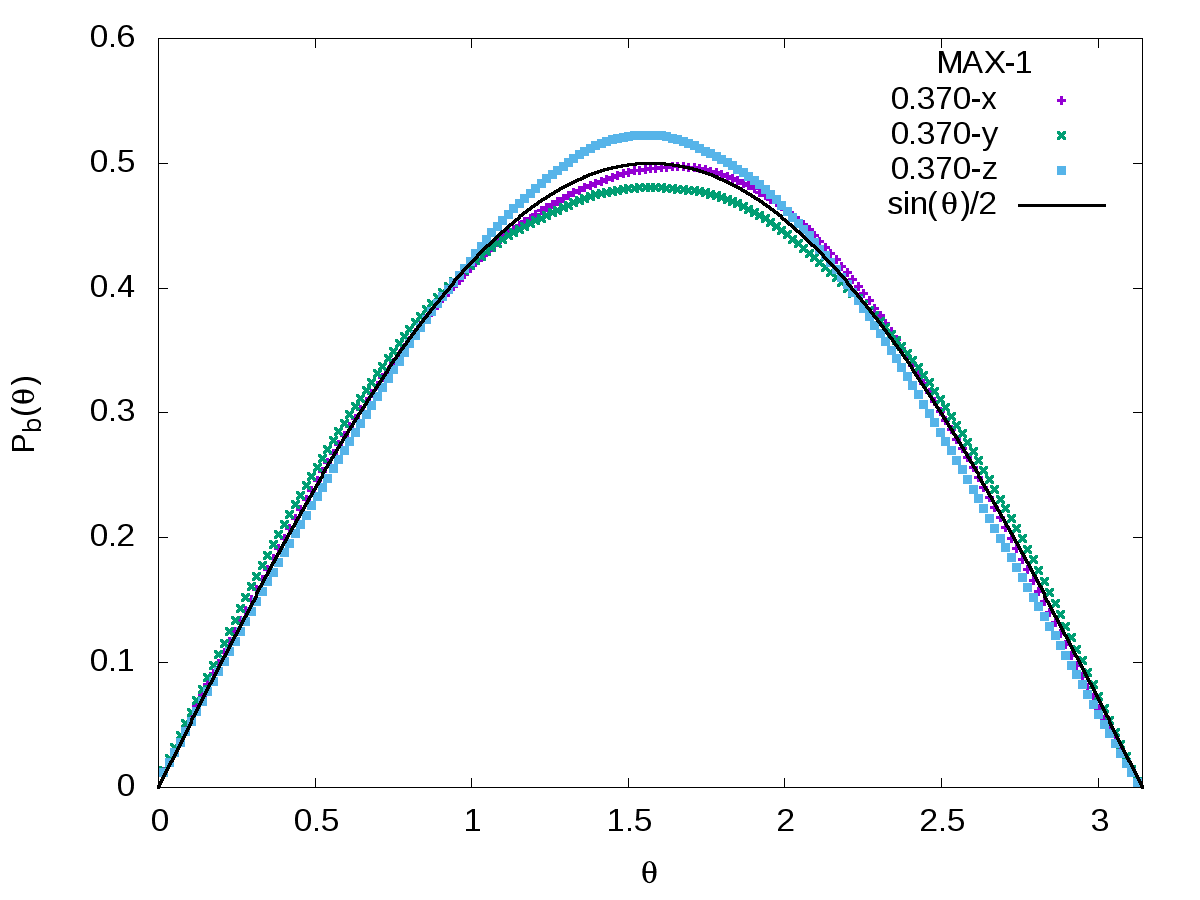}}\\
\subfigure[]{\includegraphics[width=0.48\textwidth]{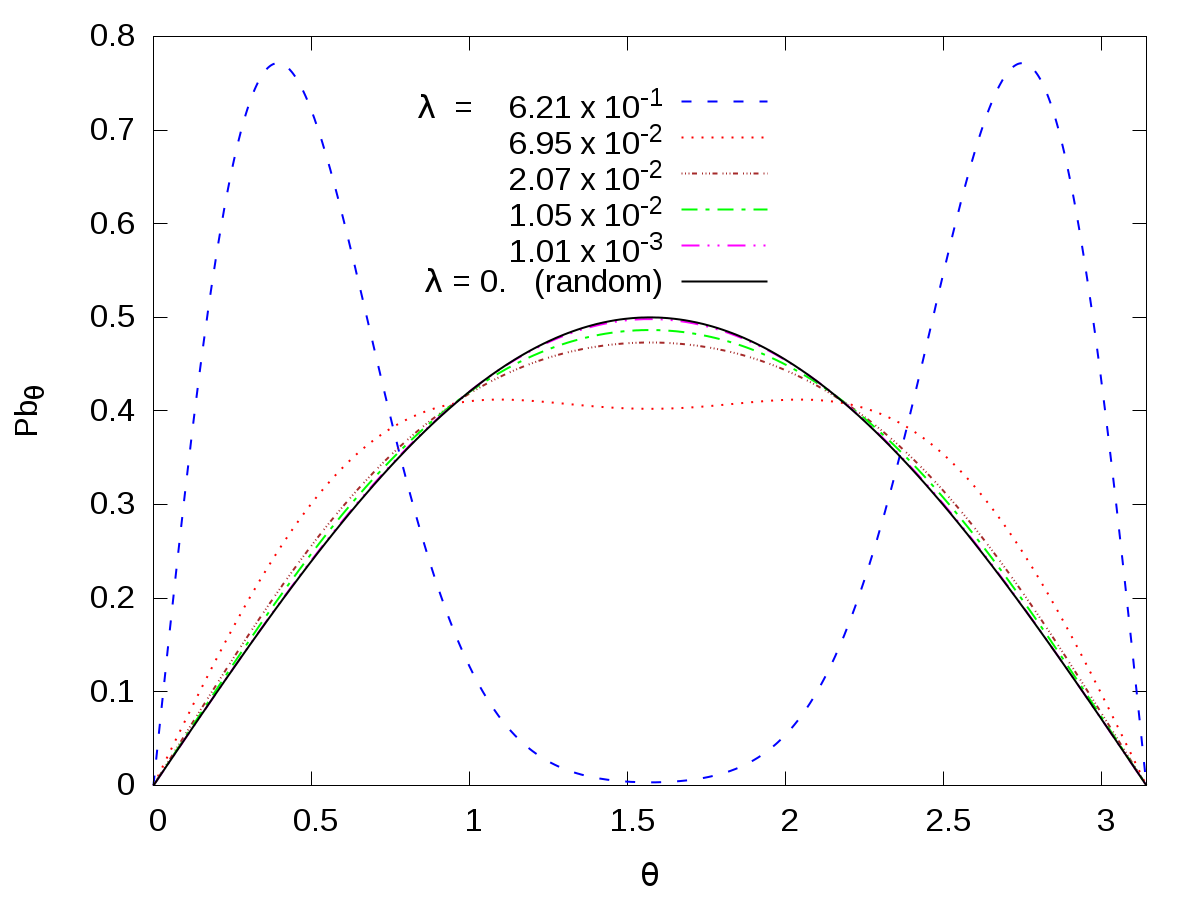}}
\caption{Pair angle distributions with respect to x, y and z for the (a) most and (b) least dense aggregates produced by MAX-1 algorithm. c) 
Polar angle distribution relative to the favorable direction in the uniaxial anisotropy case for different values of the nematic order parameter, $\lambda$ defined as the largest eigenvalue of the tensor $\bar{Q}$, equ.(\ref{q_1}) 
\label{distpair}}
\end{center}
\end{figure}

The results for the polar angle distributions relative to the axis $\hat{x}$, $\hat{y}$ and $\hat{z}$ for the most and least dense aggregates obtained by MAX-1 algorithm are presented respectively in figures \ref{distpair}.a and \ref{distpair}.b.
For the aggregates studied in this work, the deviation from isotropy as measured by the value of $\lambda_{max}$ is found to decrease when the packing fraction increases. A reliable determination of the dependence of $\lambda_{max}$ with respect to $\gamma$ is beyond the scope of this paper, all the more that the range of $\lambda_{max}$ values is quite small.
The largest eigenvalue of $\bar{Q}$ is smaller than 3~10$^{-2}$. Such a value is very small leading us to conclude that these building methods lead indeed to isotropic aggregates. 
To illustrate quantitatively this point, figure \ref{distpair}.c displays the result of the $\theta$ distribution in terms of $\lambda$ for the following model:
\begin{equation}
\label{dist_1}
P_b(\theta) = \frac{\sin(\theta)}{2} \left[ e^{(-\theta^2/(2\sigma^2))} +
e^{(-(\pi-\theta^2)/(2\sigma^2))} \right]
\end{equation}
where the polar axis is a favoured direction according to the variance $\sigma$. Clearly, the $\theta$ distributions characterized by $\lambda$ values lower than a few $10^{-2}$ can be considered isotropic.

\newpage

\subsection{$P_{ii}(r)$}

\begin{figure}[htbp]
\begin{center}
\subfigure[]{\includegraphics[width=0.48\textwidth]{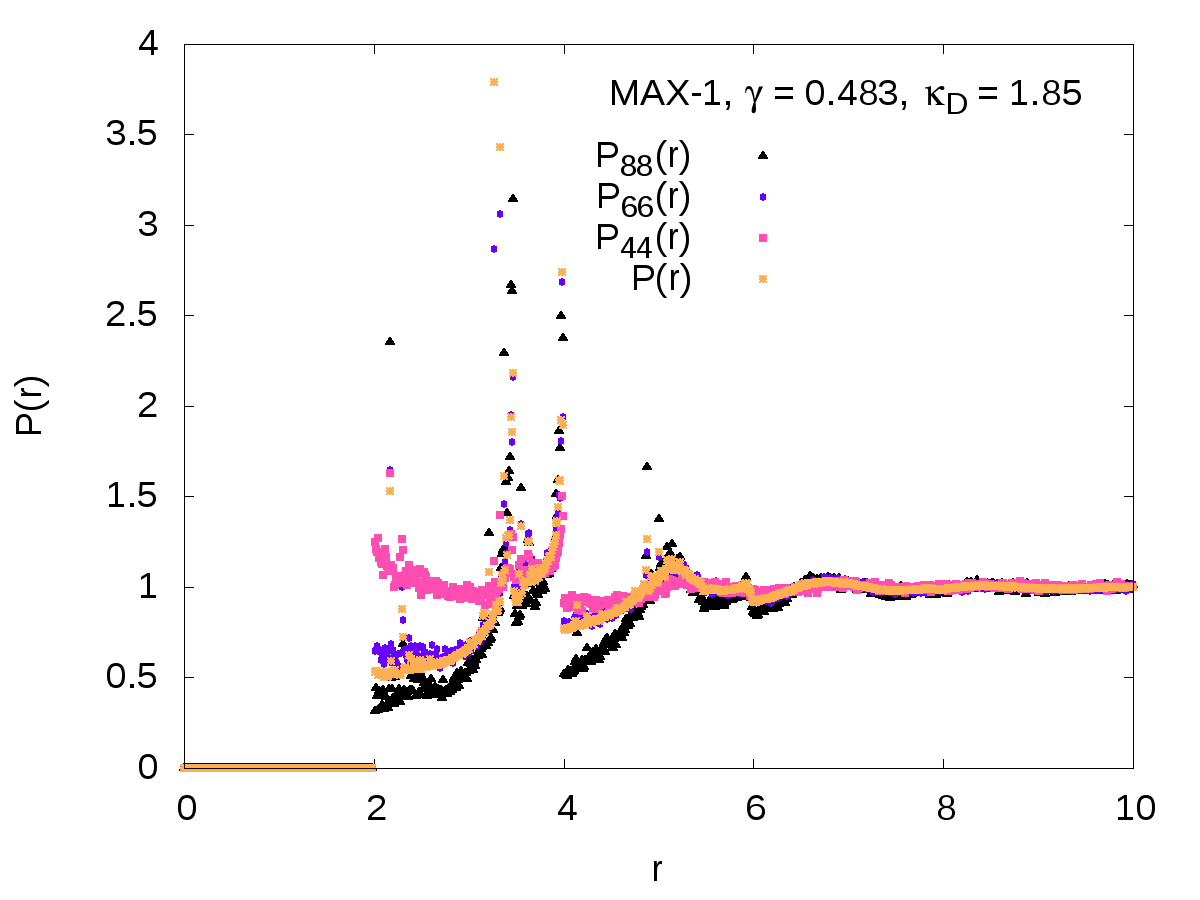}}
\subfigure[]{\includegraphics[width=0.48\textwidth]{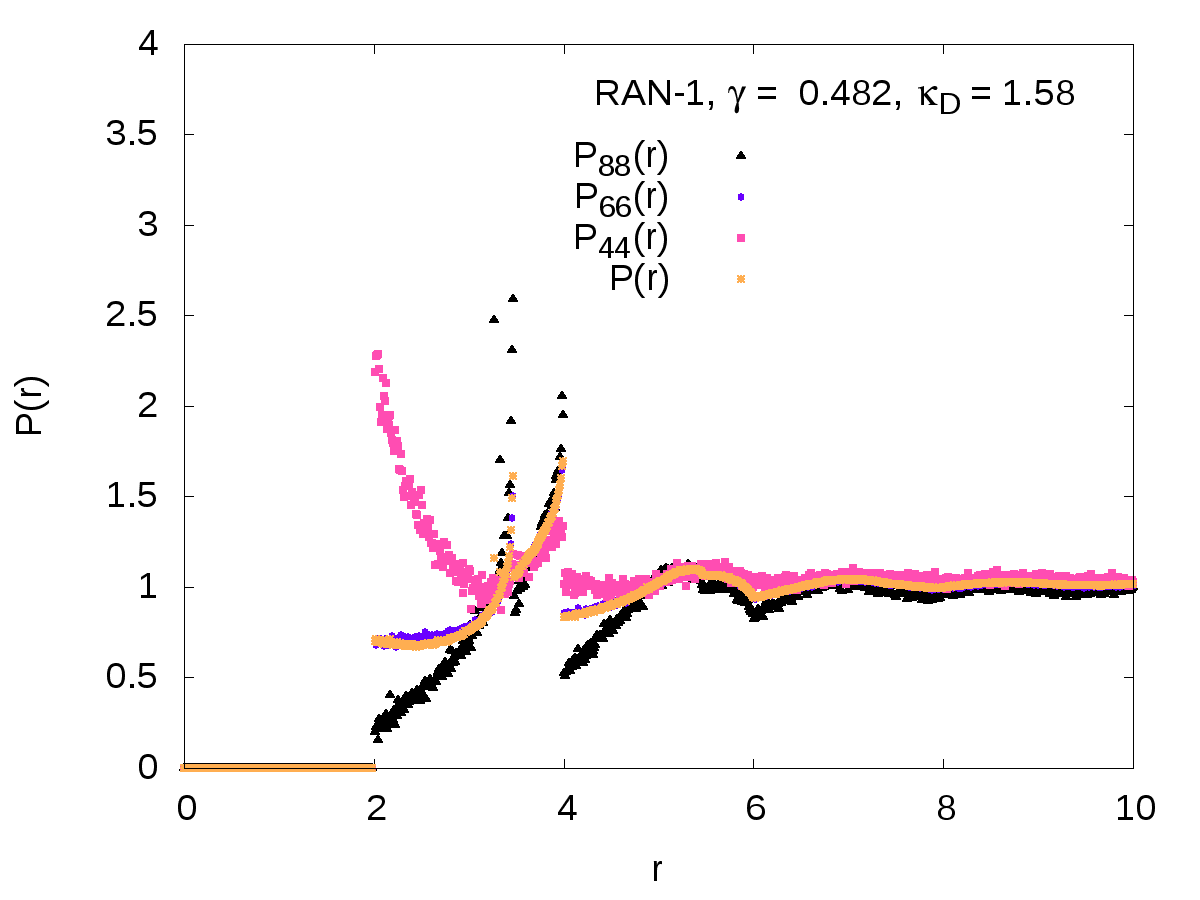}}
\subfigure[]{\includegraphics[width=0.48\textwidth]{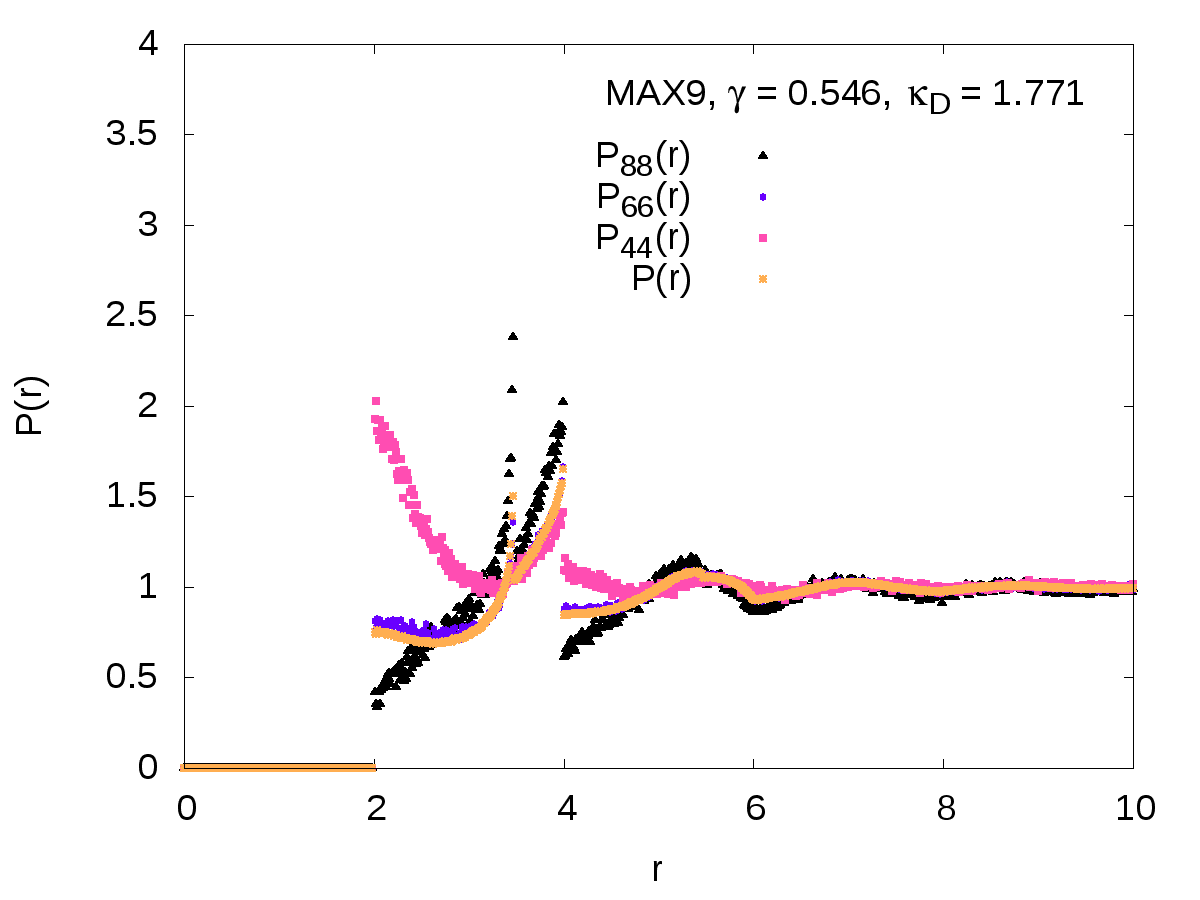}}
\subfigure[]{\includegraphics[width=0.48\textwidth]{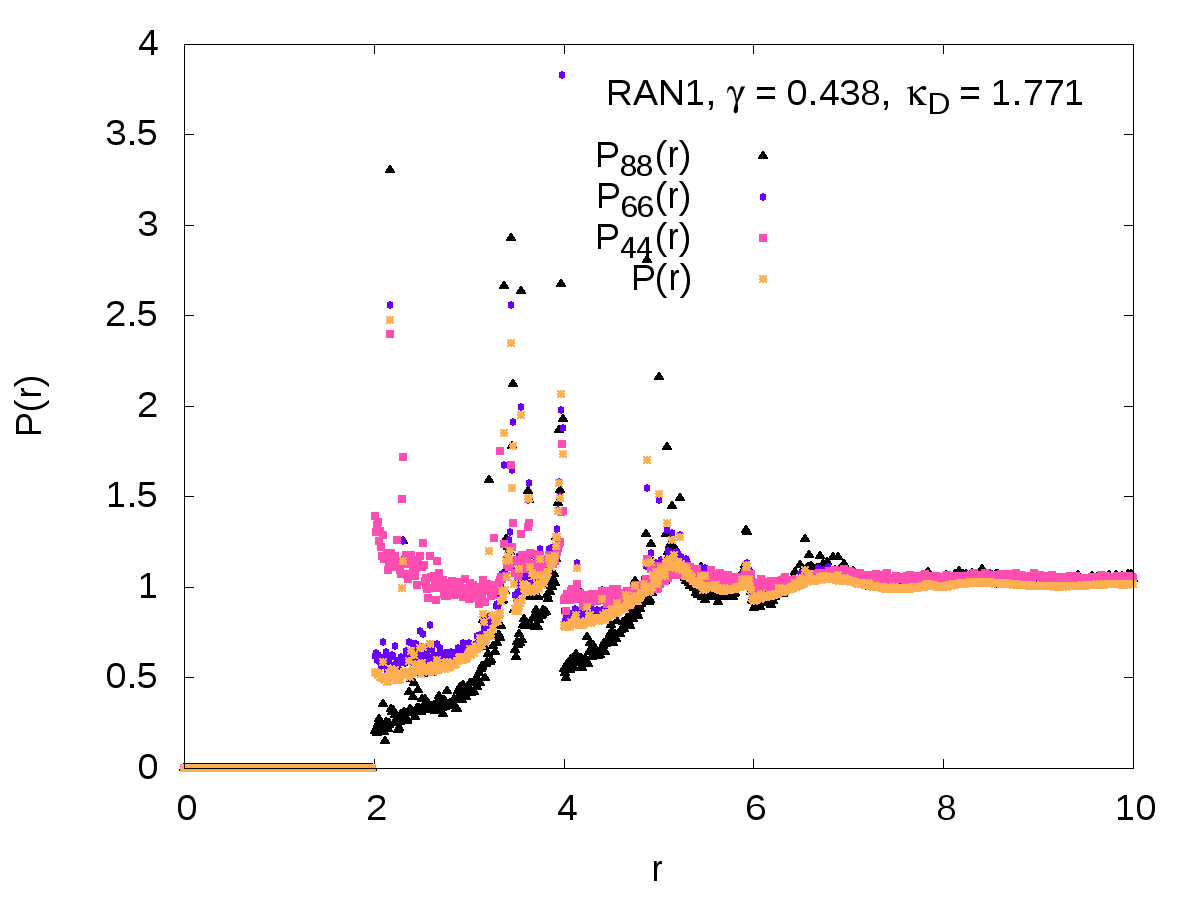}}
\subfigure[]{\includegraphics[width=0.48\textwidth]{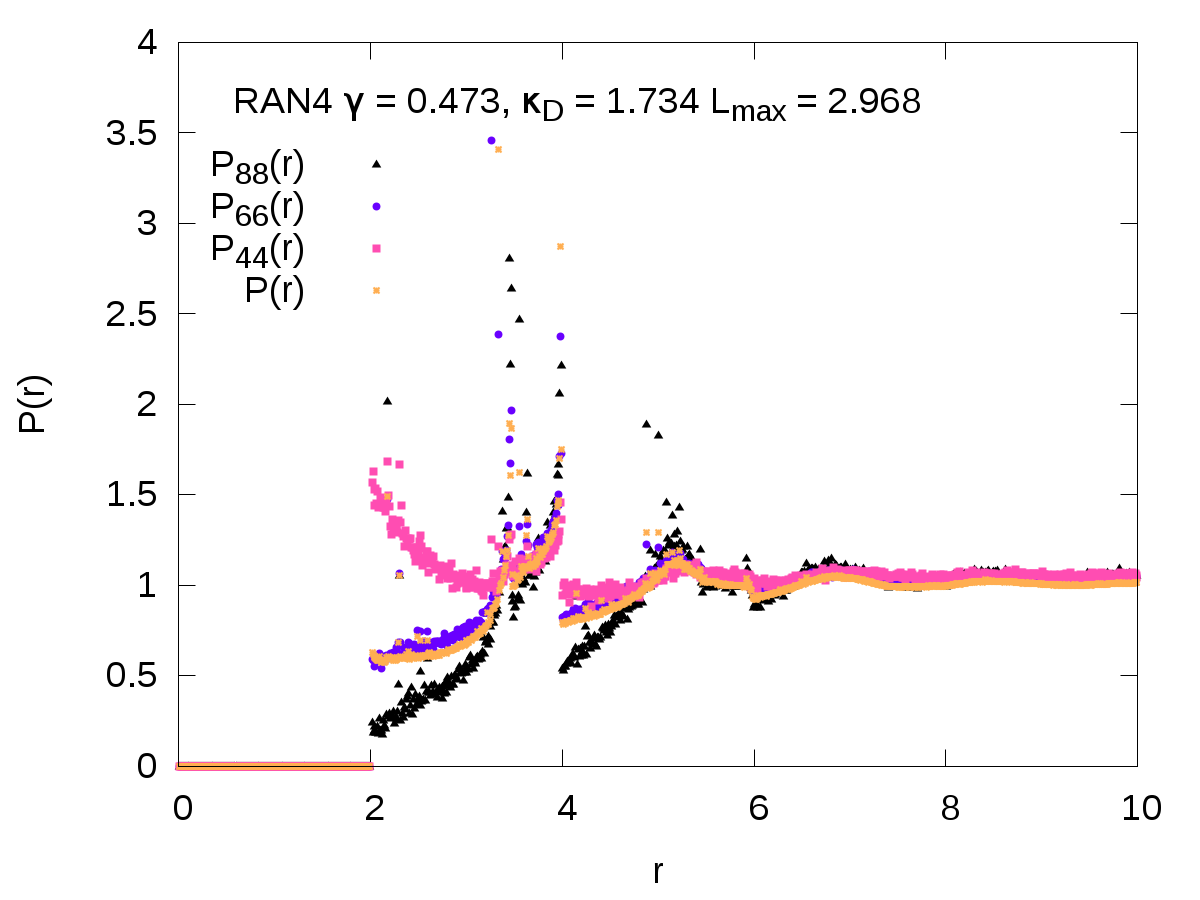}}
\subfigure[]{\includegraphics[width=0.48\textwidth]{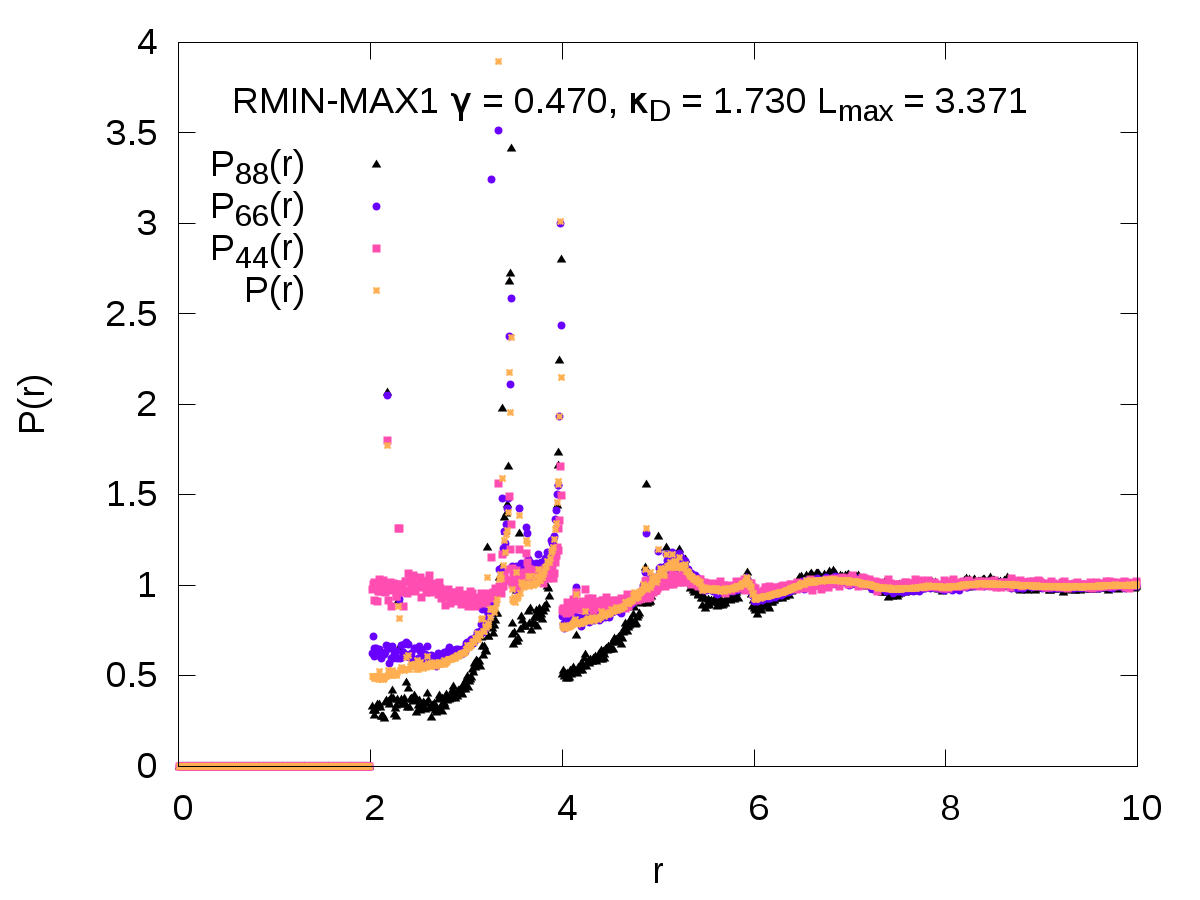}}
 \caption{Sample $P(r)$ and $P_{ii}(r)$ obtained for aggregates generated by algorithms a) MAX-1 ($\gamma = 0.483$) b) RAN1 ($\gamma = 0.482$. c) MAX9 $\betaD = 1.771$ d) RAN1 $\betaD = 1.771$ e) RAN4 f) RMIN-MAX1 .\label{pii_de_r_compaSI}}
\end{center}
\end{figure}

\newpage
\subsection{$P_{ij}(r)$}
\begin{figure}[htbp]
\begin{center}
\subfigure[]{\includegraphics[width=0.48\textwidth]{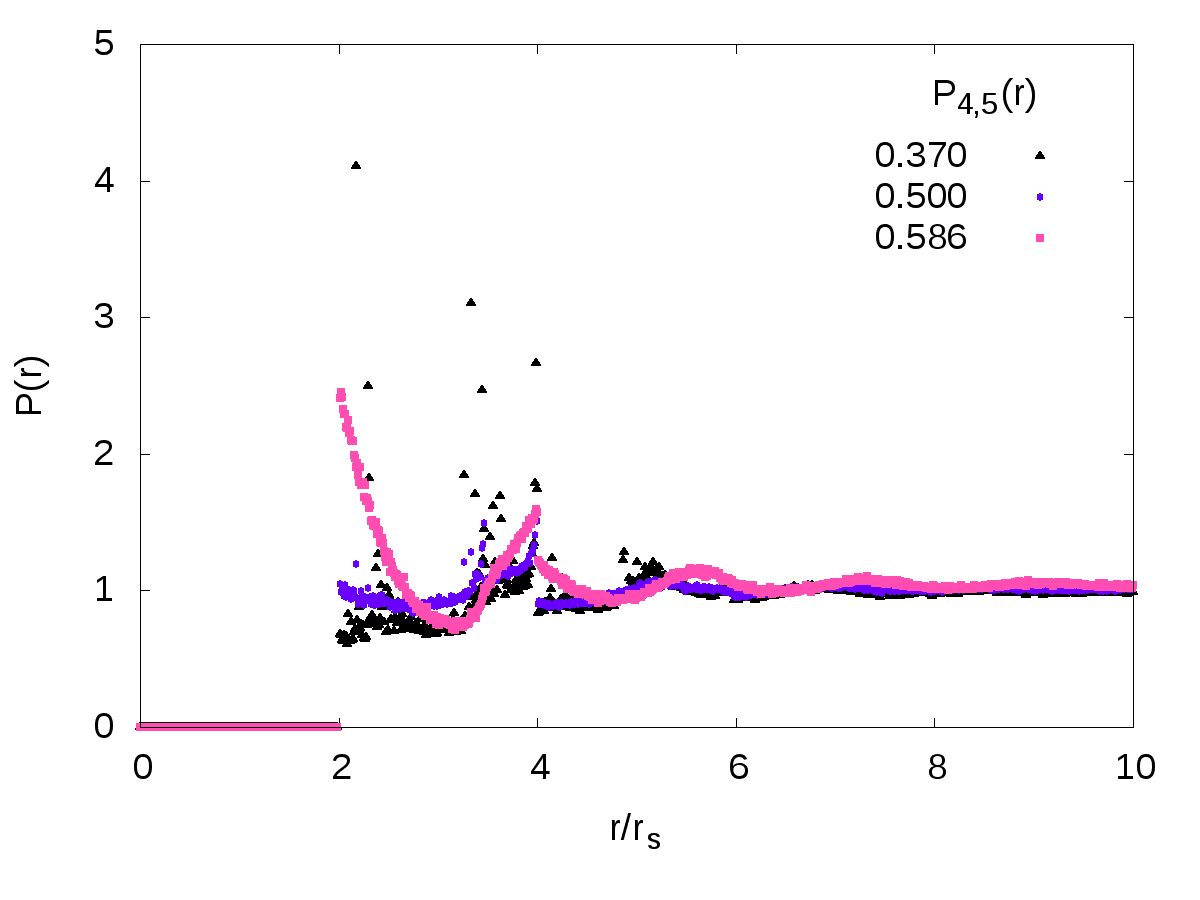}}
\subfigure[]{\includegraphics[width=0.48\textwidth]{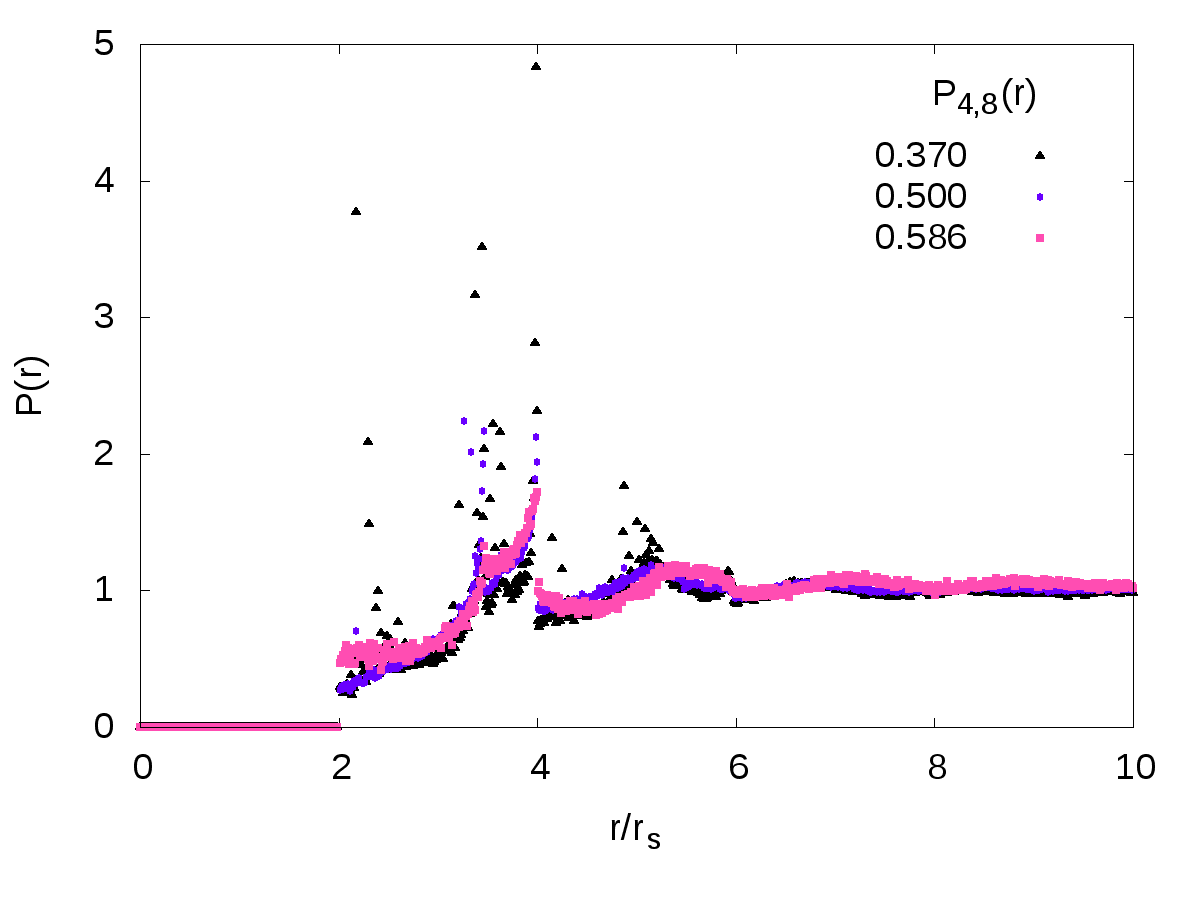}}
\subfigure[]{\includegraphics[width=0.48\textwidth]{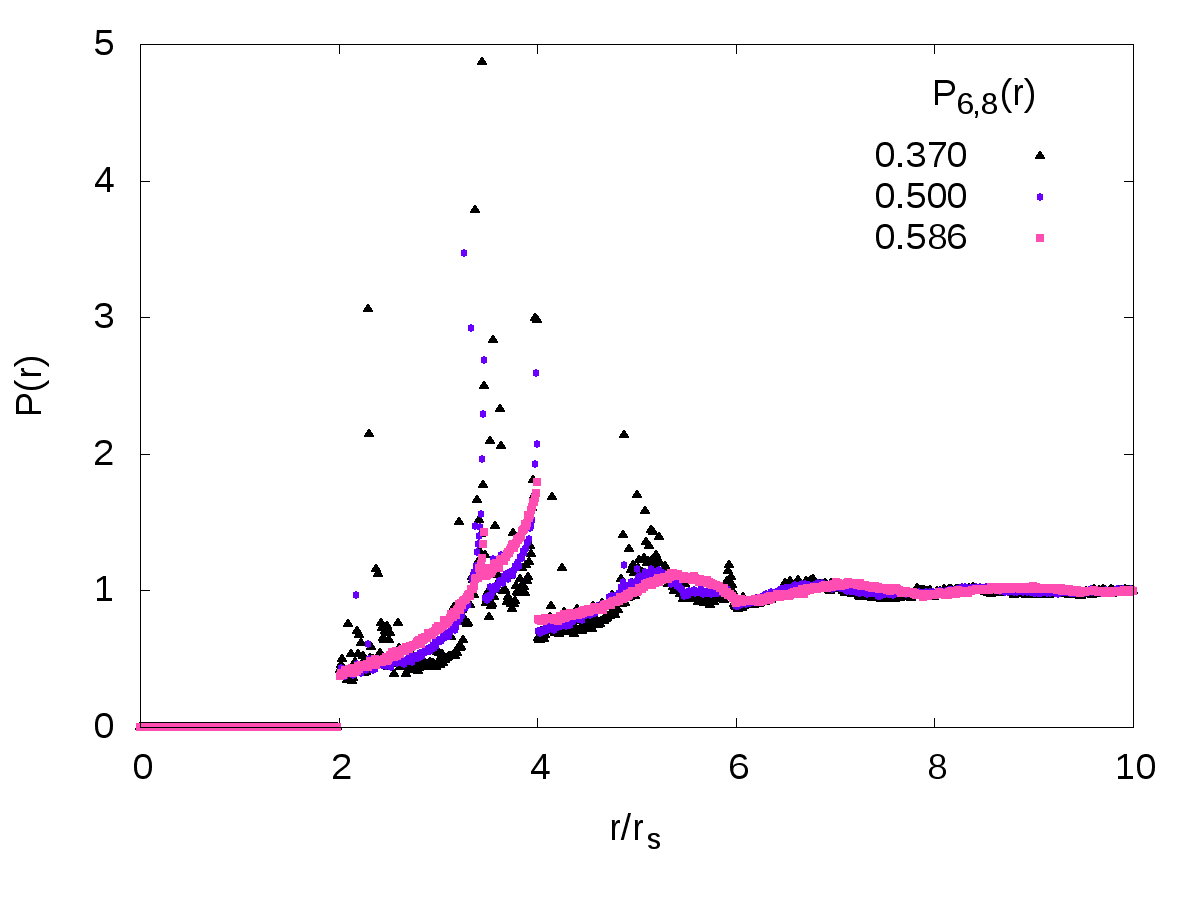}}
\subfigure[]{\includegraphics[width=0.48\textwidth]{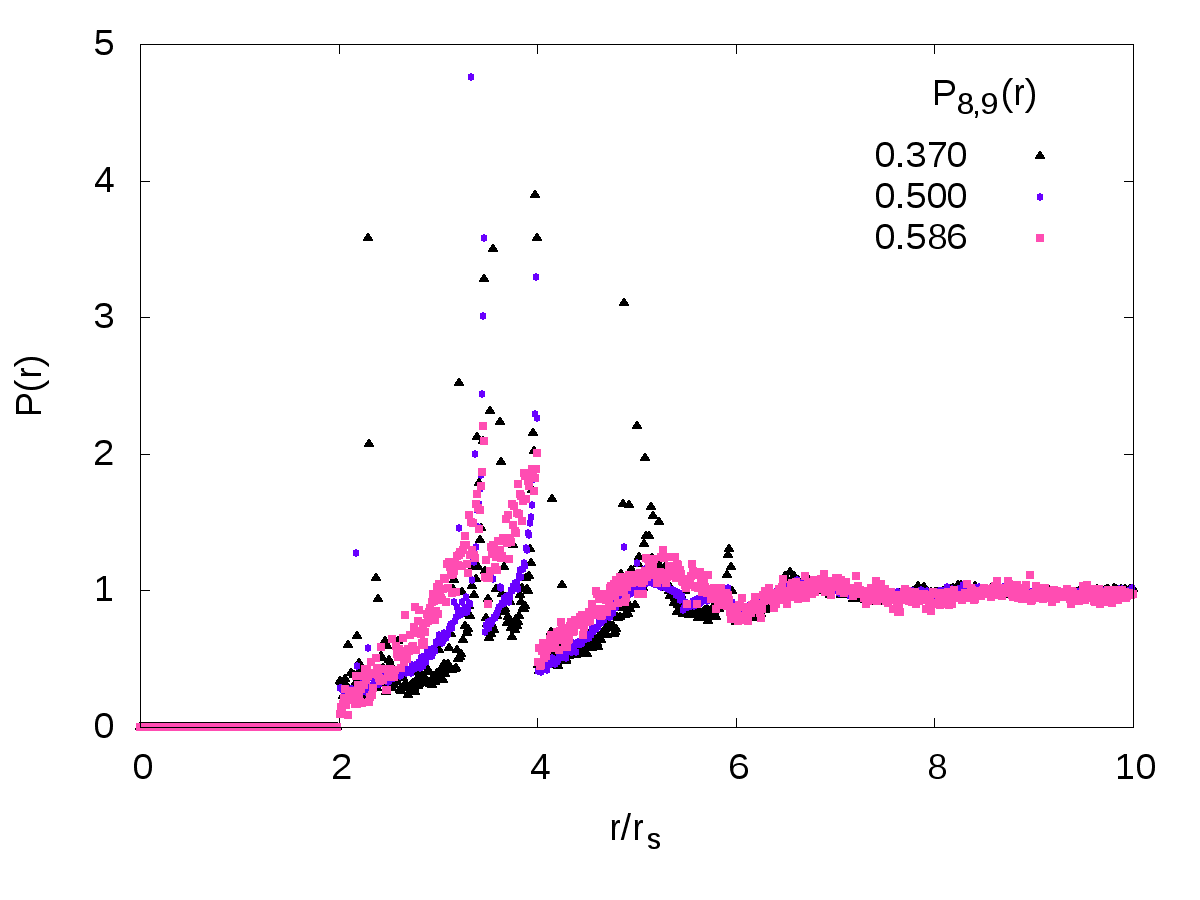}}
 \caption{A sample of $P_{ij}(r)$ for various packing fractions, algorithm MAX-1 -- a) $P_{4,5}(r)$, b) $P_{4,8}(r)$, c) $P_{6,8}(r)$, d) $P_{8,9}(r)$\label{pij_de_r}}
\end{center}
\end{figure}

\newpage
\subsection{AL structure factors}

\begin{figure}[htbp]
\begin{center}
\subfigure[]{\includegraphics[width=0.48\textwidth]{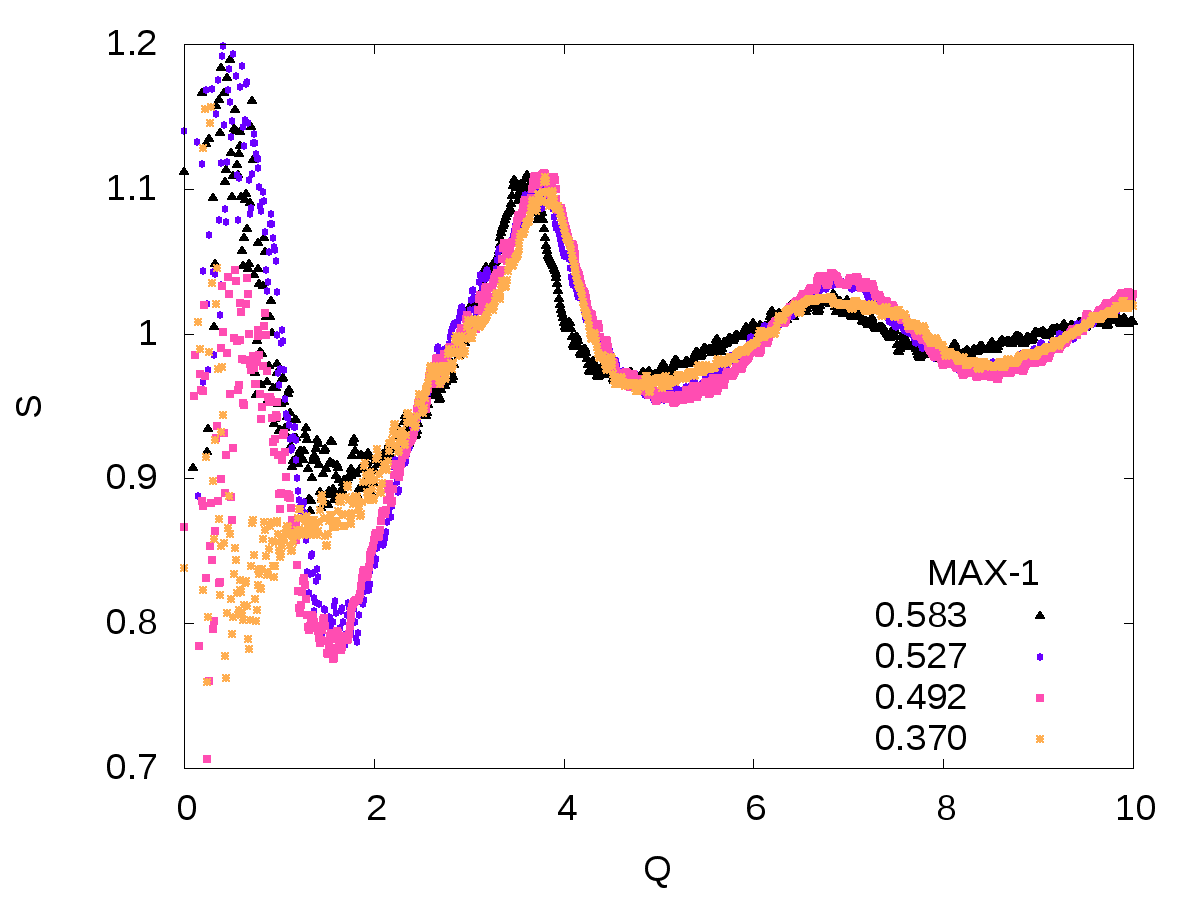}}
\subfigure[]{\includegraphics[width=0.48\textwidth]{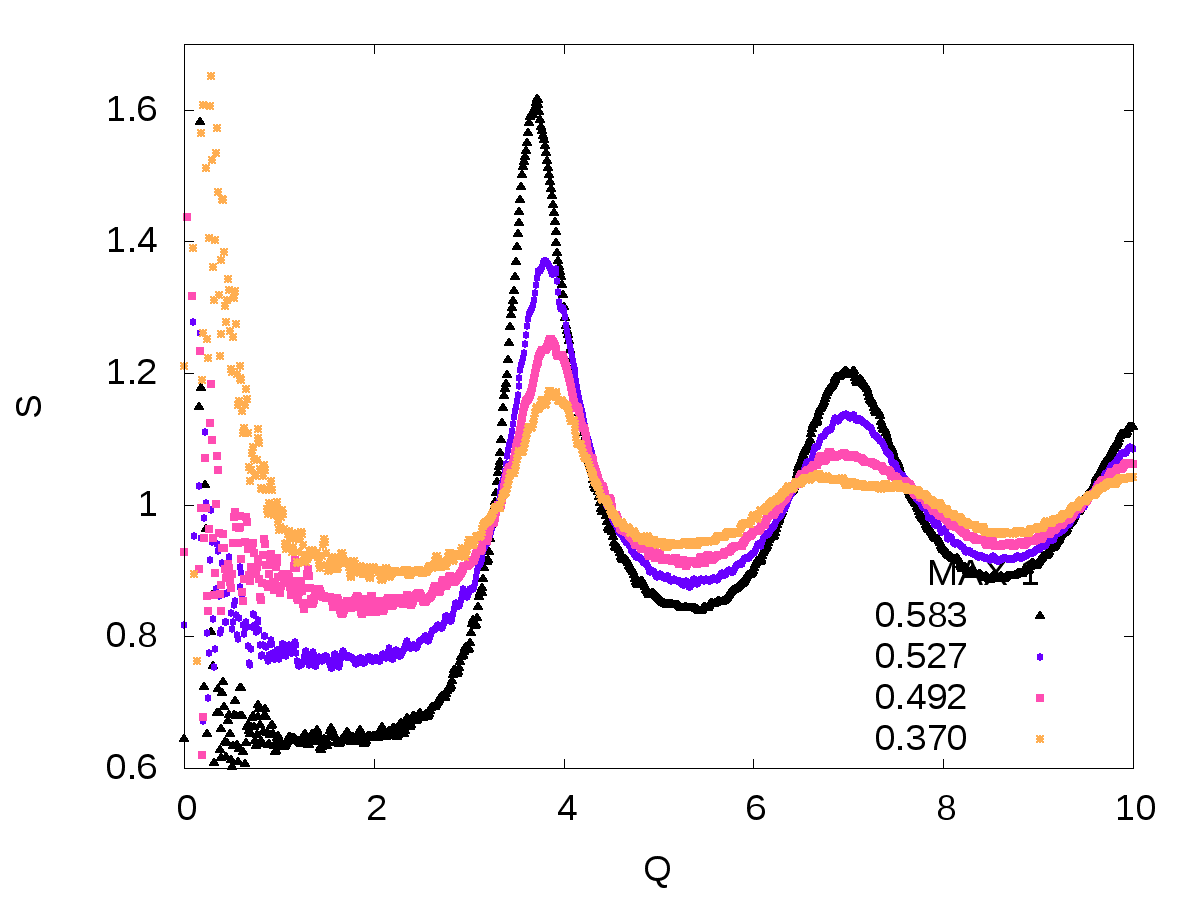}}\\
\subfigure[]{\includegraphics[width=0.48\textwidth]{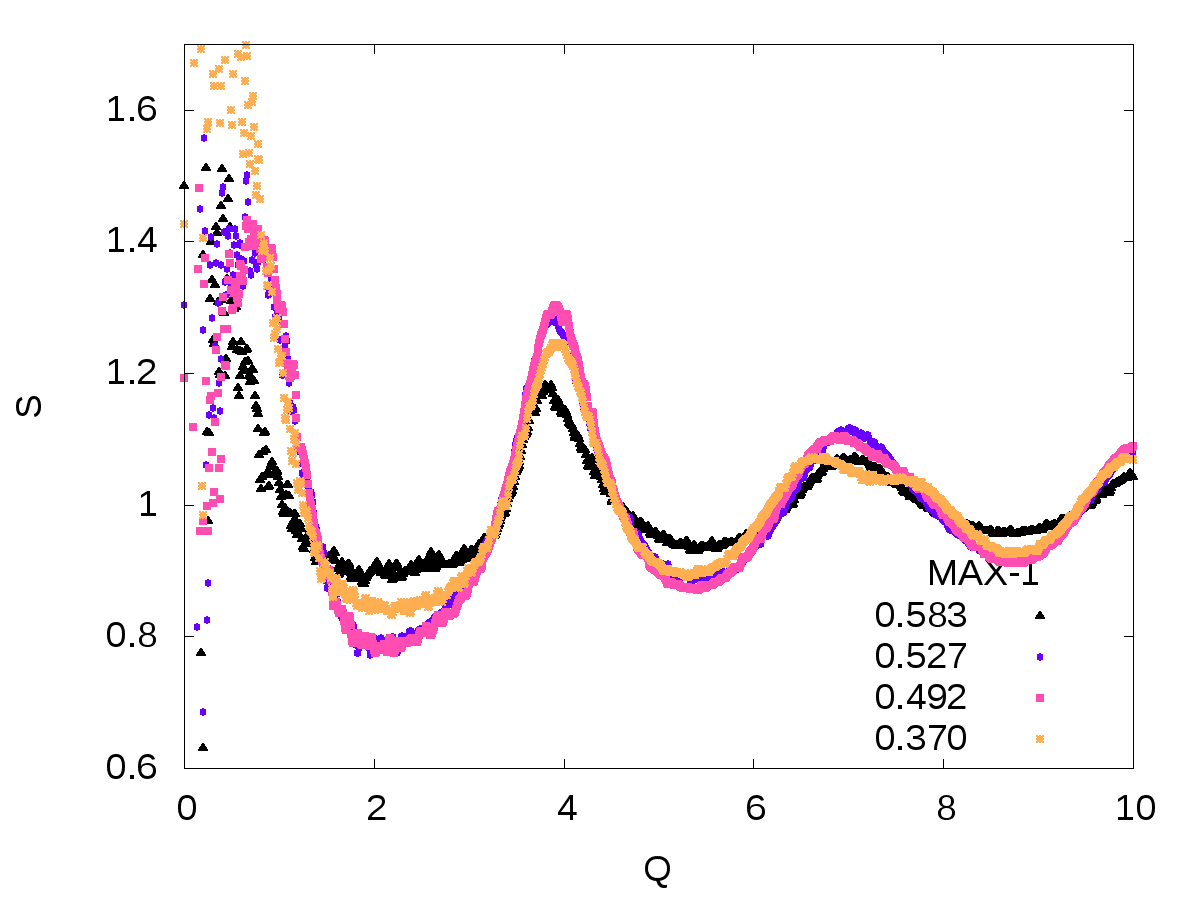}}
 \caption{Sample Ashcroft Langreth structure factors for MAX-1 aggregates. a)$S_{44}$, b) $S_{66}$, c) $S_{88}$.\label{AL_SI}}
\end{center}
\end{figure}

\newpage
\subsection{Bhatia-Thornton formalism for binary mixtures of spheres}
General BT relations then reduce to:
\begin{equation}\label{eq:snnsp}
S_{N_iN_j}(Q) = 1 + \rho' \int \Big({C'_i}^2 P_{ii} + {C'_j}^2 P_{jj} + 2{C'_i}{C'_j} P_{ij} -1\Big) \frac{\sin (Qr)}{Qr}4\pi r^2 d r
\end{equation}
where $\rho' = (N_i+N_j)/V$ and $C'_i = N_i/(N_i+N_j)$.
\begin{equation}\label{eq:sccsp}
S_{C'_iC'_j}(Q) = 1 + \rho' C'_iC'_j \int \Big[\big(P_{ii}(r)-P_{ij}(r)\big)+\big(P_{jj}(r)-P_{ij}(r)\big)\Big]\frac{\sin (Qr)}{Qr}4\pi r^2 d r
\end{equation}
\begin{equation}\label{eq:sncsp}
S_{NC'_i}(Q) = \rho' C'_j \int\Big(C'_iP_{ii}(r) - C'_jP_{jj}(r) + (C'_j-C'_i)P_{ij}(r)\Big)\frac{\sin (Qr)}{Qr}4\pi r^2 d r
\end{equation}

\begin{figure}[htbp]
\subfigure[]{\includegraphics[width=0.48\textwidth]{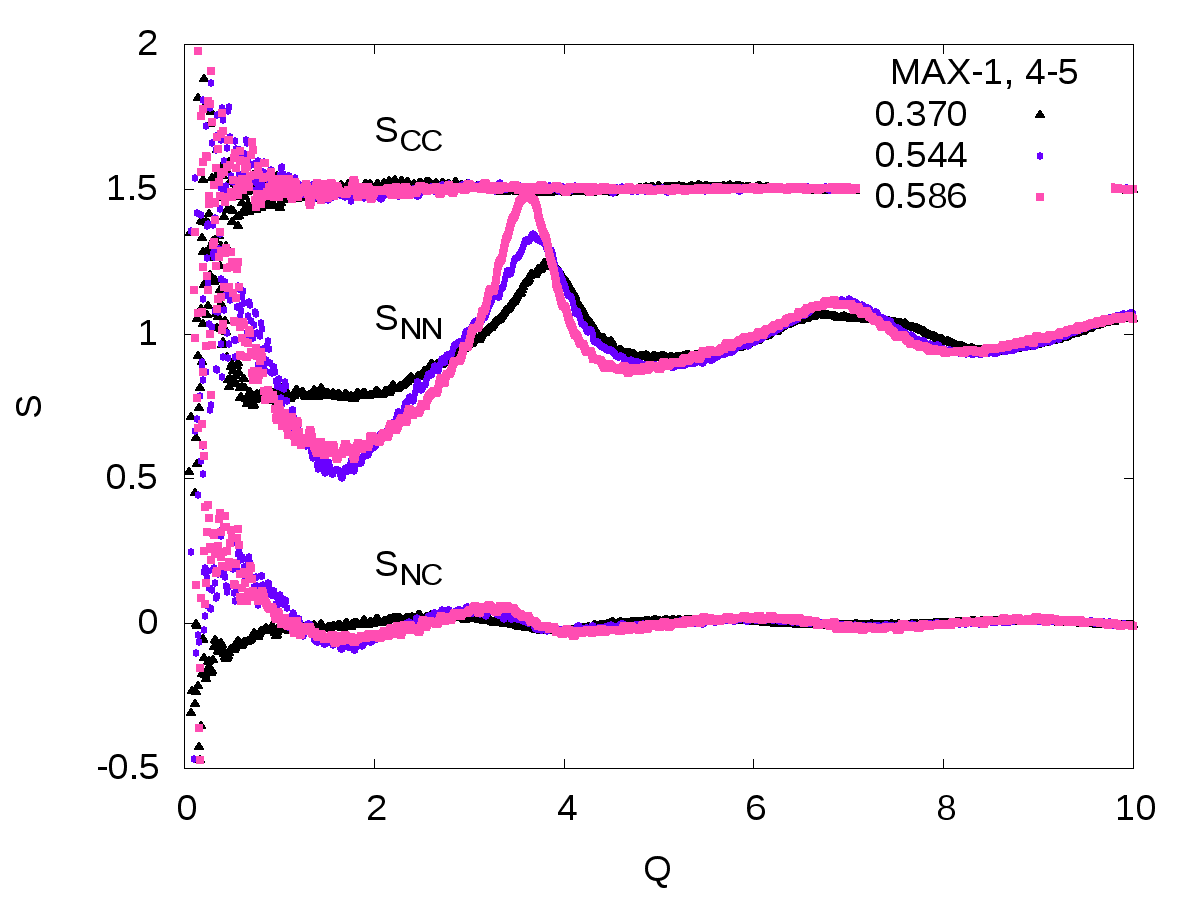}}
\subfigure[]{\includegraphics[width=0.48\textwidth]{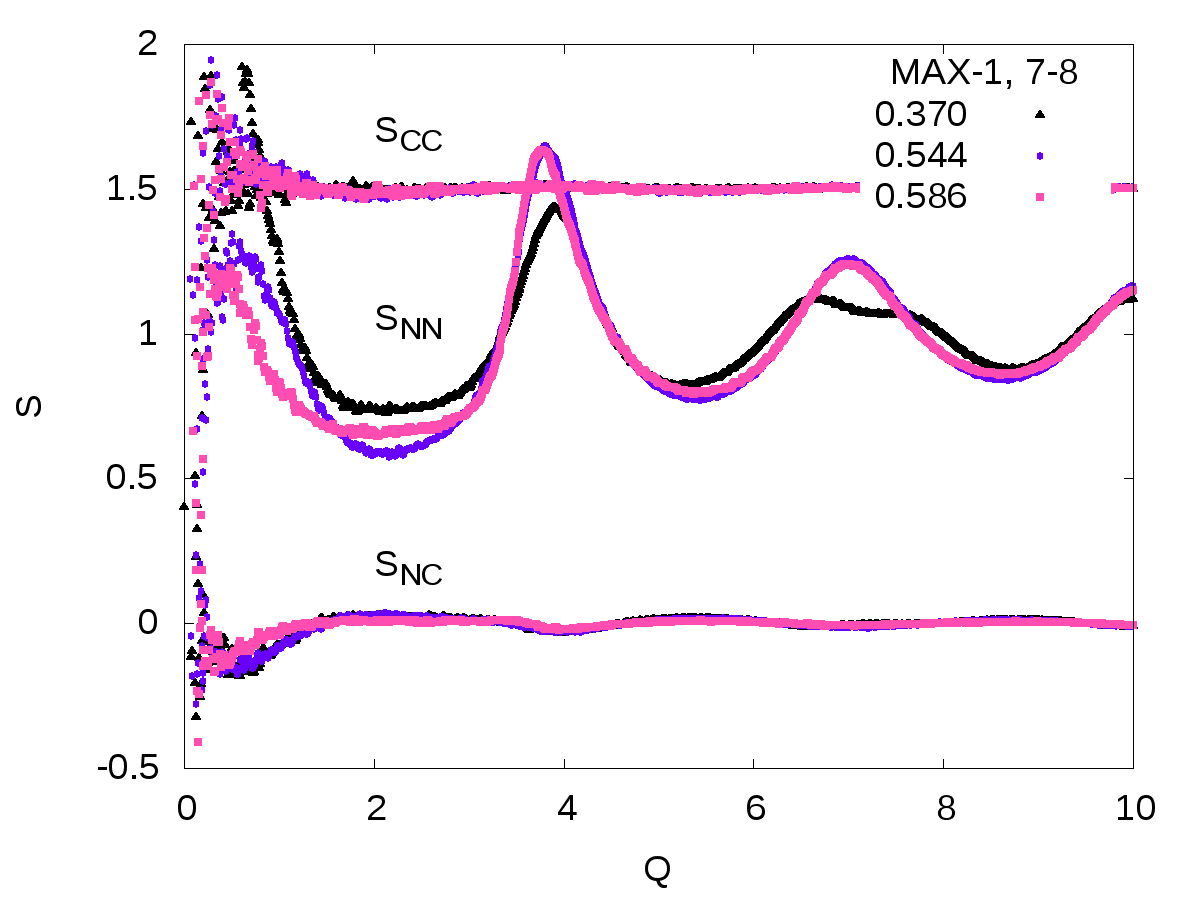}}
\subfigure[]{\includegraphics[width=0.48\textwidth]{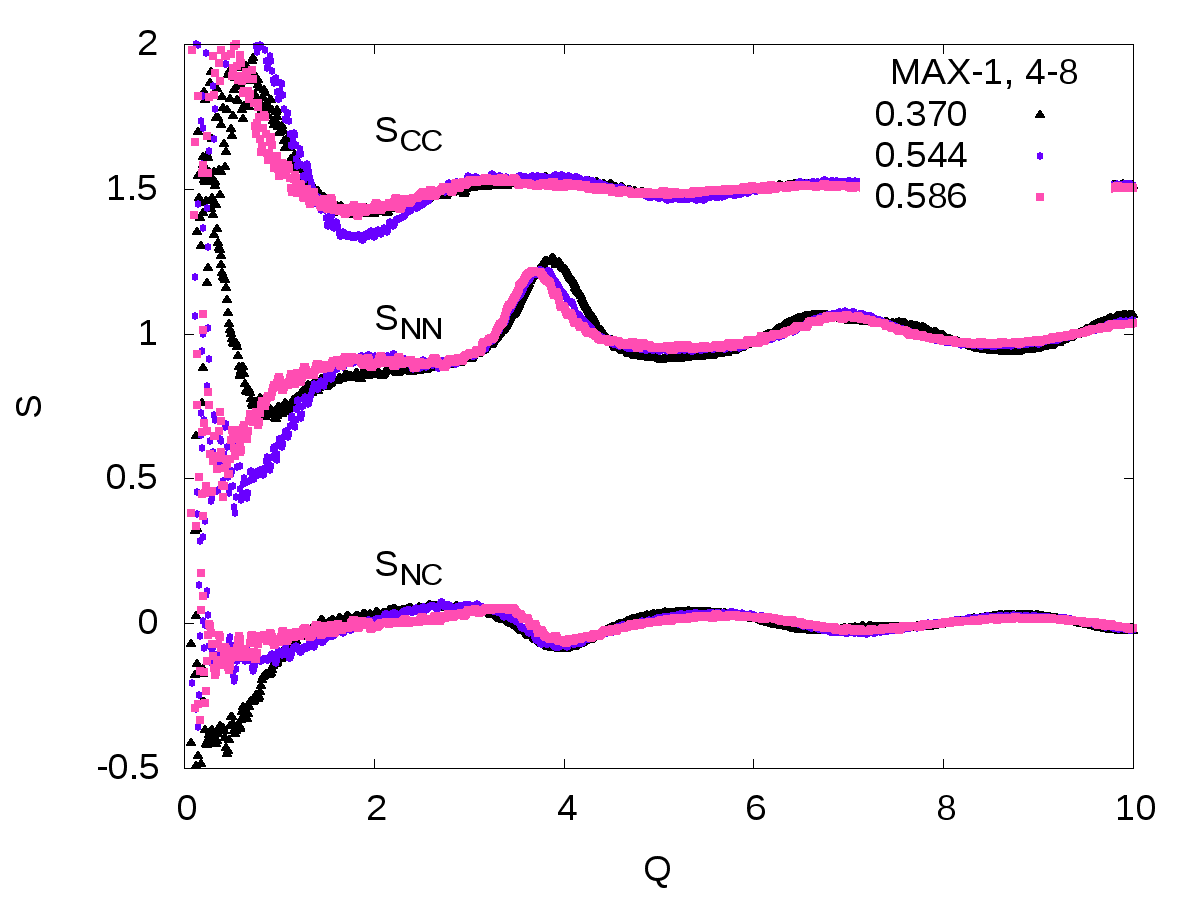}}
\subfigure[]{\includegraphics[width=0.48\textwidth]{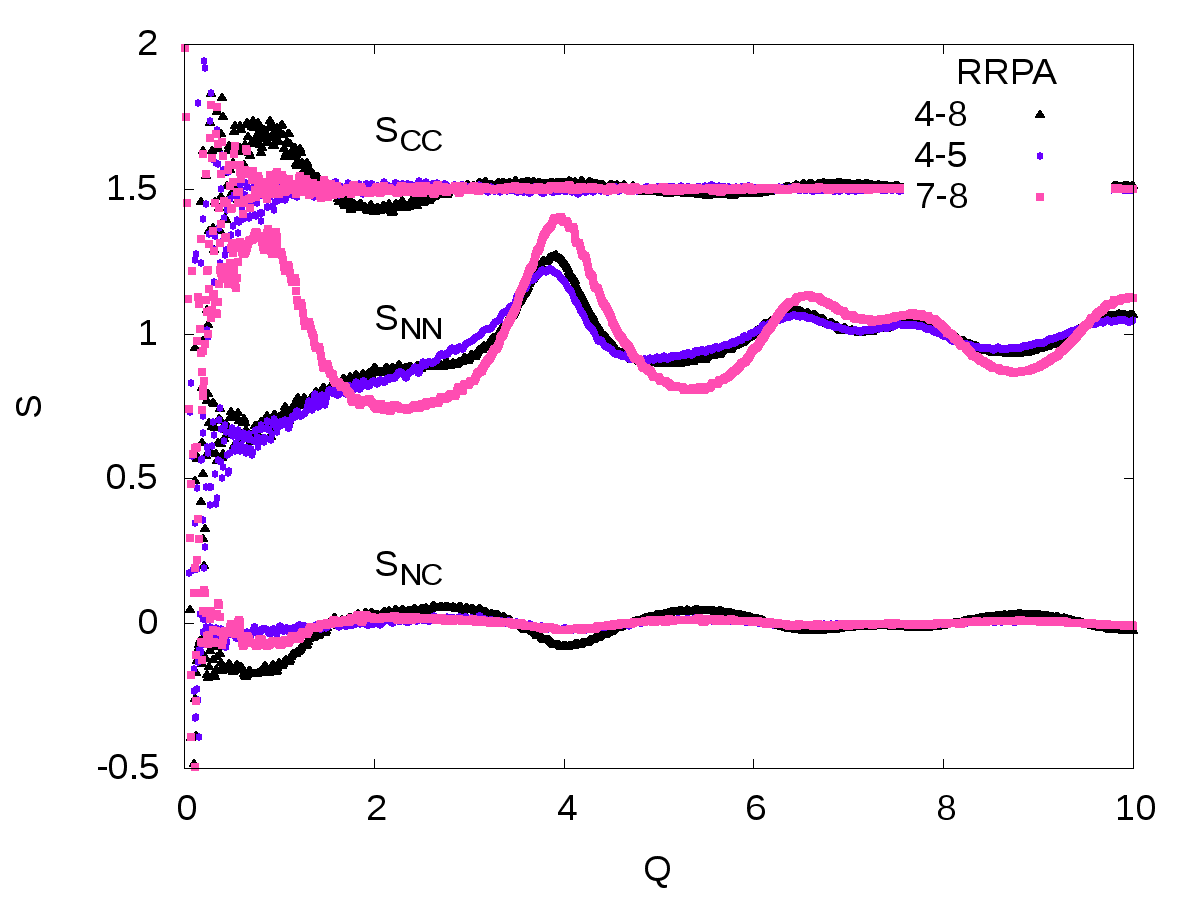}}
 \caption{Bhatia-Thornton partial structure factors for a) 4-5, b) 7-8, c) 4-8 for MAX-1 aggregates and d) RRPA aggregate. $S_{CC}$ oscillates around 1 but has been shifted to 1.5 for clarity's sake.\label{Bhathia_petQ_SI}}
\end{figure}

\end{document}